\documentclass[times]{aastex62}

\usepackage{amsmath}
\usepackage{xspace}
\usepackage{makecell}
\usepackage[figuresleft]{rotating}
\usepackage{placeins}

\usepackage[encapsulated]{CJK}
\usepackage{ucs}
\usepackage[utf8x]{inputenc}
\newcommand{\cntext}[1]{\begin{CJK}{UTF8}{gbsn}#1\end{CJK}}

\newcommand{\eg}{e.g.,~}
\newcommand{\detg}{{\sqrt{-g}}}
\newcommand{\etal}{\textit{et al.}\xspace}
\newcommand{\del}{\partial}

\newcommand{\bhac}{\texttt{BHAC}\xspace}
\newcommand{\hamr}{\texttt{H-AMR}\xspace}
\newcommand{\cosmos}{\texttt{Cosmos++}\xspace}
\newcommand{\harm}{\texttt{iharm3D}\xspace}
\newcommand{\koral}{\texttt{KORAL}\xspace}
\newcommand{\harmtwod}{\texttt{HARM}\xspace}

\newcommand{\athena}{\texttt{Athena}\xspace}
\newcommand{\athenapp}{\texttt{Athena++}\xspace}

\newcommand{\bhoss}{\texttt{BHOSS}\xspace}
\newcommand{\echo}{\texttt{ECHO}\xspace}
\newcommand{\nobleharmthreed}{\texttt{HARM3d}\xspace}
\newcommand{\nobleharm}{\texttt{HARM-Noble}\xspace}
\newcommand{\ham}{\texttt{HAM}\xspace}
\newcommand{\IGM}{\texttt{IllinoisGRMHD}\xspace}
\newcommand{\cyverse}{\texttt{cyverse}\xspace}

\received{April 5, 2019}
\revised{May 17, 2019}
\accepted{May 28, 2019}

\shorttitle{Event Horizon Code Comparison}
\shortauthors{GRMHD community and the EHTC}

\begin{document}

\title{The Event Horizon General Relativistic Magnetohydrodynamic Code Comparison Project}

\correspondingauthor{Oliver Porth}
\email{o.porth@uva.nl}

\author[0000-0002-4584-2557]{Oliver Porth}
\affiliation{Anton Pannekoek Institute for Astronomy, University of Amsterdam, Science Park 904, 1098 XH, Amsterdam, The Netherlands}
\affiliation{Institut f{\"u}r Theoretische Physik, Goethe-Universit{\"a}t Frankfurt, Max-von-Laue-Stra{\ss}e 1, D-60438 Frankfurt am Main, Germany}

\author{Koushik Chatterjee}
\affiliation{Anton Pannekoek Institute for Astronomy, University of Amsterdam, Science Park 904, 1098 XH, Amsterdam, The Netherlands}

\author[0000-0002-1919-2730]{Ramesh Narayan}
\affiliation{Center for Astrophysics $\vert$ Harvard \& Smithsonian, 60 Garden Street, Cambridge, MA 02138, USA}
\affiliation{Black Hole Initiative at Harvard University, 20 Garden Street, Cambridge, MA 02138, USA}

\author[0000-0001-7451-8935]{Charles F. Gammie}
\affiliation{Department of Astronomy; Department of Physics; University of Illinois, Urbana, IL 61801 USA}

\author[0000-0002-8131-6730]{Yosuke Mizuno}
\affiliation{Institut f{\"u}r Theoretische Physik, Goethe-Universit{\"a}t Frankfurt, Max-von-Laue-Stra{\ss}e 1, D-60438 Frankfurt am Main, Germany}

\author{Peter Anninos}
\affiliation{Lawrence Livermore National Laboratory, Livermore, CA 94550, USA}

\author{John G.~Baker}
\affiliation{Gravitational Astrophysics Laboratory, NASA Goddard Space Flight Center, Greenbelt, MD 20771, USA}
\affiliation{Joint Space-Science Institute, University of Maryland, College Park, MD 20742, USA}

\author[0000-0002-7834-0422]{Matteo Bugli}
\affiliation{IRFU/D\'epartement d’Astrophysique, Laboratoire AIM, CEA/DRF-CNRS-Universit\'e Paris Diderot, CEA-Saclay F-91191, France}	

\author[0000-0001-6337-6126]{Chi-kwan Chan}
\affiliation{Steward Observatory and Department of Astronomy, University of Arizona, 933 N. Cherry Ave., Tucson, AZ 85721, USA}
\affiliation{Data Science Institute, University of Arizona, 1230 N. Cherry Ave., Tucson, AZ 85721, USA}

\author[0000-0002-2685-2434]{Jordy Davelaar}
\affiliation{Department of Astrophysics, Institute for Mathematics, Astrophysics and Particle Physics (IMAPP), Radboud University, P.O. Box 9010, 6500 GL Nijmegen, The Netherlands}

% ...

\collaboration{(The Event Horizon Telescope Collaboration)}

\begin{abstract}
Recent developments in compact object astrophysics, especially
the discovery of merging neutron stars by LIGO, the
imaging of the black hole in M87 by the Event
Horizon Telescope (EHT) and high precision astrometry of the Galactic Center at close to the event horizon scale by the GRAVITY experiment 
motivate the development of numerical source models
that solve the equations of general relativistic magnetohydrodynamics (GRMHD).  
Here we compare GRMHD solutions for the evolution of a magnetized accretion 
flow where turbulence is promoted by the magnetorotational instability from a 
set of nine GRMHD codes:  \athenapp, \bhac, \cosmos, \echo, \hamr, \harm, \nobleharm, \IGM and \koral.
Agreement between the codes improves as
resolution increases, as measured by a consistently applied, specially
developed set of code performance metrics.   We conclude that the 
community of GRMHD codes is mature, capable, and consistent on
these test problems.  
\end{abstract}


\section{Introduction} \label{sec:intro}
Fully general relativistic models of astrophysical sources are in high demand, not only since the discovery of gravitational waves emitted from merging stellar mass black holes \citep{AbbottAbbott2016a}.  
The need for an accurate description of the interplay between strong gravity, matter and electromagnetic fields is further highlighted by the recent detection of electromagnetic counterpart radiation to the coalescence of a neutron star binary \citep{Abbott2017_etal}.   
Our own effort is motivated by the Event Horizon Telescope (EHT) project, which allows direct imaging of hot, luminous plasma near a black hole event horizon \citep{DoelemanWeintroub2008,GoddiFalcke2017,CollaborationAkiyamaEtAl2019}.  The main targets of the EHT are the black hole at the center of the Milky Way (also known by the name of the associated radio source, Sgr A*, e.g. \cite{LuKrichbaumEtAl2018}) and the black hole at the center of the galaxy M87 with associated central radio source M87* \citep{doeleman2012, AkiyamaLuEtAl2015}.  
In order to extract information on the dynamics of the plasma that lies within a few $GM/c^2$ of the event horizon ($M \equiv$ black hole mass) as well as information about the black hole's gravitational field, it is necessary to develop models of the accretion flow, associated winds and relativistic jets, and the emission properties in each of the components.

Earlier semi-analytic works \citep{NarayanYi1995, NarayanMahadevanEtAl1998a, YuanMarkoff2002a} have provided with the general parameter regime of the galactic center by exploiting spectral information.  For example, \cite{MahadevanQuataert1997} demonstrated that the electrons and ions are only weakly collisionally coupled and unlikely in thermal equilibrium.  Also key parameters like the accretion rate are typically estimated based on simple one-dimensional models \citep{MarroneMoran2007}.  
They have solidified the notion that the accretion rate in Sgr A* is far below the Eddington limit $\dot{M}_{\rm Edd}=L_{\rm Edd}/(0.1 c^2)\simeq 2\, M/(10^8 M_{\odot})\, M_\odot \rm yr^{-1}$ where $L_{\rm Edd}=4\pi G M c/\sigma_{\rm T}$ is the Eddington luminosity (with $\sigma_{\rm T}$ being the Thomson electron cross section).  
New observational capabilities like mm- and IR- interferometry, as provided by the EHT and GRAVITY \citep[][]{GravityCollaborationAbuterEtAl2018} collaborations  now allow to go much closer to the source which requires a description of general relativistic and dynamical (hence time-dependent) effects.  

The most common approach to dynamical relativistic source modeling uses the ideal general relativistic magnetohydrodynamic (GRMHD) approximation.  It is worth reviewing the nature and quality of the two approximations inherent in the GRMHD model.  First, the plasma is treated as a fluid rather than a collisionless plasma.  Second, the exchange of energy between the plasma and the radiation field is neglected. 

The primary EHT sources Sgr A* and M87* fall in the class of low-luminosity active galactive nuclei (AGN) and accrete with $\dot{M}/\dot{M}_{\rm Edd}\lesssim 10^{-6}$ \citep{MarroneMoran2007} and $\dot{M}/\dot{M}_{\rm Edd}\lesssim 10^{-5}$ \citep{KuoAsada2014} far below the Eddington limit.  In both cases the accretion flow is believed to form an optically thin disk that is geometrically thick and therefore has temperature comparable to the virial temperature (see \citealt{YuanNarayan2014} for a review).   The plasma is at sufficiently high temperature and low density that it is collisionless: ions and electrons can travel $\gg GM/c^2$ along magnetic field lines before being significantly deflected by Coulomb scattering, while the effective mean free path perpendicular to field lines is the gyroradius, which is typically $\ll GM/c^2$.  A rigorous description of the accreting plasma would thus naively require integrating the Boltzmann equation at far greater expense than integrating the fluid equations.   Full Boltzmann treatments of accretion flows are so far limited to the study of localized regions within the source \citep[e.g.][]{Hoshino2015, KunzStone2016}.  Global models that incorporate nonideal effects using PIC-inspired closure models suggest, however, that the effects of thermal conduction and pressure anisotropy (viscosity) are small \citep{Chandra2015, ChandraFoucart2017, FoucartChandra2017}, and thus that one would not do too badly with an ideal fluid prescription.

For Sgr A*, radiative cooling is negligible \citep{Dibi+2012}.   For M87 radiative cooling is likely important \citep[e.g.][]{Moscibrodzka2011,RyanRessler2018, ChaelNarayan2018}.  Cooling through the synchrotron process and via inverse Compton scattering primarily affects the electrons, which are weakly coupled to the ions and therefore need not be in thermal equilibrium with them.  To properly treat the radiation field for the non-local process of Compton scattering requires solving the Boltzmann equation for photons (the radiative transport equation) in full \citep[e.g.][]{RyanDolence2015} or in truncated form with ``closure''. 
A commonly employed closure is to assume the existence of a frame in which the radiation field can be considered isotropic, yielding the ``M1'' closure \citep{levermore84} for which a general relativistic derivation is shown for example in \cite{sadowski13}.  As expected, the computational demands imposed by the additional ``radiation fluid'' are considerable.  
It may however be possible to approximate the effects of cooling by using a suitable model to assign an energy density (or temperature) to the electrons \citep{MoscibrodzkaFalcke2016a}.  Again an ideal fluid description, which automatically satisfies energy, momentum, and particle number conservation laws is not a bad place to start.   

It is possible to write the GRMHD equations in conservation form.  This enables one to evolve the GRMHD equations using techniques developed to evolve other conservation laws such as those describing nonrelativistic fluids and magnetized fluids.   Over the last decades, a number of GRMHD codes have been developed, most using conservation form, and applied to a large variety of astrophysical scenarios \citep{Hawley84a, KoideShibata1999, DeVilliers03a, Gammie2003, Baiotti04, Duez05MHD0, Anninos2005, Anton05, Mizuno06, DelZanna2007, Giacomazzo:2007ti, TchekhovskoyNarayan2011, RadiceRezzollaProceedings2013, Radice2013b, sadowski14, McKinneyTchekhovskoy2014, Etienne:2015cea, White2016, Zanotti2015, Meliani2016, LiskaHesp2018}.
  
Despite the conceptual simplicity of the MHD equations, the non-linear properties which allow for shocks and turbulence render their treatment difficult.  
This is particularly true for the case study considered here:  in state-of-the-art simulations of black hole accretion, angular momentum transport is provided by Maxwell- and Reynolds- stresses of the orbiting plasma. MHD turbulence is seeded by the magnetorotational instability (MRI) in the differentially rotating disk \citep{Balbus1991,BalbusHawley1998} and gives rise to chaotic behavior which hinders strict convergence of the solutions.  
Nonetheless, it can be demonstrated that certain global properties of the solutions exhibit signs of convergence.  

Another challenge is posed by the ``funnel'' region near the polar axis where low angular momentum material will be swallowed up by the black hole \citep[e.g.][]{mckinney2006}.  The strong magnetic fields that permeate the black hole (held in place by the equatorial accretion flow) are able to extract energy in the form of Poynting flux from a rotating black hole, giving rise to a relativistic ``jet'' \citep{BlandfordZnajek1977,TakahashiNittaEtAl1990}.  The ensuing near-vacuum and magnetic dominance are traditionally difficult to handle for fluid-type simulations, but analytic calculations \citep[e.g.][]{PuNakamura2015} and novel kinetic approaches \citep{ParfreyPhilippovEtAl2019} can be used to validate the flow in this region.  

Due to their robustness when dealing e.g. with super-sonic jet outflows, current production codes typically employ high-resolution shock-capturing schemes in a finite-volume or finite-difference discretization  \citep{Font2008, Rezzolla_book:2013, 2015LRCA....1....3M}.  However, new schemes, for example based on discontinuous Galerkin finite-element methods, are continuously being developed \citep{AnninosBryant2017, FambriDumbser2018}.  

Given the widespread and increasing application of GRMHD simulations, it is critical for the community to evaluate the underlying systematics due to different numerical treatments and demonstrate the general robustness of the results.  
Furthermore, at the time of writing, all results on the turbulent black hole accretion problem under consideration are obtained without explicitly resolving dissipative scales in the system (called the implicit large eddy simulation (ILES) technique).  Hence differences are in fact expected to prevail even for the highest resolutions achievable.  
Quantifying how large the systematic differences are is one of the main objectives in this first comprehensive code comparison of an accreting black hole scenario relevant to the EHT science case.   
This work has directly informed the generation of the simulation library utilized in the modeling of the EHT 2017 observations \citep{CollaborationAkiyamaEtAl2019d}.  
We use independent production codes that differ widely in their algorithms and implementation.  In particular, the codes that are being compared are: \athenapp{}, \bhac, \cosmos, \echo, \hamr, \harm, \nobleharm, \IGM and \koral.  These codes are described further in section \ref{sec:codes} below.  

The structure of this paper is as follows: in section \ref{sec:equations} we introduce the GRMHD system of equations, define the notation used throughout this paper and briefly discuss the astrophysical problem.  Code descriptions with references are given in section \ref{sec:codes}.  The problem setup is described in section \ref{sec:setup}, where code-specific choices are also discussed.  Results are presented in section \ref{sec:results} and we close with a discussion in section \ref{sec:discussion}.  

\section{Astrophysical Problem}\label{sec:problem}

Let us first give a brief overview of the problem under investigation and the main characteristics of the accretion flow.  

\subsection{GRMHD system of equations} \label{sec:equations}

For clarity and consistency of notation, we here give a brief summary 
of the ideal GRMHD equations in a coordinate basis $(t, x^i)$ with 
four-metric $(g_{\mu\nu})$ and metric determinant $g$.  As customary, Greek indices run through $[0,1,2,3]$ while Roman indices span $[1,2,3]$.  The equations are solved in (geometric) code units (i.e. setting the gravitational constant and speed of light to unity $G=c=1$) where, compared to the standard Gauss cgs system, the factor $1/\sqrt{4\pi}$ is further absorbed in the definition of the magnetic field.  Hence in the following, we will report times and radii simply in units of the mass of the central object $\rm M$. 

The equations describe particle number conservation;
\begin{equation}\label{PARTNOCONS}
\partial_t (\detg \rho u^t) = -\del_i (\detg \rho u^i)
\end{equation}
where $\rho$ is the rest-mass density and $u^\mu$ is the
four-velocity; conservation of energy-momentum:
\begin{equation}\label{ENER_MOM}
\partial_t \left( \detg \, {T^t}_\nu \right)
	= -\partial_i \left( \detg \, {T^i}_\nu
	\right) + \detg \, {T^\kappa}_\lambda \, {\Gamma^\lambda}_{\nu \kappa},
\end{equation}
where ${\Gamma^\lambda}_{\nu \kappa}$ is
the metric connection; the definition of the stress-energy tensor for ideal
MHD:
\begin{equation}\label{TMUNUMHD}
T^{\mu\nu}_\mathrm{MHD} = (\rho + u + p + b^2) u^\mu u^\nu 
+ \left(p + \frac{1}{2} b^2\right) g^{\mu\nu} - b^\mu b^\nu
\end{equation}
where $u$ is the fluid internal energy, $p$ the fluid pressure, and $b^\mu$
is the magnetic field four-vector; the definition of $b^\mu$
in terms of magnetic field variables $B^i$, which are commonly
used as ``primitive'' or dependent variables:
\begin{equation}\label{BTSOL}
b^t = B^i u^\mu g_{i\mu},
\end{equation}
\begin{equation}\label{BISOL}
b^i = (B^i + b^t u^i)/u^t;
\end{equation}
and the evolution equation for $B^i$, which follows from the
source-free Maxwell equations:
\begin{equation}\label{INDUCTION}
\del_t (\detg B^i) = -\del_j (\detg \, (b^j u^i - b^i u^j)).
\end{equation}
The system is subject to the additional no-monopoles constraint
\begin{equation}\label{DIVB}
\frac{1}{\detg} \del_i (\detg\, B^i) = 0,
\end{equation}
which follows from the time component of the homogeneous 
Maxwell equations.  For closure, we adopt the equation of state of an ideal gas.  This takes the form $p = (\hat{\gamma} - 1) u$, where $\hat{\gamma}$ is the adiabatic index.  
More in-depth discussions of the ideal GRMHD system of equations can be found in the various publications of the authors, e.g. \cite{Gammie2003, Anninos2005, DelZanna2007, White2016, PorthOlivares2017}.  

To establish a common notation for use in this paper, we note the following definitions:  the magnetic field strength as measured in the fluid-frame is given by $B:=\sqrt{b^\alpha b_\alpha}$.  This leads to the definition of the magnetization $\sigma:=B^2/\rho$ and the plasma-$\beta$ $\beta:=2p/B^2$.  In addition, we denote with $\Gamma$ the Lorentz factor with respect to the normal observer frame.

\subsection{The magnetised black hole accretion problem}

\begin{figure}[htbp]
\begin{center}
\includegraphics[height=7.5cm]{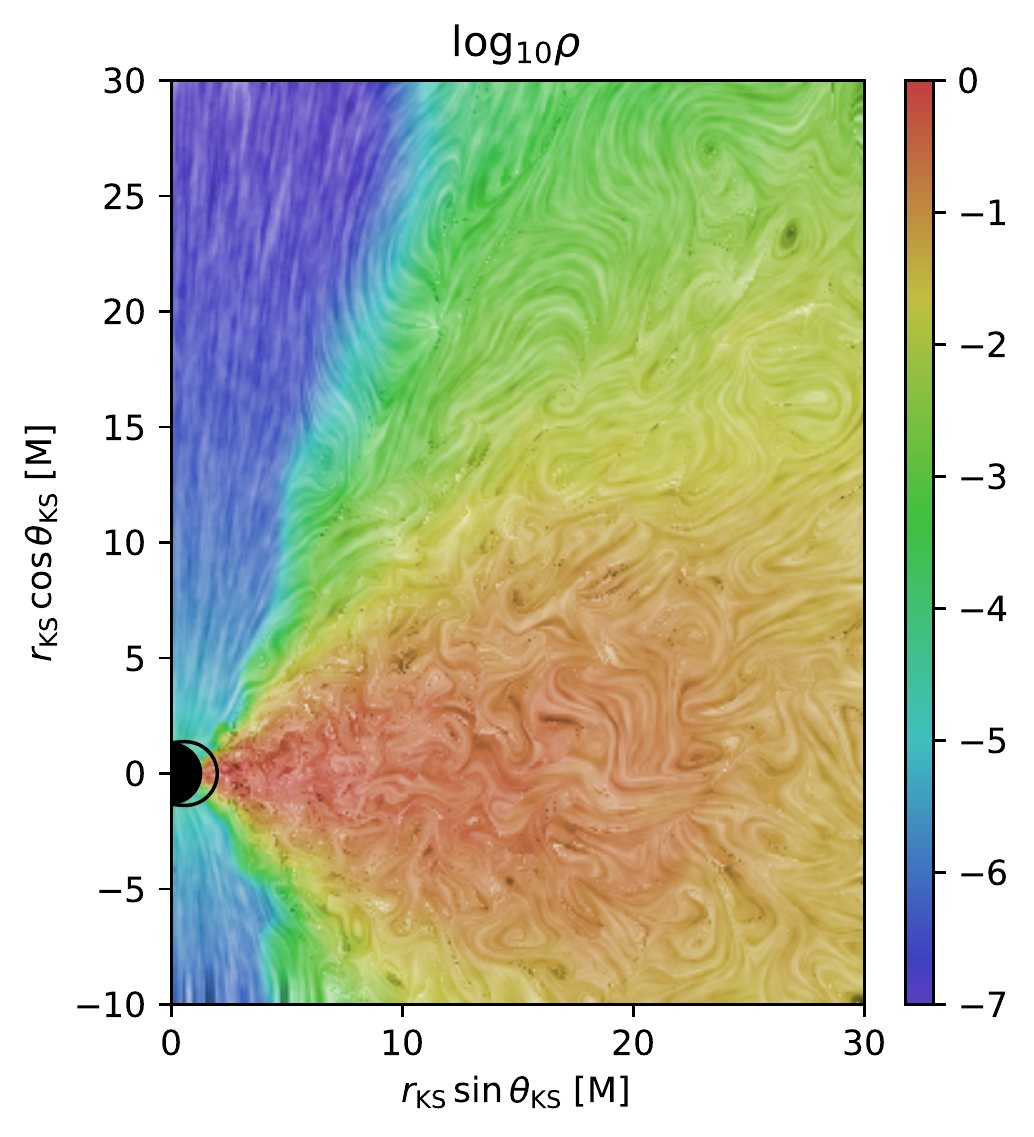}
\includegraphics[height=7.5cm]{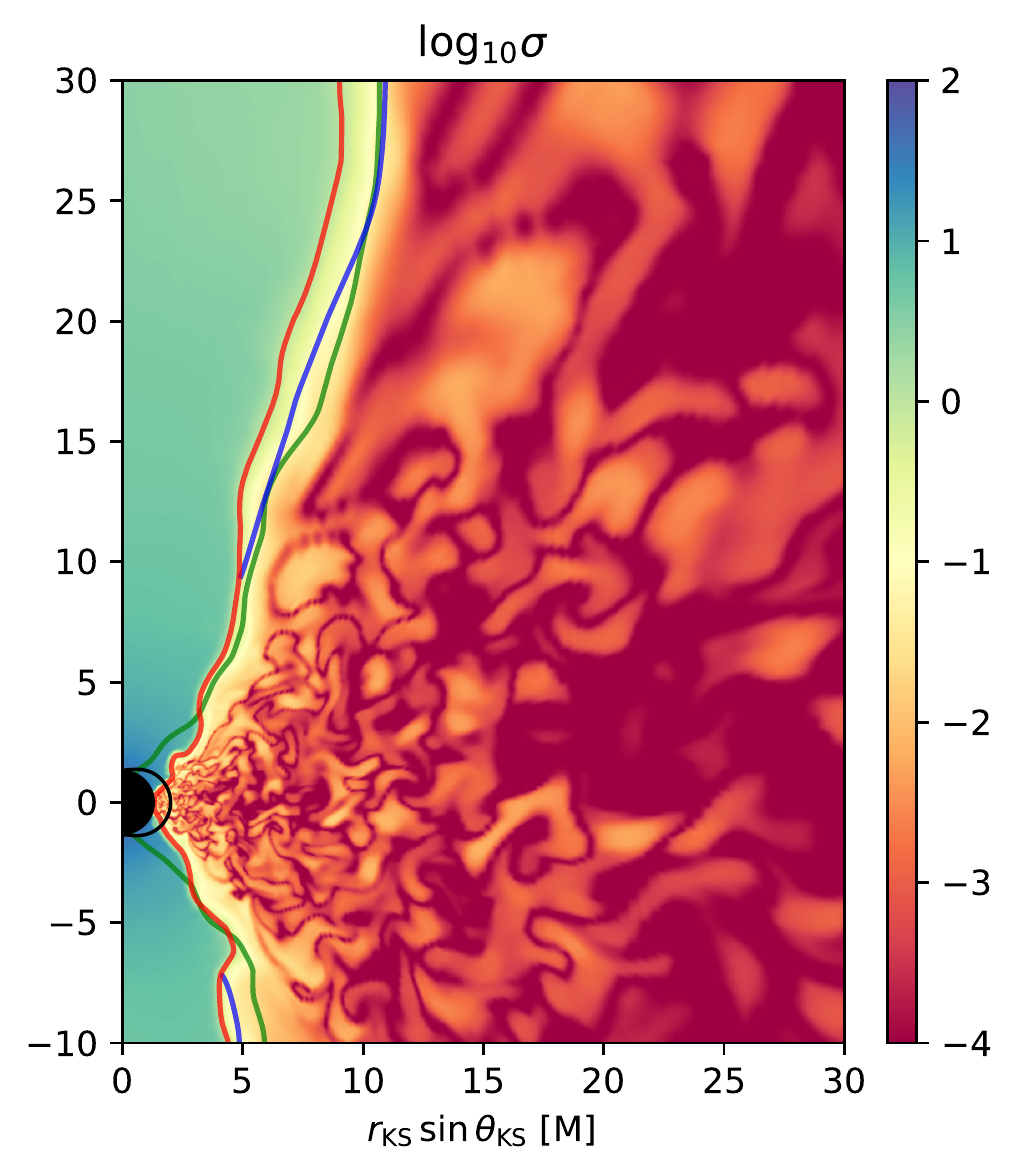}
\includegraphics[height=7.5cm]{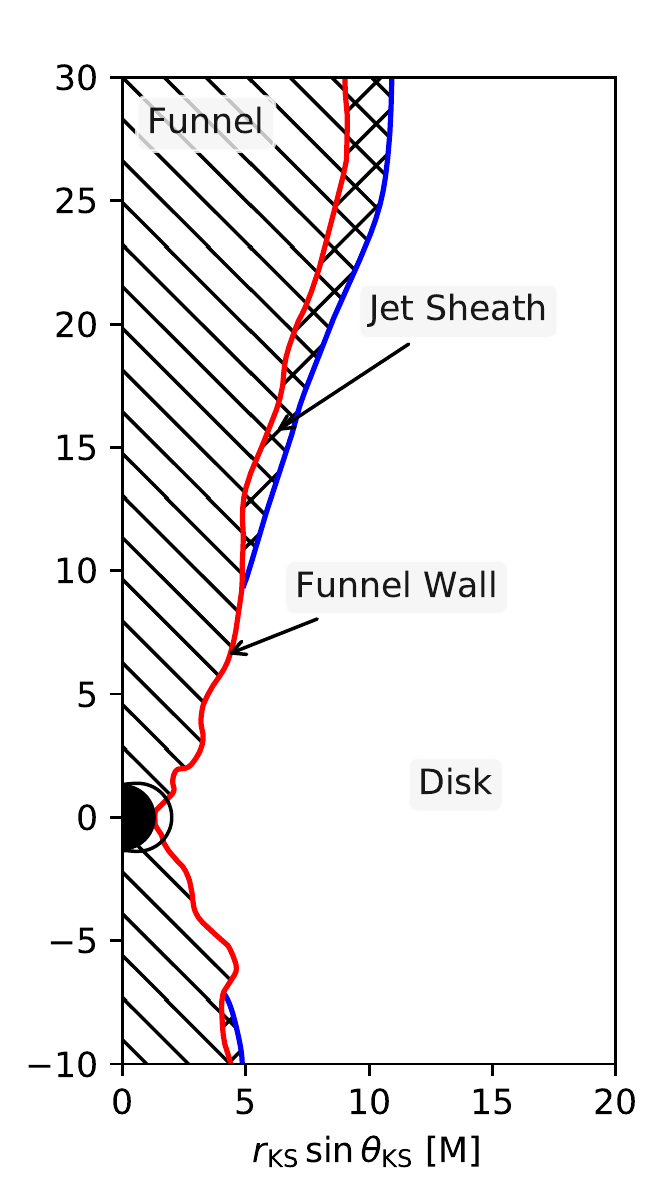}
\caption{Views of the radiatively inefficient turbulent black hole accretion problem at $t_{\rm KS}=10\, 000 \, \rm M$ against the Kerr-Schild coordinates (subscript KS).  
\textit{Left:} logarithmic rest-frame density (hue) and rendering of the magnetic field structure using line-integral convolution (luminance), showing ordered field in the funnel region and turbulence in the disk.  \textit{Center:} the logarithm of the magnetization with colored contours indicating characteristics of the flow.  The magnetized funnel is demarcated by $\sigma=1$, (red lines), the disk is indicated by $\beta=1$ (green lines) and the geometric Bernoulli criterion ($u_t=-1$) is given as blue solid line in the region outside of the funnel.
\textit{Right:} schematic of the main components.  In these plots, the black hole horizon is the black disk and the ergosphere is shown as black contour.
The snapshot was obtained from a simulation with \bhac.
}
\label{fig:schematic}
\end{center}
\end{figure}

We will now discuss the most important features of the problem at hand and introduce the jargon that has developed over the years.  A schematic overview with key aspects of the accretion flow is given in Figure \ref{fig:schematic}.  

At very low Eddington rate $\dot{M}\sim10^{-6} \dot{M}_{\rm Edd}$, the radiative cooling timescale becomes longer than the accretion timescale.  In such radiatively inefficient accretion flows (RIAF), dynamics and radiation emission effectively decouple.  For the primary EHT targets, Sgr A* and M87*, this is a reasonable first approximation and hence purely non-radiative GRMHD simulations neglecting cooling can be used to model the data.  
For a RIAF, the protons assume temperatures close to the virial one which leads to an extremely ``puffed-up'' appearance of the tenuous accretion \textit{disk}.  

In the polar regions of the black hole, plasma is either sucked in or expelled in an outflow, leaving behind a highly magnetized region called the \textit{funnel}.  
The magnetic field of the funnel is held in place by the dynamic and static pressure of the disk. 
Since in ideal MHD, plasma cannot move across magnetic field lines (due to the frozen-in condition), there is no way to re-supply the funnel with material from the accretion disk and hence the funnel would be completely devoid of matter if no pairs were created locally.  In state-of-the-art GRMHD calculations, this is the region where numerical \textit{floor} models inject a finite amount of matter to keep the hydrodynamic evolution viable.

The general morphology is separated into the components of 
\textit{i)} the \textit{disk} which contains the bound matter 
\textit{ii)} the evacuated \textit{funnel} extending from the polar caps of the black hole and the
\textit{iii)} \textit{jet sheath} which is the remaining outflowing matter.
In Figure \ref{fig:schematic}, the regions are indicated by commonly used discriminators in a representative simulation snapshot: the blue contour shows the bound/unbound transition defined via the geometric Bernoulli parameter $u_t = -1$
\footnote{There are various ways to define the unbound material in the literature.  For example, \cite{McKinneyGammie2004} used the \textit{geometric} Bernoulli parameter $u_t$.  The \textit{hydrodynamic} Bernoulli parameter used for example by \cite{MoscibrodzkaFalcke2016} is given by $-h u_t$ where $h=c^2+p+u$ denotes the specific enthalpy. \cite{ChaelNarayan2018,NarayanSadowski2012} included also \textit{magnetic} and \textit{radiative} contributions. Certainly the geometric and hydrodynamic prescriptions underestimate the amount of outflowing material.}
, the red contour demarcates the funnel-boundary $\sigma=1$ and the green contour the equipartition $\beta=1$ which is close to the bound/unbound line along the disk boundary (consistent with \cite{McKinneyGammie2004}).  
In \cite{McKinneyGammie2004} also a disk-\textit{corona} was introduced for the material with $\beta\in[1,3]$, however as this choice is arbitrary, there is no compelling reason to label the corona as separate entity in the RIAF scenario.  

Since plasma is evacuated within the funnel, it has been suggested that unscreened electric fields in the charge starved region can lead to particle acceleration which might fill the magnetosphere via a pair cascade \citep[e.g.][]{BlandfordZnajek1977,BeskinIstomin1992,HirotaniOkamoto1998,LevinsonRieger2011,BroderickTchekhovskoy2015}.  The most promising alternative mechanism to fill the funnel region is by pair creation via $\gamma\gamma$ collisions of seed photons from the accretion flow itself \citep[e.g.][]{StepneyGuilbert1983,Phinney1995}.  Neither of these processes is included in current state of the art GRMHD simulations, however the efficiency of pair formation via $\gamma\gamma$ collisions can be evaluated in post-processing as demonstrated by \cite{Moscibrodzka2011}.  

Turning back to the morphology of the RIAF accretion, Figure \ref{fig:schematic}, one can see that between evacuated funnel demarcated by the \textit{funnel wall} (red) and bound disk material (blue), there is a strip of outflowing material often also referred to as the \textit{jet sheath} \citep{DexterMcKinneyEtAl2012,MoscibrodzkaFalcke2013,MoscibrodzkaFalcke2016,DavelaarMoscibrodzkaEtAl2018}.  
As argued by \cite{HawleyKrolik2006}, this flow emerges as plasma from the disk is driven against the centrifugal barrier by magnetic and thermal pressure (which coined the alternative term \textit{funnel wall jet} for this region). 
In current GRMHD based radiation models as utilized e.g. in \cite{CollaborationAkiyamaEtAl2019d}, as the density in the funnel region is dominated by the artificial floor model, the funnel is typically excised from the radiation transport.  The denser region outside the funnel wall remains which naturally leads to a limb-brightened structure of the observed M87 ``jet'' at radio frequencies \citep[e.g.][]{MoscibrodzkaFalcke2016,ChaelNarayan2018,2019arXiv190610065D}.  In the mm-band \citep[][]{CollaborationAkiyamaEtAl2019}, the horizon scale emission originates either from the body of the disk or from the region close to the funnel wall, depending on the assumptions on the electron temperatures \citep[][]{CollaborationAkiyamaEtAl2019d}.

In RIAF accretion, a special role is played by the horizon penetrating magnetic flux $\Phi_{\rm BH}$: normalized by the accretion rate $\phi:=\Phi_{\rm BH}/\sqrt{\dot{M}}$, it was shown that a maximum for the magnetic flux $\phi_{\rm max}\approx 15$ (in our system of units) exists which depends only mildly on black hole spin, but somewhat on the disk scale height (with taller disks being able to hold more magnetic flux, \citealt{TchekhovskoyMcKinney2012}). 
Once the magnetic flux reaches $\phi_{\rm max}$, accretion is brought to a near-stop by the accumulation of magnetic field near the black hole \citep[][]{TchekhovskoyNarayan2011,McKinneyTchekhovskoy2012} leading to a fundamentally different dynamic of the accretion flow and maximal energy extraction via the \cite{BlandfordZnajek1977} process.  This state is commonly referred to as Magnetically Arrested Disk (MAD, \citealt{Bisnovatyi-KoganRuzmaikin1976, NarayanIgumenshchev2003}) to contrast with the Standard and Normal Evolution (SANE) where accretion is largely unaffected by the black hole magnetosphere (here $\phi\sim \rm few$).  
While the MAD case is certainly of great scientific interest, in this initial code comparison we focus on the SANE case for two reasons:  \textit{i)} the SANE case is already extensively discussed in the literature and hence provides the natural starting point \textit{ii)} the MAD dynamics poses additional numerical challenges (and remedies) which render it ill-suited to establish a baseline agreement of GMRHD accretion simulations.

\section{Code descriptions} \label{sec:codes}

In this section, we give a brief, alphabetically ordered overview of the codes participating in this study, with notes on development history and target applications.  Links to public release versions are provided, if applicable. 

\subsection{\athenapp{}}

\athenapp{} is a general-purpose finite-volume astrophysical fluid dynamics framework, based on a complete rewrite of \athena{} \citep{2008ApJS..178..137S}. It allows for adaptive mesh refinement in numerous coordinate systems, with additional physics added in a modular fashion. It evolves magnetic fields via the staggered-mesh constrained transport algorithm of \citet{2005JCoPh.205..509G} based on the ideas of \citet{evans1988}, exactly maintaining the divergence-free constraint. The code can use a number of different time integration and spatial reconstruction algorithms. \athenapp{} can run GRMHD simulations in arbitrary stationary spacetimes using a number of different Riemann solvers \citep{White2016}.   Code verification is described in \cite{White2016} and a public release can be obtained from \url{ https://github.com/PrincetonUniversity/athena-public-version}

\subsection{\bhac}

The BlackHoleAccretionCode (\bhac) first presented by \cite{PorthOlivares2017} is a multidimensional GRMHD module for the MPI-AMRVAC framework \citep{Keppens2012718,PorthXia2014,XiaTeunissenEtAl2018}. \bhac has been designed to solve the equations of general-relativistic magnetohydrodynamics in arbitrary spacetimes/coordinates and exploits adaptive mesh refinement techniques with an oct-tree block-based approach.  The algorithm is centred on second order finite volume methods and various schemes for the treatment of the magnetic field update have been implemented, on ordinary and staggered grids. More details on the various $\mathbf{\nabla\cdot B}$ preserving schemes and their implementation in \bhac can be found in \cite{OlivaresPorth2018,2019arXiv190610795O}.  
Originally designed to study black hole accretion in ideal MHD, \bhac has been extended to incorporate nuclear equations of state, neutrino leakage, charged and purely geodetic test particles \citep{BacchiniRipperda2018a,2019ApJS..240...40B} and non-black hole fully numerical metrics.  In addition, a non-ideal resistive GRMHD module is under development \cite[e.g.][]{2019MNRAS.485..299R,2019arXiv190707197R}.  
Code verification is described in \cite{PorthOlivares2017} and a public release version can be obtained from \url{https://bhac.science}.

\subsection{\cosmos}

\cosmos \citep{Anninos2005,Fragile2012,Fragile2014} is a parallel, multidimensional, fully covariant, modern object-oriented (C++) radiation hydrodynamics and MHD code for both Newtonian and general relativistic astrophysical and cosmological applications. Cosmos++ utilizes unstructured meshes with adaptive ($h$-) refinement \citep{Anninos2005}, moving-mesh ($r$-refinement) \citep{Anninos2012}, and adaptive order ($p$-refinement) \citep{AnninosBryant2017} capabilities, enabling it to evolve fluid systems over a wide range of spatial scales with targeted precision. It includes numerous hydrodynamics solvers (conservative and non-conservative), magnetic fields (ideal and non-ideal), radiative cooling and transport, geodesic transport, generic tracer fields, and full Navier-Stokes viscosity \citep{FragileEtheridgeEtAl2018}. For this work, we utilize the High Resolution Shock Capturing scheme with staggered magnetic fields and Constrained Transport as described in \citet{Fragile2012}.
Code verification is described in \citet{Anninos2005}.

\subsection{\echo}
The origin of the \emph{Eulerian Conservative High-Order} (\texttt{ECHO}) code dates back to the year 2000 \citep{Londrillo2000,londrillo2004divergence}, when it was first proposed a shock-capturing scheme for classical MHD based on high-order finite-differences reconstruction routines, one-wave or two-waves Riemann solvers, and a rigorous enforcement of the solenoidal constraint for staggered electromagnetic field components (the \emph{Upwind Constraint Transport}, UCT). The GRMHD version of \echo used in the present paper is described in \cite{DelZanna2007} and preserves the same basic characteristics. Important extensions of the code were later presented for dynamical spacetimes \citep{BucciantiniDel-Zanna2011} and non-ideal Ohm equations \citep{BucciantiniDel-Zanna2013,2016MNRAS.460.3753D,Del-Zanna2018}. Specific recipes for the simulation of accretion tori around Kerr black holes can be found in \cite{BugliDel-Zanna2014,Bugli2018}. Further references and applications may be found at \url{www.astro.unifi.it/echo}.
Code verification is described in \cite{DelZanna2007}.  

\subsection{\hamr}
\hamr is a 3D GRMHD code which builds upon \harmtwod \citep{Gammie2003,Noble2006} and the public code \texttt{HARM-PI} (\href{https://github.com/atchekho/harmpi}{https://github.com/atchekho/harmpi}) 
and has been extensively rewritten to increase the code's speed and add new features \citep{LiskaHesp2018,ChatterjeeEtAl2019}. \hamr makes use of GPU acceleration in a natively developed hybrid CUDA-OpenMP-MPI framework with adaptive mesh refinement (AMR) and locally adaptive timestepping (LAT) capability. LAT is superior to the 'standard' hierarchical timestepping approach implemented in other AMR codes since the spatial and temporal refinement levels are decoupled, giving much greater speedups on logarithmic spaced spherical grids. These advancements bring GRMHD simulations with hereto unachieved grid resolutions for durations exceeding $10^5M$ within the realm of possibility.

\subsection{\nobleharm}

The \nobleharm code \citep{Gammie2003,Noble2006,Noble2009} is a
flux-conservative, high-resolution shock-capturing GRMHD code that
originated from the 2D GRMHD code called \harmtwod
\citep{Gammie2003,Noble2006} and the 3D version
(Gammie 2006, private communication) of the 2D Newtonian MHD code called \ham
\citep{guan2008,hamcode}.  Because of its shared history, \nobleharm is very similar to the \harm code.
\footnote{We use the name \nobleharm to differentiate this code from the other \harmtwod-related codes referenced herein.  
However, \nobleharm is more commonly referred to as \nobleharmthreed in papers using the code.  }
Numerous features and changes were made
from these original sources, though.  Some additions include piecewise
parabolic interpolation for the reconstruction of primitive variables
at cell faces and the electric field at the cell edges for the
constrained transport scheme, and new schemes for ensuring a physical
set of primitive variables is always recovered.  \nobleharm  was also written 
to be agnostic to coordinate and spacetime choices, making
it in a sense generally covariant.  This feature was most extensively demonstrated 
when dynamically warped coordinate systems were implemented 
\citep{ZilhaoNoble2014}, and time-dependent spacetimes were incorporated \citep[e.g.][]{Noble2012,Bowen2018}.

\subsection{\harm}

The \harm code \citep{Gammie2003,Noble2006,Noble2009}, (also J. Dolence, private communication) is a conservative, 3D GRMHD code. The equations are solved on a logically Cartesian grid in arbitrary coordinates. Variables are zone-centered, and the divergence constraint is enforced using the Flux-CT constrained transport algorithm of \cite{toth2000}. Time integration uses a second-order predictor-corrector step. Several spatial reconstruction options are available, although linear and WENO5 algorithms are preferred. Fluxes are evaluated using the local Lax-Friedrichs (LLF) method \citep{Rusanov1961}. Parallelization is achieved with a hybrid MPI/OpenMP domain decomposition scheme. \harm has demonstrated convergence at second order on a suite of problems in Minkowski and Kerr spacetimes \citep{Gammie2003}.  Code verification is described in \cite{Gammie2003} and a public release 2D version of the code  (which differs from that used here) can be obtained from \url{ http://horizon.astro.illinois.edu/codes}.

\subsection{\IGM}

The \IGM code~\citep{Etienne:2015cea} is an open-source,
vector-potential-based, Cartesian AMR code in the Einstein
Toolkit~\citep{Loffler:2011ay}, used primarily for dynamical-spacetime GRMHD
simulations. For the simulation presented here, spacetime and grid
dynamics are disabled. \IGM exists as a complete rewrite of (yet
agrees to roundoff-precision with) the long-standing GRMHD
code~\citep{Duez05MHD0,Etienne:2010ui,Etienne:2011re} developed by
the Illinois Numerical Relativity group to model
a large variety of dynamical-spacetime GRMHD phenomena (see, e.g.,
\cite{Etienne:2006am,Paschalidis:2011ez,Paschalidis:2012ff,Paschalidis:2014qra,Gold:2014dta}
for a representative sampling).
Code verification is described in \cite{Etienne:2015cea} and a public release 
can be obtained from \url{https://math.wvu.edu/\textasciitilde zetienne/ILGRMHD/}

\subsection{\koral}

\koral \citep{sadowski13,sadowski14} is a multidimensional GRMHD code
which closely follows, in its treatment of the MHD conservation
equations, the methods used in the \harm code
(\citealt{Gammie2003,Noble2006}, see description above). \koral can be
run with various first-order reconstruction schemes (Minmod,
Monotonized Central, Superbee) or with the higher-order PPM
scheme. Fluxes can be computed using either the LLF or the
HLL method. There is an option to include an artifical magnetic dynamo
term in the induction equation \citep{sadowski15a}, which is helpful
for running 2D axisymmetric simulations for long durations (not possible without this term since the MRI dies away in 2D).

Although \koral is suitable for pure GRMHD simulations such as the ones
discussed in this paper, the code was developed with the goal of
simulating general relativistic flows with radiation
\citep{sadowski13,sadowski14} and multi-species fluid. Radiation is
handled as a separate fluid component via a moment formalism using M1
closure \citep{levermore84}. A radiative viscosity term is included
\citep{sadowski15a} to mitigate ``radiation shocks'' which can
sometimes occur with M1 in optically thin regions, especially close to
the symmetry axis. Radiative transfer includes continuum opacity from
synchrotron free-free and atomic bound-free processes, as well as
Comptonization \citep{sadowski15b,sadowski17}. In addition to radiation density
and flux (which are the radiation moments considered in the M1
scheme), the code also separately evolves the photon number density,
thereby retaining some information on the effective temperature of the
radiation field.  Apart from radiation, \koral can handle
two-temperature plasmas, with separate evolution equations
(thermodynamics, heating, cooling, energy transfer) for the two
particle species \citep{sadowski17}, and can also evolve an isotropic
population of nonthermal electrons \citep{chael17}.
Code verification is described in \cite{sadowski14}.

\section{Setup description} \label{sec:setup}

As initial condition for our 3D GRMHD simulations, we consider a hydrodynamic equilibrium torus threaded by a single weak ($\beta\gg1$) poloidal magnetic field loop. The particular equilibrium torus solution with constant angular momentum $l:=u_\phi u^t$ was first presented by \cite{Fishbone76} and
\cite{Kozlowski1978} and is now a standard test for GRMHD simulations
\citep[see e.g.][]{Font02a,Zanotti03,Gammie2003,Anton05,Rezzolla_book:2013,White2016,PorthOlivares2017}.
Note that there exist two possible choices for the constant angular momentum, the alternative being $-u_\phi u_t$ which was used e.g. by \cite{Kozlowski1978} throughout most of their work.  
For ease of use, the coordinates $(r,\theta,\phi)$ noted in the following are standard spherical Kerr-Schild coordinates, however each code might employ different coordinates internally. 
To facilitate cross-comparison, we set the initial conditions in the torus close to those adopted by the more recent works of \cite{Shiokawa2012, White2016}. Hence the spacetime is a Kerr black hole (hereafter BH) with dimensionless spin parameter $a = 0.9375$. The inner radius of the torus is set to $r_{\rm in} = 6\,M$ and the density maximum is located at $r_{\rm max}=12\,M$.  
We adopt an ideal gas equation of state with an adiabatic index of $\hat{\gamma}=4/3$. A weak single magnetic field loop defined by the vector potential
\begin{align}
A_{\phi} \propto {\rm max} (\rho/\rho_{\rm max} - 0.2, 0) \,,
\end{align}
is added to the stationary solution. The field strength is set such that
$2 p_{\rm max}/(B^2)_{\rm max}=100$ 
, where global maxima of pressure
$p_{\rm max}$ and magnetic field strength $B^2_{\rm max}$ do not necessarily coincide. 
With this choice for initial magnetic field geometry and strength, the simulations are anticipated to progress according to the SANE regime, although this can only be verified a-posteriori.  

In order to excite the MRI inside the torus, the thermal pressure is perturbed by white noise of amplitude  $4\%$.  More precisely:
\begin{align}
  p^* = p\,(1+X_p)
\end{align}
and $X_p$ is a uniformly distributed random variable between $-0.02$ and $0.02$.  

To avoid density and pressures dropping to zero in the funnel region, floor models are customarily employed in fluid codes.  Since the  strategies differ significantly between implementations, only a rough guideline on the floor model was given.  
The following floor model was suggested: $\rho_{\mathrm{fl}} = 10^{-5}r^{-3/2}$ and $p_{\mathrm{fl}} =1/3\times10^{-7}\ r^{-5/2}$ which corresponds to the one used by \cite{McKinneyGammie2004} in the powerlaw indices.  Thus for all cells which satisfy $\rho \le \rho_{\mathrm{fl}}$, set $\rho=\rho_{\mathrm{fl}}$, in addition if $p \le p_{\mathrm{fl}}$, set $p = p_{\mathrm{fl}}$.  
It is well known that occasionally unphysical cells are encountered with e.g. negative pressures and high Lorentz factor in the funnel. 
For example, it can be beneficial to enforce that the Lorentz factor stay within reasonable bounds.  This delimits the momentum and energy of faulty cells and thus aids to keep the error localized.  The various failsafes and checks of each code are described in more detail in Section \ref{sec:setup-code}. 
The implications of the different choices will be discussed in Sections \ref{sec:t-phi-averages} and \ref{sec:discussion}.  

In terms of coordinates and gridding, we deliberately gave only loose guidelines.  The reasoning is two-fold: first, this way the results can inform on the \emph{typical} setup used with a particular code, thus allowing to get a feeling for how existing results compare.  The second reason is purely utilitarian, as settling for a common grid setup would incur extra work and likely introduce unnecessary bugs.  
For spherical setups which are the majority of the participants, a form of logarithmically stretched Kerr-Schild coordinates with optional warping/cylindrification in the polar region was hence suggested.

Similarly, the positioning and nature of the boundary conditions has been left free for each group with only the guideline to capture the domain of interest $r\in [r_h,50]$, $\theta\in[0,\pi]$, $\phi\in[0,2\pi]$.  The implications of the different choices will be discussed in Sections \ref{sec:disk} and \ref{sec:discussion}.  
Three rounds of resolutions are suggested in order to judge the convergence between codes.  These are low-res: $96^3$, mid-res: $128^3$ and high-res: $192^3$ where the resolution corresponds to the domain of interest mentioned above.  

To make sure the initial data is setup correctly in all codes, a stand-alone Fortran 90 program was supplied and all participants have provided radial cuts in the equatorial region.  This has proven to be a very effective way to validate the initial configuration.  

An overview of the algorithms employed for the various codes can be found in table \ref{tab:settings}.  Here, the resolutions $(\Delta r_{\rm p}, \Delta \theta_{\rm p}, \Delta \phi_{\rm p})$ refer to the proper distance between grid cells at the density maximum in the equatorial plane for the low resolution realization (typically $96 \times 96 \times 96$).  Specifically, we define 
\begin{align}
    \Delta r_p := \sqrt{g_{rr}(12 {\rm M}, \pi/2)}\, \Delta r
\end{align}
and analogue for the other directions.  For the two Cartesian runs, we report the proper grid-spacings at the same position (in the $xz-$plane) for the x,y and z-directions respectively and treat the $\Delta z_{\rm p}$ as representative for the out-of-plane resolution $\Delta \theta_{\rm p}$ in the following sections.  

\begin{table}[htp]
\caption{Algorithmic details of the code comparison runs for the low-resolution realizations. \\
The columns are: code name, order of the time-integration, approximate Riemann solver, spatial reconstruction scheme, scheme for magnetic field evolution, proper distance between grid-cells at the midplane and radial and meridional extents of the computational domain.  The azimuthal direction spans $[0,2\pi]$ in all cases. 
\label{tab:settings} 
}
\begin{center}
\scriptsize
  \begin{tabular}{cccccccc}
Code & Timestepper & Riemann s. & Rec. & Mag. field & $(\Delta r_{\rm p}, \Delta \theta_{\rm p}, \Delta \phi_{\rm p})$ & Domain: r & Domain: $\theta$\\
\hline \hline
  \athenapp & 2nd Order  & HLL & PPM & CT \citep{2005JCoPh.205..509G} & $(0.506,0.393,0.788)$ & $[1.152, 50]$& $[0, \pi]$\\
  
  \bhac & 2nd Order & LLF & PPM & UCT \citep{DelZanna2007} & $(0.402,0.295,0.788)$ & $[1.185,3333]$& $[0,\pi]$\\

  \bhac Cart. & 2nd Order & LLF & PPM & UCT \citep{DelZanna2007} & \makecell[tl]{Cartesian AMR \\ $(0.176,0.163,0.163)$} & See Sect.~\ref{sec:setup-bhac}.  & --\\
     
  \cosmos & SSPRK, 3rd Order & HLL & PPM & \citep{Fragile2012} & $(0.508,0.375,0.788)$ & $[1.25, 354]$ & $[0.070,3.072]$\\
  
  \echo & 3rd Order & HLL & PPM & UCT \citep{DelZanna2007} & $(0.460,0.382,0.752)$ & $[1.048,2500]$& $[0.060,3.082]$\\
  
  \hamr & 2nd Order & HLL & PPM & UCT \citep{2005JCoPh.205..509G} & $(0.523,0.379,0.785)$  &$[1.169,500]$ &$[0,\pi]$ \\
  
  \nobleharm & 2nd Order & LLF & PPM & \makecell[tl]{PPM+Flux-CT \\ \citep{toth2000},\\\citep{Noble2009}} &$(0.578,0.182,0.784)$  & $[1.090,80]$ & $[0.053,3.088]$ \\

  \harm & 2nd Order  & LLF  & PLM  & Flux-CT \citep{toth2000} & $(0.519,0.118,0.788)$ & $[1.073, 50]$ & $[0,\pi]$\\

  \IGM & RK4 & HLL & PPM & \makecell[tl]{Vector potential-based \\PPM+Flux-CT \\\citep{toth2000},\\\citep{Etienne:2010ui},\\\citep{Etienne:2011re} } & \makecell[tl]{Cartesian AMR \\ $(0.246,0.228,0.229)$} & See 
Sect.~\ref{sec:setup-illinois}.  & --\\

  \koral & 2nd order & LLF & PPM & Flux-CT \citep{toth2000} & $(0.478,0.266,0.785)$ & $[1.15,50]$ & $[0,\pi]$\\
 
\end{tabular}
\end{center}
\label{default}
\end{table}

\subsection{Code specific choices} \label{sec:setup-code}

\subsubsection{\athenapp{}}

For these simulations, \athenapp{} uses the second-order van~Leer integrator \citep{vanLeer2006} with third-order PPM reconstruction \citep{Colella1984}. Magnetic fields are evolved on a staggered mesh as described in \citet{2005JCoPh.205..509G} and generalized to GR in \citet{White2016}. The two-wave HLL approximate Riemann solver is used to calculate fluxes \cite{harten1983upstream}. The coordinate singularity at the poles is treated by communicating data across the pole in order to apply reconstruction, and by using the magnetic fluxes at the pole to help construct edge-centered electric fields in order to properly update magnetic fields near the poles. Mass and/or internal energy are added in order to ensure $\rho > 10^{-5} r^{-3/2}$, $p > 1/3 \times 10^{-7} r^{-5/2}$, $\sigma < 100$, and $\beta > 10^{-3}$. Additionally, the normal-frame Lorentz factor is kept under $50$ by reducing the velocity if necessary.

All \athenapp{} simulations are done in Kerr--Schild coordinates. The grids are logarithmically spaced in $r$ and uniformly spaced in $\theta$ and $\phi$. They use the fiducial resolution but then employ static mesh refinement to derefine in all coordinates toward the poles, stepping by a factor of $2$ each time. The $96^3$ grid achieves the fiducial resolution for $\pi/4 < \theta < 3\pi/4$ at all radii and for $\pi/12 < \theta < 11\pi/12$ when $r > 19.48$, derefining twice; the $128^3$ grid achieves this resolution for $3\pi/16 < \theta < 13\pi/16$ at all radii and for $\pi/16 < \theta < 15\pi/16$ when $r > 19.68$, derefining twice; and the $192^3$ grid achieves this resolution for $7\pi/24 < \theta < 17\pi/24$ when $r > 1.846$ and for $\pi/8 < \theta < 7\pi/8$ when $r > 31.21$, derefining three times. The outer boundary is always at $r = 50$, where the material is kept at the background initial conditions. The inner boundaries are at radii of $1.152$, $1.2$, and $1.152$, respectively, ensuring that exactly one full cell at the lowest refinement level is inside the horizon. Here the material is allowed to freely flow into the black hole, with the velocity zeroed if it becomes positive.

\subsubsection{\bhac}\label{sec:setup-bhac}
In \bhac, we employ the LLF fluxes in combination with PPM reconstruction \citep{Colella1984} and a two-step predictor corrector scheme.  
The setup analyzed here was run with the staggered upwind constrained transport (UCT) scheme of 
\cite{DelZanna2007}.  
The simulations are performed in modified Kerr-Schild coordinates \citep{McKinneyGammie2004} with $\theta$-coordinate stretching parameter $h=0.75$. 
In the staggered case, two to three grid levels are utilized (three for the high-resolution run) with static mesh refinement chosen such that the polar axis at small radii is fully de-refined and the torus is fully resolved on the highest grid level.  This allows to significantly increase the timestep which is otherwise dominated by the cells close to the singular axis.  Hence compared to the brute force uniform grid setup, the timestep in the 3-level run is increased by a factor of 16.
We adopt a floor model in the Eulerian frame as suggested, however, since staggered fields need to be interpolated to the cell centers which introduces an additional error, we have to increase the floors to $\rho_{\mathrm{fl}} = 10^{-4}r^{-3/2}$ and $p_{\mathrm{fl}} =1/3\times10^{-6}\ r^{-5/2}$.  Floors that are lower by an order of magnitude were no problem for centered field FluxCT runs \citep{toth2000}.

On the singular axis, we explicitly zero out the fluxes as well as the $\phi-$ and $r-$ components of the electric fields.  Furthermore, we employ $\pi$-periodic boundary conditions hence fill the axial ghost-cells with their diagonal counterpart.  No further pole-fix was required for stable evolution of the setup.

To increase the variety of the setups considered in the comparison and to introduce a case for which the difficulties related to the polar axis are not present, we also carry out a simulation in Cartesian Kerr-Schild (CKS) coordinates.
For this run (referred to as \bhac Cart. in the following), we use a combination of adaptive and fixed mesh refinement in an attempt to simultaneously resolve the MRI turbulence within the disk and follow the propagation of the jet. Adaptive mesh refinement is triggered by variations in the plasma magnetization $\sigma$ and the density, quantified by means of the L\"ohner scheme \citep{Loehner87}.
We structure the domain as a set of nested boxes such that the highest AMR level achievable for each box increases inwards, and the highest level in the simulation can be achieved only by the innermost box, containing the event horizon.
The simulation employs 8 such levels and a base resolution of $N_x\times N_y\times N_z = 96 \times 96 \times 192$, and extends over $x,y \in [-500\,M,500\,M]$, and $z \in [-1000\,M,1000\,M]$.  In order to prevent an unphysical outflow from the black hole interior, we apply cut-off values for the density and pressure in the region $r<0.5(r_{\rm H- } + r_{\rm H+ })$ where $r_{\rm H- }$ and $r_{\rm H+ }$ are the location of the inner and outer event horizons. Specifically, we set $\rho_{\rm cut} = 10^{-2}$ or $p_{\rm cut} = 10^{-4}$.  Other than that, the evolution in the interior of the event horizon is followed with the same algorithm as in the exterior, in particular the magnetic field update is obtained by the UCT algorithm providing zero divergence of the magnetic field throughout.  

For all simulations, to more accurately treat the highly magnetised polar region, we employ the entropy strategy discussed in \cite{PorthOlivares2017}, that is, each time the plasma-$\beta$ drops below the threshold value of $10^{-2}$, the advected entropy is used for primitive variable recovery instead of the conserved energy.  

\subsubsection{\cosmos}
In \cosmos, we use the HLL Riemann solvers with PPM reconstruction \citep{Colella1984} and a five-stage, strong-stability-preserving Runge-Kutta time integration scheme \citep{Spiteri2002}, set to third order.  
The magnetic fields were evolved using the staggered constrained-transport scheme described in \citet{Fragile2012}. A uniform $\theta$-coordinate was used with small cut-outs near the poles ($\theta < 4^\circ$) to keep the timestep reasonable. The outer radial boundary was placed at $(50)^{1.5} M$ to reduce boundary effects. We then increased the number of zones in the radial direction by $N_r^{1.5}$ to maintain the desired resolution in the region of interest. Outflow boundary conditions (copying scalar fields to ghost zones, while ensuring the velocity component normal to the boundary points outward) were used on the radial and polar boundaries. For the primitive inversion step, we primarily used the ``2D'' option from \citet{Noble2006}, with a 5D numerical inversion scheme (similar to the 9D inversion described in \citet{Fragile2014}) as a backup. In cases where both solvers failed, we ignored the conserved energy and instead used the entropy to recover the primitive variables. Otherwise, the default suggestions, as laid out in section 4 were used.

\subsubsection{\echo}
The time-integration performed in \echo uses the third order accurate IMplicit-EXplicit (IMEX) strong-stability preserving scheme \citep{Pareschi2005}, which in the case of ideal GRMHD reduces effectively to a third-order Runge-Kutta. 
The upwind fluxes are computed with the HLL Riemann solver with PPM reconstruction \citep{Colella1984}, using the  upwind constrained transport scheme of \cite{DelZanna2007} for the treatment of the magnetic field.

The numerical grid is logarithmically stretched in radius and uniform in both $\theta$ and $\phi$, excluding from the polar angle domain the regions close to the rotation axis to avoid excessively small time steps. At the radial and polar boundaries we impose outflow boundary conditions, i.e. we copy the value of the primitive variables and set the velocity normal to the boundary to zero whenever it has a positive (negative) value at the inner (outer) boundary.

As suggested, we adopt the floor model for rest-mass density and pressure used by \cite{McKinneyGammie2004}.
For the primitive variable recovery we first use three-dimensional scheme described in \cite{BucciantiniDel-Zanna2013}, and in case of failure we apply the 1D scheme from \cite{PorthOlivares2017}. Should none of these routines be successful, we then attempt to retrieve the primitives using the advected entropy instead of the total energy. In case of persisting failures, we finally reset the value of density and pressure to the atmospheric floor values and set the Eulerian velocity to zero, without modifying the magnetic field.

\subsubsection{\hamr}\label{sec:setup-hamr}
Like \harmtwod \citep{Gammie2003}, \hamr evolves the GRMHD equations in arbitrary (fixed) spacetimes. \hamr is $3^{rd}$ order accurate in space, by using PPM \citep{Colella1984} reconstruction of primitive variables at cell faces, and $2^{nd}$ order accurate in time. The fluxes at cell faces are calculated using an approximate HLL Riemann solver, while the magnetic field is evolved on a staggered grid, where the electric fields have been velocity upwinded to add dissipation \citep{2005JCoPh.205..509G}.  Since the funnel is devoid of matter, \hamr artificially inserts mass in the drift frame \citep{Ressler2017}. This does not lead to a runaway in velocity, which occurs when mass is inserted in the fluid frame \citep{Gammie2003}, or to artificial drag on the field lines, which occurs when mass is inserted in the ZAMO frame \citep{McKinneyTchekhovskoy2012}. We enforce floor values on the rest-mass density $\rho_{\rm fl}=$MAX$[B^2/20,10^{-5}(r/M)^{-3/2}]$ and  internal energy $u_{\rm fl}=$MAX$[B^2/750,3.33\times10^{-8}(r/M)^{-5/2}]$. To provide a backup inversion method for primitive variable recovery if all other primary inversion method(s) fail \citep{Noble2006, HamlinNewman2013},  \hamr also advects the conserved entropy \citep{Noble2009}. We typically use $3-4$ levels of local adaptive timestepping to speed up the code by an additional factor $3-5$ to $3-5 \times 10^8$ zone-cycles/s/GPU \citep{ChatterjeeEtAl2019}.

We use a (close to) uniformly spaced logarithmic spherical grid combined with outflow boundary conditions in the radial direction, transmissive boundary conditions in the $\theta$-direction and periodic boundary conditions in the $\phi$-direction. Since cells get squeezed near the pole, the timestep in all spherical grids is reduced by an additional factor proportional to the resolution in the $\phi$-direction. To remedy this issue, we stretch out cells immediately adjacent to the pole in the $\theta$-direction \citep{TchekhovskoyNarayan2011} and use multiple levels of static mesh de-refinement in the $\phi$-direction to keep the cell's aspect ratio close to uniform at high latitudes. As an example, the very high resolution (effectively, $1608\times1056\times1024$) run in this work uses a $\phi$-resolution of 64 cells for $0^{\circ}<\theta<3.75^{\circ}$, 128 cells for $3.75^{\circ}<\theta<7.5^{\circ}$, 256 cells for $7.5^{\circ}<\theta<15^{\circ}$, 512 cells for $15^{\circ}<\theta<30^{\circ}$ and the full 1024 cells for $30^{\circ}<\theta<90^{\circ}$. This method prevents the squeezing of cells near the pole from reducing the global timestep, while maintaining high accuracy in all 3 dimensions.

\subsubsection{\nobleharm}

The results using \nobleharm given here used the LF approximate
Riemann solver as defined in \citep{Gammie2003}, RK2 time integration,
the FluxCT method of \cite{toth2000}, and PPM reconstruction
\citep{Colella1984} for the primitive variables at cell faces and the
electric fields at cell edges (for the sake of the FluxCT method).  We
used a ``modified Kerr-Schild'' coordinate system specified by
\begin{align}
  r(x^{1^\prime})  = \exp\left(x^{1^\prime}\right)  \quad , \quad 
\theta(x^{2^\prime}) = \theta_c +  h  \sin\left(2 \pi x^{2^\prime}\right)  \label{noble-harm-coords}
\end{align}
with $\theta_c = 0.017 \pi$, and $h = 0.25$.  Zeroth-order
extrapolation was used to set values in the ghost zones at the radial
and poloidal boundaries, outflow conditions were enforced at the inner
and outer radial boundaries.  The poloidal vector components were
reflected across the poloidal boundary (\eg $B^\theta \rightarrow
-B^\theta$, $v^\theta \rightarrow -v^\theta$).

Instead of using $96$ cells (for instance for the low resolution run)
within $r=50\rm M$ as many of the others used, the \nobleharm runs used
this number over the entire radial extent it used, out to $r=80\rm M$.  This means
that a \nobleharm run has lower radial resolution than another code's run
with the same cell count.

Recovery of the primitive variables from the conserved variables was
performed with the ``2D'' and ``1$\mathrm{D}_W$'' methods of
\citet{Noble2006}, where the latter method is used if the former one
fails to converge to a sufficiently accurate, physical solution.

The \nobleharm runs also used the so-called ``entropy fix'' of \cite{Balsara1999b} as
described in \citep{Noble2009}, wherein the entropy EOM replaces the
energy EOM in the primitive variable method whenever it 
fails to converge, the resultant primitive variables are unphysical, or 
$u < 10^{-2} \, B^2$. 
\footnote{During the production stage of publication, Noble discovered
the \nobleharm simulations errantly used the same seed across all MPI
processes for the random number generator used to perturb the initial
conditions.  Since the azimuthal dimension was decomposed and because
the other dimensions are decomposed uniformly, using the same seed
meant the initial conditions were azimuthally periodic with period
equal to $\Delta \phi = 2 \pi / N$ where $N$ is the number of
azimuthal decompositions. 
Due to the symmetry preserving implementation of \nobleharm, these modes do not appreciably grow out of accumulated round-off error during the course of the simulation.  
In other words, the
\nobleharm simulations are missing the $m=1,\ldots,\left(N-1\right)$
azimuthal modes of dynamics, where $N=\left(3,6,6\right)$ for runs
$96^3, 128^3, 192^3$, respectively. }

\subsubsection{\harm}

\harm is an unsplit method-of-lines scheme that is spatially and temporally second order. A predictor-corrector scheme is used for timestepping.  For models presented here, spatial reconstruction was carried out with a piecewise linear method using the monotonized central slope limiter. The divergence-free condition on the magnetic field was maintained to machine precision with the Flux-CT scheme of \cite{toth2000}.

The simulations used ``funky'' modified Kerr-Schild (FMKS) coordinates $t, X^1, X^2, X^3$
that are similar to the MKS coordinates of \citet{McKinneyGammie2004}, with $R_0 = 0$, $h = 0.3$, but with an additional modification to MKS to enlarge grid zones near the polar axis at small radii and increase the timestep. For FMKS, then, $X^2$ is chosen to smoothly interpolate from KS $\theta$ to an expression that deresolves the poles:
\begin{equation}
\theta = N \, y \, \left(1 + \frac{(y/x_t)^\alpha}{\alpha + 1}\right) + \frac{\pi}{2}
\end{equation}
where N is a normalization factor, $y \equiv 2 X^2 - 1$, and we choose $x_t = 0.82$ and $\alpha = 14$.

{\tt iharm3d} imposes floors on rest-mass density $\rho$ and internal energy $u$ to enforce $\rho > 10^{-5} r^{-2}$ and $u > 10^{-7} r^{-2 \hat{\gamma}}$, where $\hat{\gamma}$ is the adiabatic index. It also requires that $\rho > B^2/50$, $u > B^2/2500$, and subsequently that $\rho > u/50 $ \citep{Ressler2017}. At high magnetizations, mass and energy created by the floors are injected in the drift frame \citep{Ressler2017}; otherwise, they are injected in the normal observer frame. The fluid velocity primitive variables are rescaled until the lorentz factor in the normal observer frame is $< 50$. If inversion from conserved to primitive variables fails, the primitive variables are set to an average over a stencil of neighboring cells in which the inversion has not failed.

The radial boundaries use outflow-only conditions. The polar axis boundaries are purely reflecting, and the $X^1$ and $X^3$ fluxes of the $X^2$ component of the magnetic field are antiparallel across the boundary. The $X^3$ boundaries are periodic.

\subsubsection{\IGM}\label{sec:setup-illinois}

\IGM simulations presented here adopt a Cartesian FMR grid (using the
Cactus/Carpet infrastructure), in which four overlapping cubes with
half-sidelength 27.34375$M$ are centered on the $x$--$y$ plane at
positions $(x/M,y/M)=(25,25);(25,-25);(-25,25);(-25,-25)$. These
cubes, all at resolution $\Delta x=\Delta y=\Delta z\approx
M/4.388571$, constitute the highest refinement level, and a total of
six factor-of-two derefinements (with approximate resolutions
$M/2.19$, $M/1.10$, $1.82M$, $3.65M$, $7.29M$, and $14.58\bar{3}M$
[exact]) are placed outside this high-resolution level so that the
outer boundary is a cube with half-sidelength $1750M$, centered at
$(x,y,z)\approx (M/8.78,M/8.78,M/8.78)$. This ensures full
cell-centering on the finest refinement level, which maximally avoids
the $z$-axis and $r=0$ coordinate singularities when mapping initial
data to the Cartesian grids. 

\IGM adopts the same formalism as \harm~\citep{Gammie2003} for the GRMHD
field equations, with the exception of the magnetic field evolution;
\IGM evolves the vector potential
directly~\citep{Etienne:2010ui,Etienne:2011re}. Evolving the vector
potential enables {\it any} interpolation scheme to be used at AMR
refinement boundaries, and the formulation \IGM adopts reduces to the
standard staggered Flux-CT scheme on uniform-resolution grids. As for
GRMHD field reconstruction, \IGM makes use
of the PPM algorithm and HLL for its approximate
Riemann solver. The conservative-to-primitives solver in \IGM is based
on the open-source Noble \etal code~\citep{Noble2006}, but has
been extended to adjust conservative variables that violate
physicality inequalities prior to the solve~\citep{Etienne:2011ea}.

\subsubsection{\koral}

The simulations presented here used PPM reconstruction and
Lax-Friedrichs fluxes. Modified Kerr-Schild coordinates were employed,
using the technique developed in \citet{TchekhovskoyNarayan2011}, whereby the
$\theta$-grid was concentrated modestly towards the equator, and was
moved away from the poles at small radii to avoid very small time
steps. The following floors and ceilings were applied for numerical
stability (they mostly affect the polar low-density regions, which are
not the focus of the present paper): $10^{-8} \leq u/\rho c^2 \leq
10^2$, $B^2/u \leq 10^5$, $B^2/\rho c^2 \leq 50$, $\Gamma \leq 20$. Outflow boundary conditions were used at the outer radius, and reflecting boundary conditions at the poles.

\section{Results} \label{sec:results}

A consistent set of diagnostics focusing both on horizon scale quantities and on the global evolution of the accretion flow is performed with all codes.  For ease of implementation, all diagnostics are performed in the standard Kerr-Schild coordinates.  We now first describe the quantifications and then move on to the inter-code comparison.  

\subsection{Diagnostics} \label{sec:diag}

\begin{itemize}
\item
  \textbf{Horizon penetrating fluxes.}  
  The relevant quantities: mass, magnetic, angular momentum and (reduced) energy fluxes are defined as follows:
  \begin{align}
    \dot{M} &:= \int_0^{2\pi}\int_{0}^{\pi} \rho u^r\sqrt{-g} \, d\theta \, d\phi \,, \label{eq:mdot} \\ 
    \Phi_{\rm BH} &:= \frac{1}{2} \int_{0}^{2\pi}\int_{0}^{\pi} |^*F^{rt}| \sqrt{-g} \, d\theta \, d\phi \,, \label{eq:phib} \\
    \dot{L}&:= \int_0^{2\pi}\int_{0}^{\pi} T^r_\phi \sqrt{-g} \, d\theta \, d\phi \,, \label{eq:ldot} \\ 
    \dot{E}&:= \int_0^{2\pi}\int_{0}^{\pi} (- T^r_t)  \sqrt{-g} \, d\theta \, d\phi  \label{eq:edot} 
  \end{align}
  where all quantities are evaluated at the outer horizon $r_{\mathrm{h}}$.  A cadence of $1M$ or less is chosen.  In practise these quantities are non-dimensionalised with the accretion rate, yielding for example the normalized magnetic flux $\phi=\Phi_{\rm BH}/\sqrt{\dot{M}}$ also know as the ``MAD parameter'' which, for spin $a=0.9375$ and torus scale height $H/R\approx0.3$ has the critical value $\phi_{\rm max}\approx 15$ (within the units adopted here, \cite{TchekhovskoyMcKinney2012}).  
  
\item
  \textbf{Disk-averaged quantities.}  We compare averages of the basic flow variables $q\in\{ \rho, p, u^\phi, \sqrt{b_\alpha b^\alpha}, \beta^{-1}\}$.  
  These are defined similar to \cite[][]{Shiokawa2012,White2016,Beckwith2008}. Hence for a quantity $q(r,\theta,\phi,t)$, the shell average is defined as
  \begin{align}
    \langle q \rangle(r,t) := \frac{\int_0^{2\pi}\int_{\theta{\rm min}}^{\theta{\rm max}} q(r,\theta,\phi,t) \sqrt{-g} \, d\phi \, d\theta}{\int_0^{2\pi}\int_{\theta{\rm min}}^{\theta{\rm max}} \sqrt{-g} \, d\phi \, d\theta} \,, \label{eq:averaged}
  \end{align}
  which is then further averaged over the time interval $t_{\rm KS}\in [5\,000,10\, 000] M$ to yield
  $\langle q \rangle(r)$ (note that we omit the weighting with the density
  as done by \citet{Shiokawa2012,White2016}). The limits $\theta_{\rm
    min}=\pi/3$, $\theta_{\rm max}=2\pi/3$ ensure that only material from the disk is taken into account in the averaging. 
  
  \item 
    \textbf{Emission proxy.}  To get a feeling for the code-to-code variations in synthetic optically thin light-curves, we also integrate the pseudo emissivity for thermal synchrotron radiation following an appropriate non-linear combination of flow variables.  
    \begin{align}
      \mathcal{L}(t) := \int_0^{2\pi}\int_{\theta{\rm min}}^{\theta{\rm max}}\int_{r_{\rm h}}^{r_{\rm max}} j (B,p,\rho) \sqrt{-g} \, d\phi \, d\theta \, dr \label{eq:j}
    \end{align}
    where again $\theta_{\rm min}=\pi/3$ and $\theta_{\rm max}=2\pi/3$ are and $r_{\rm max}=50 M$ are used.
    The emissivity $j$ is here defined as follows: $j=\rho^{3}p^{-2} \exp(-C (\rho^2/(B p^2))^{1/3})$, which captures the high-frequency limit of the thermal synchrotron emissivity $\nu\gg \nu_c \Theta_e^2\sin(\theta)$, (compare with Equation (57) of \cite{Leung2011}).   The constant $C$ is chosen such that the radiation is cutoff after a few gravitational radii, resembling the expected $mm$-emission from the galactic center, that is $C=0.2$ in geometrised units.    This emission proxy is optimized for the science case of the EHT where near optically thin synchrotron emission is expected.\footnote{In this the proxy is distinct from the prescription often used for optically thick, geometrically thin disks which is based on an estimation of the turbulent dissipation, (e.g. \cite{HubenyHubeny1998,ArmitageReynolds2003}).}  In order to compare the variability, again a cadence of $1M$ or less is chosen in most cases. 

  \item
    \textbf{$t,\phi$ - averages.}  Finally, we compare temporally and azimuthally averaged data for a more global impression of the disk and jet system.  

    \begin{align}
      \langle q \rangle (r,\theta) := \frac{\int_{t_{\rm beg}}^{t_{\rm end}} \int_0^{2\pi} q(r,\theta,\phi,t) \, d\phi\, dt}{2\pi (t_{\rm end}-t_{\rm beg})}
    \end{align}
    for the quantities $q\in\{\rho,\beta^{-1},\sigma\}$ with the averaging interval ranging from $5000M$ to $10000M$.  

\end{itemize}

\FloatBarrier 
\subsection{Time series}\label{sec:ts}

\begin{figure}[htbp]
\begin{center}
\includegraphics{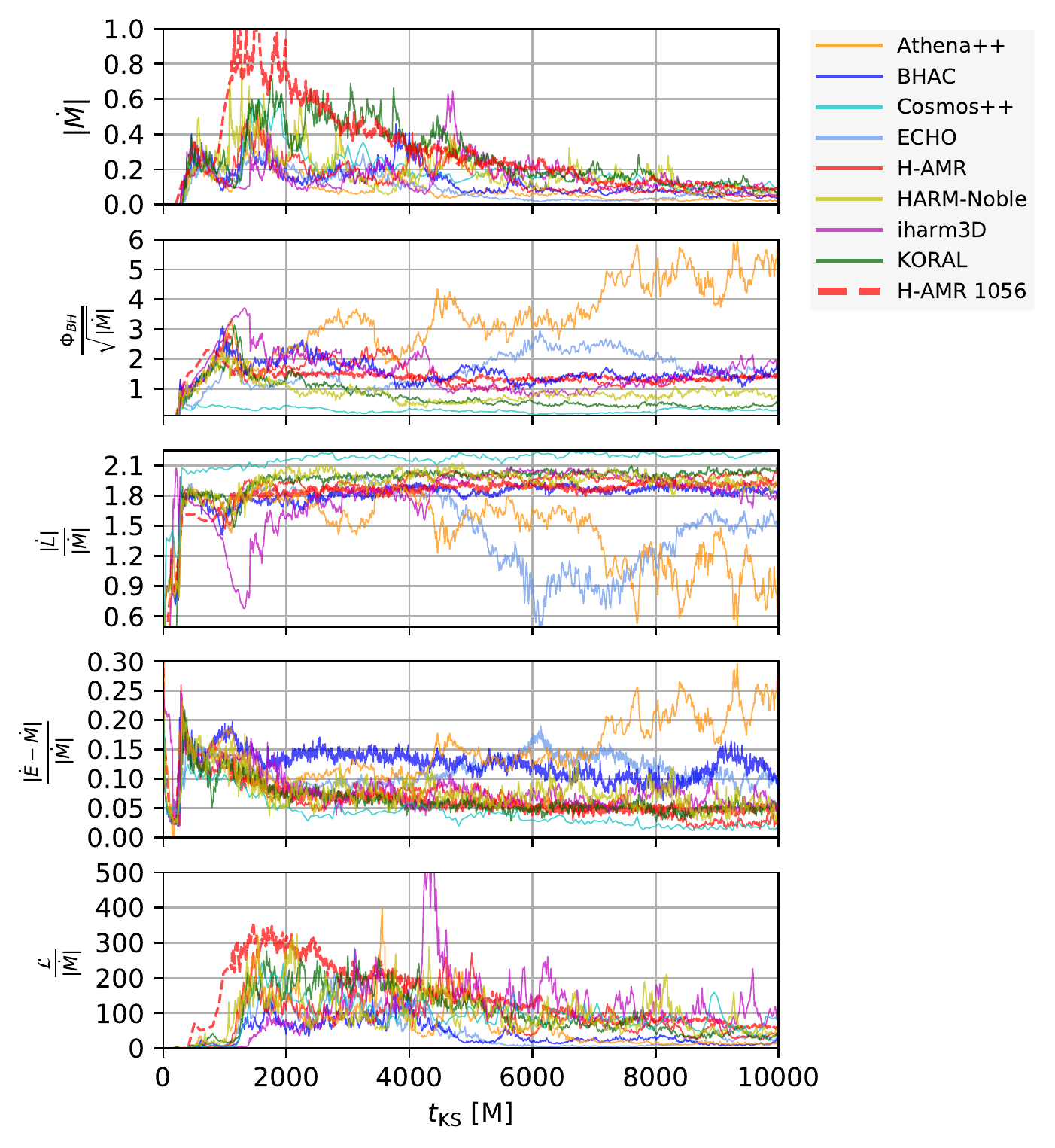}
\caption{Time-series at resolution of $96^3$ in the domain of interest for all codes.  From top to bottom, the panels show mass-accretion rate, horizon penetrating magnetic flux, accretion of angular momentum and energy as well as the pseudo luminosity according to Eq. (\ref{eq:j}).  The data in panels 2--5 have been non-dimensionalized with the accretion rate $\dot{M}$. For reference, the \hamr $N_{\theta}=1056$ solution was added (thick dashed red lines).  
}
\label{fig:ts96}
\end{center}
\end{figure}

\begin{figure}[htbp]
\begin{center}
\includegraphics{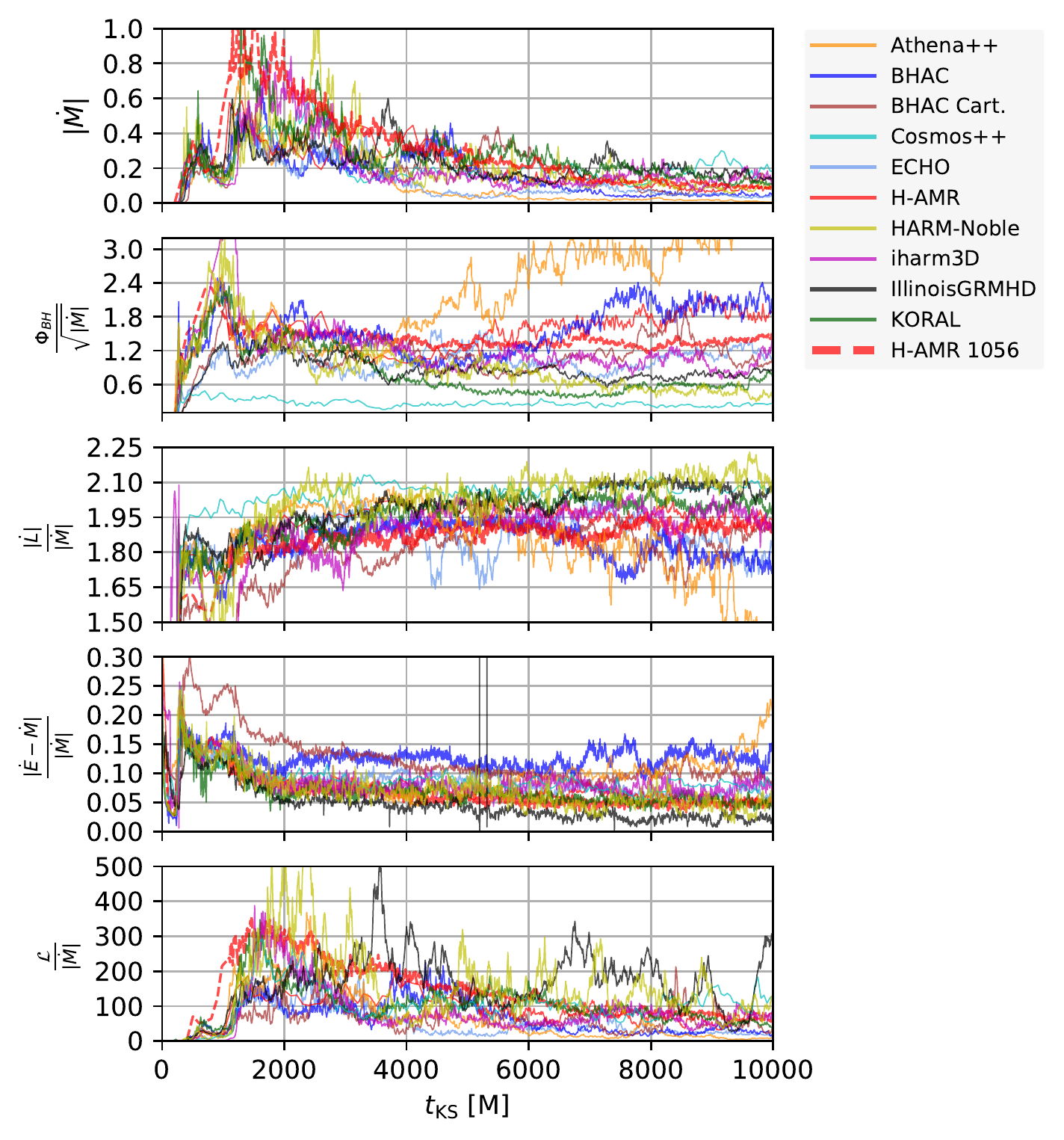}
\caption{As figure \ref{fig:ts96} for the $128^3$ data and the cartesian runs ``\IGM'' and ``\bhac Cart.''
}
\label{fig:ts128}
\end{center}
\end{figure}

\begin{figure}[htbp]
\begin{center}
\includegraphics{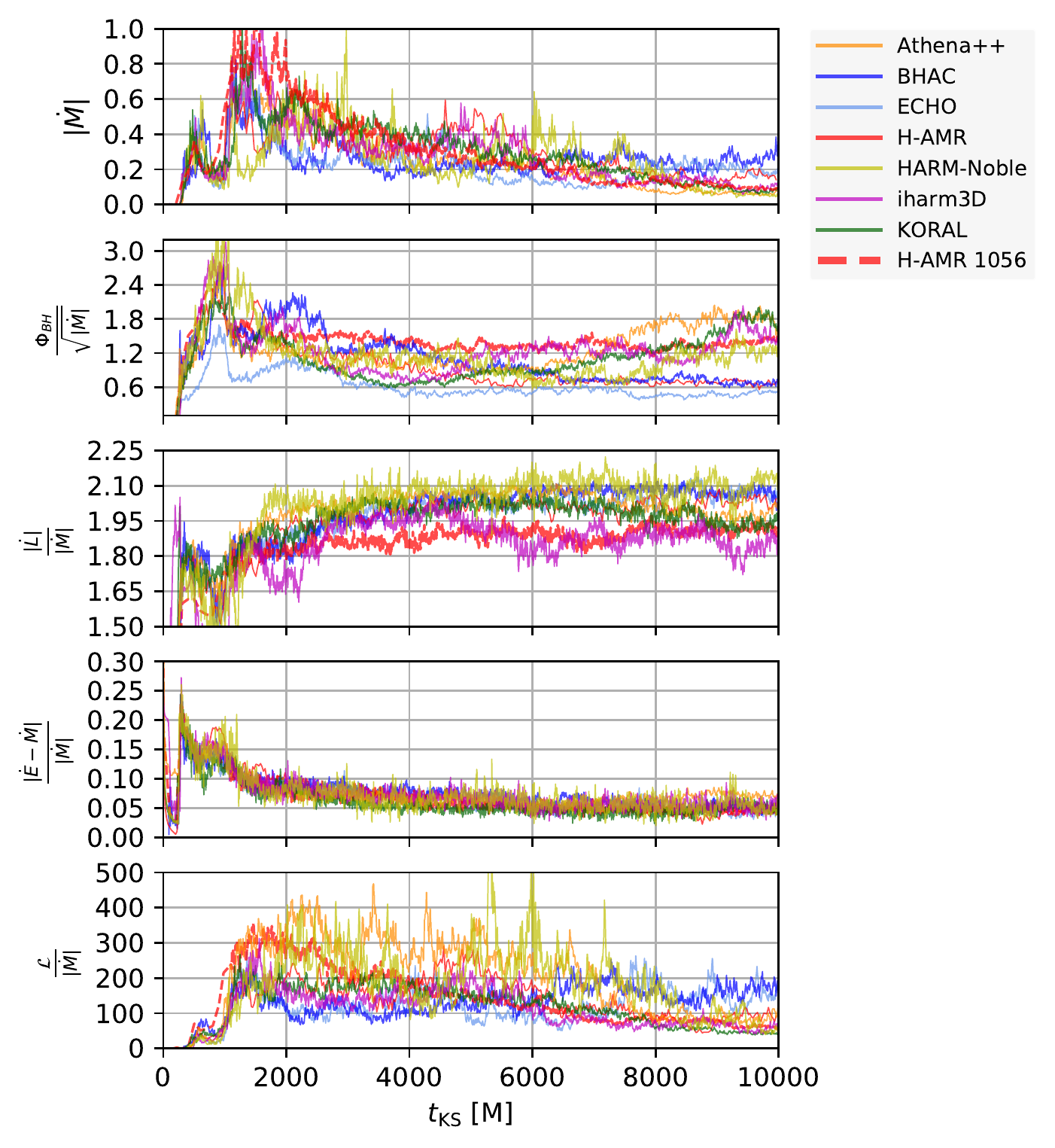}
\caption{As figure \ref{fig:ts96} for the $192^3$ data. 
}
\label{fig:ts192}
\end{center}
\end{figure}

Time series data of horizon penetrating fluxes is presented in Figures \ref{fig:ts96} to \ref{fig:ts192} for low, medium and high resolutions respectively.  Since the mass accretion governs the behavior of these fluxes, the data has been appropriately normalized by the accretion rate.
All codes capture accurately the linear phase of the MRI leading to an onset of accretion to the BH at $t\simeq300M$.  While there is still a good correspondence of $\dot{M}$ for early times $<1000M$, the chaotic nature of the problem fully asserts itself after $2000M$.  At low resolution, there exist order-of-magnitude variations in the data, most notably for the normalized horizon penetrating magnetic fluxes $\phi$ and the energy fluxes.  The low value of $\phi$ for the \cosmos-data is caused by the choice of boundary conditions near the polar region as will be discussed in Section \ref{sec:t-phi-averages} in more detail.  
As indicated by $\phi\sim0.5-6<\phi_{\rm max}$, all simulations are in the SANE regime of radiatively inefficient black hole accretion.  
As a measure for the variance between codes, in Table \ref{tab:quantify-ts} we quantify the peak fluxes as well as the average values far in the non-linear regime.

It is worthwhile to point out the good agreement in angular momentum- and energy fluxes at high resolution for the codes employing spherical grids, where the highest average values differ from the lowest ones by $12\%$ and $40\%$.  

In Figure~\ref{fig:ts128}, we additionally show the Cartesian realizations next to the medium resolution cases.  It should be kept in mind though that the resolution in the Cartesian cases is typically much worse near the horizon but better at larger radii which makes it hard to directly compare the simulations.  
Qualitatively there is very good agreement of the Cartesian runs with the spherical data in all quantities except for the energy fluxes where \bhac Cart. is systematically higher and \IGM slightly lower than all spherical codes.  

\begin{figure}[htbp]
\begin{center}
\includegraphics[width=14cm]{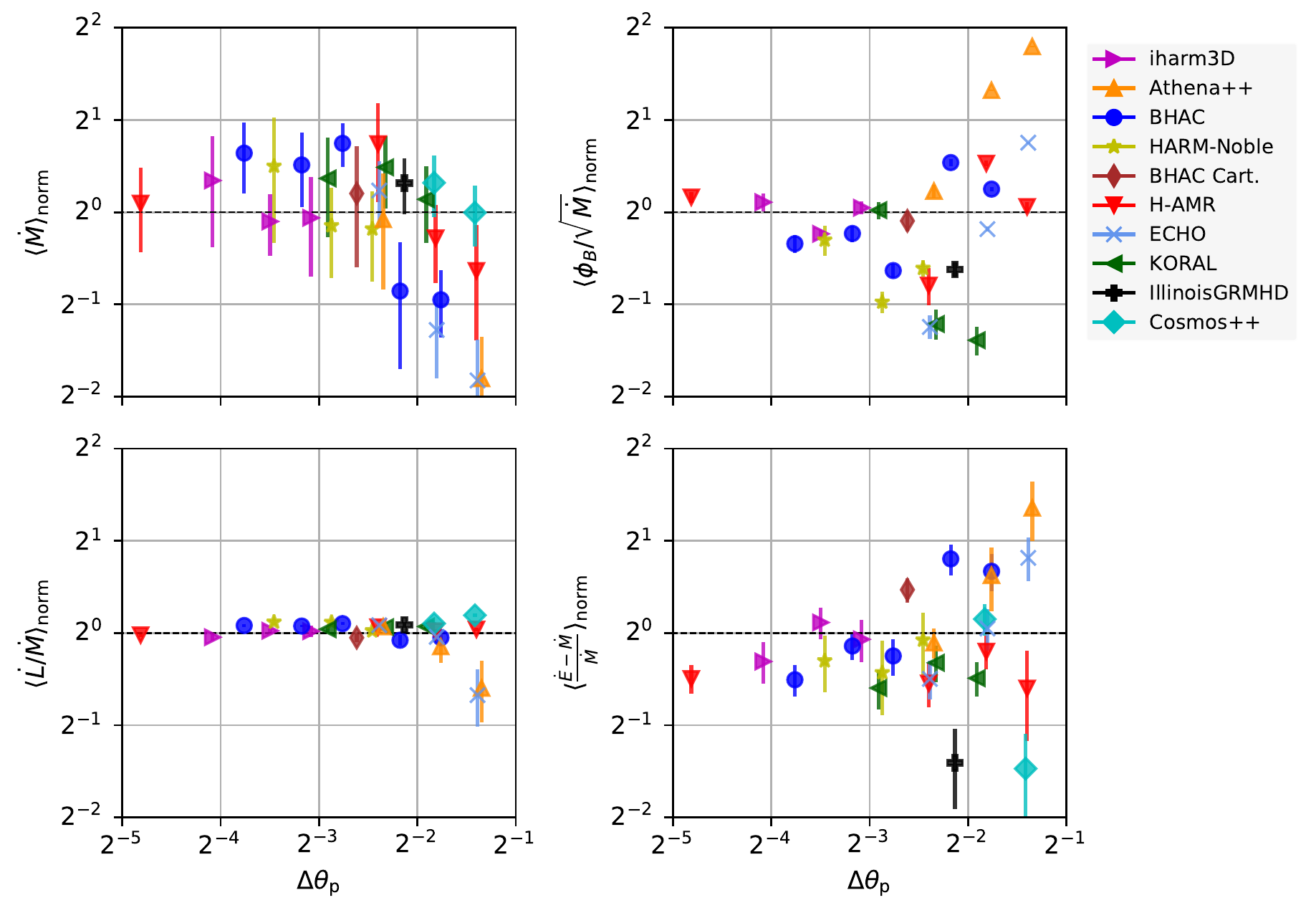}
\caption{Resolution dependence of the averaged quantities for all runs of Table \ref{tab:quantify-ts}.  The figure has been normalized to the mean of the distribution allowing to judge the overall spread and errors mark the standard deviation over the averaging interval.  Upon increased resolution the spread in the quantities decreases and one can make out a critical resolution $\Delta \theta_p = 0.0125-0.25$ below which accretion rates and horizon penetrating fluxes tend to converge to the same answer.  
The accreted angular momentum flux $\dot{L}/\dot{M}$ is very well captured at all resolutions with minimal spread.  Note that the data of the magnetic fluxes (top right panel) for \cosmos lie below the plotting window.  
}
\label{fig:quantify-ts}
\end{center}
\end{figure}

Since meshes with various amounts of compression in the mid-plane were employed by the different groups, a comparison purely based on the number of grid cells is however hardly fair even in the spherical cases.  Hence in Figure \ref{fig:quantify-ts} we show the data of Table \ref{tab:quantify-ts} against the proper grid spacing at the location of the initial density maximum $\Delta\theta_{\rm p}$.  
Ordered by $\Delta\theta_{\rm p}$, the trends in the data becomes clearer: for example, the mid-plane resolution of the \harm run at $N_{\theta}=96$ is higher than the resolutions employed e.g. in \bhac and \koral at $N_{\theta}=192$.  Hence even the lowest resolution \harm case is able to properly resolve the MRI at late times 
(see Section \ref{sec:further}) yielding consistent results across all runs.  Taken for itself also \koral and \nobleharm show very good agreement among the three resolution instances (with the exception of horizon penetrating magnetic flux for \koral), however, the horizon penetrating flux in \cosmos are consistently off by a factor of $\sim 4$ compared to the distribution average.  The trend can be explained by noting that \cosmos used a polar cutout with open outflow boundary conditions.  This allows magnetic flux to effectively leak out of the domain which further reduces the magnetization of the funnel region (see also Section \ref{sec:t-phi-averages})

\begin{table}[htp]
\caption{Quantifications -- Time series data.   Quantities in angular brackets $\langle\cdot\rangle$ denote time averages in the interval $t\in[5000,10000]\rm M$ with error given by the standard deviation.  For further convergence testing, another two \bhac runs ($256^*$ using $256\times256\times128$ cells and $384\times384\times384$ cells) and a \hamr run with $1608\times1056\times1024$ cells are listed here, where the resolutions correspond to effective $N_r\times N_\theta\times N_\phi$ in the disk region.  
\label{tab:quantify-ts}}
\begin{center}
\begin{tabular}{clllllllll}
$N_{\theta}$ & 
Code & 
$\dot{M}_{\rm Peak}$ & 
$\langle\dot{M}\rangle$ & 
$\left.\frac{\Phi_{\rm BH}}{\sqrt{\dot{M}}}\right|_{\rm Peak}$ & 
$\langle\Phi_{\rm BH}/\sqrt{\dot{M}}\rangle$ & 
$\left.\dot{L}/\dot{M}\right|_{\rm Peak}$ & 
$\langle\dot{L}/\dot{M}\rangle$ & 
$\left.\frac{\dot{E}-\dot{M}}{\dot{M}}\right|_{\rm Peak}$ & 
$\langle\frac{\dot{E}-\dot{M}}{\dot{M}}\rangle$ \\
  \hline \hline
96  & \athenapp  & $ 0.461 $ & $ 0.041  \pm  0.019 $ & $ 5.951 $ & $ 4.164  \pm  0.869 $ & $ 1.95 $ & $ 1.276  \pm  0.326 $ & $ 0.332 $ & $ 0.18  \pm  0.044 $ \\
  
  &  \bhac & $ 0.507 $ & $ 0.074  \pm  0.021 $ & $ 3.099 $ & $ 1.426  \pm  0.16 $ & $ 1.964 $ & $ 1.865  \pm  0.033 $ & $ 0.33 $ & $ 0.112  \pm  0.017 $ \\

  & \cosmos  & $ 0.602 $ & $ 0.142  \pm  0.036 $ & $ 0.481 $ & $ 0.239  \pm  0.065 $ & $ 2.268 $ & $ 2.207  \pm  0.025 $ & $ 0.232 $ & $ 0.025  \pm  0.009 $ \\
  
  & \echo  & $ 0.331 $ & $ 0.04  \pm  0.018 $ & $ 2.949 $ & $ 2.021  \pm  0.377 $ & $ 2.005 $ & $ 1.211  \pm  0.289 $ & $ 0.33 $ & $ 0.124  \pm  0.022 $ \\
  
  &  \hamr  & $ 0.536 $ & $ 0.091  \pm  0.048 $ & $ 2.908 $ & $ 1.249  \pm  0.131 $ & $ 2.066 $ & $ 1.985  \pm  0.028 $ & $ 0.336 $ & $ 0.046  \pm  0.018 $ \\

  & \nobleharm & $ 0.797 $ & $ 0.125  \pm  0.05 $ & $ 2.236 $ & $ 0.786  \pm  0.119 $ & $ 2.206 $ & $ 1.964  \pm  0.051 $ & $ 0.327 $ & $ 0.067  \pm  0.017 $ \\
  
  & \harm  & $ 0.645 $ & $ 0.136  \pm  0.06 $ & $ 3.705 $ & $ 1.239  \pm  0.346 $ & $ 2.083 $ & $ 1.958  \pm  0.087 $ & $ 0.331 $ & $ 0.067  \pm  0.011 $ \\
  
  &  \koral  & $ 0.738 $ & $ 0.157  \pm  0.051 $ & $ 3.13 $ & $ 0.458  \pm  0.072 $ & $ 2.1 $ & $ 2.027  \pm  0.026 $ & $ 0.328 $ & $ 0.05  \pm  0.007 $ \\

  & max/min  &  2.406  &  3.903  &  12.374  &  17.408  &  1.163  &  1.823  &  1.448  &  7.096  \\
   \hline
   
128  &  \athenapp{}  & $ 0.847 $ & $ 0.023  \pm  0.012 $ & $ 5.385 $ & $ 2.995  \pm  0.694 $ & $ 2.082 $ & $ 1.743  \pm  0.208 $ & $ 0.332 $ & $ 0.108  \pm  0.029 $ \\
  
  &  \bhac  & $ 0.66 $ & $ 0.078  \pm  0.046 $ & $ 2.492 $ & $ 1.74  \pm  0.407 $ & $ 1.974 $ & $ 1.829  \pm  0.074 $ & $ 0.331 $ & $ 0.123  \pm  0.015 $ \\

  &  \bhac Cart.  & $ 0.57 $ & $ 0.163  \pm  0.091 $ & $ 2.137 $ & $ 1.123  \pm  0.277 $ & $ 1.947 $ & $ 1.864  \pm  0.061 $ & $ 0.331 $ & $ 0.097  \pm  0.009 $ \\
  
  &  \cosmos  & $ 0.642 $ & $ 0.177  \pm  0.045 $ & $ 0.484 $ & $ 0.245  \pm  0.032 $ & $ 2.133 $ & $ 2.073  \pm  0.019 $ & $ 0.232 $ & $ 0.078  \pm  0.01 $ \\

  &  \echo  & $ 0.507 $ & $ 0.059  \pm  0.021 $ & $ 1.575 $ & $ 1.056  \pm  0.182 $ & $ 2.028 $ & $ 1.876  \pm  0.083 $ & $ 0.331 $ & $ 0.073  \pm  0.012 $ \\
  
  &  \hamr  & $ 0.588 $ & $ 0.117  \pm  0.039 $ & $ 2.378 $ & $ 1.729  \pm  0.204 $ & $ 2.04 $ & $ 1.961  \pm  0.031 $ & $ 0.336 $ & $ 0.061  \pm  0.008 $ \\
  
  &  \nobleharm  & $ 1.28 $ & $ 0.128  \pm  0.051 $ & $ 3.23 $ & $ 0.61  \pm  0.179 $ & $ 2.259 $ & $ 2.093  \pm  0.054 $ & $ 0.328 $ & $ 0.052  \pm  0.017 $ \\

  &  \harm  & $ 0.841 $ & $ 0.132  \pm  0.035 $ & $ 3.464 $ & $ 1.019  \pm  0.135 $ & $ 2.089 $ & $ 1.964  \pm  0.037 $ & $ 0.325 $ & $ 0.076  \pm  0.01 $ \\

  &  \IGM  & $ 0.6 $ & $ 0.177  \pm  0.041 $ & $ 1.477 $ & $ 0.779  \pm  0.08 $ & $ 2.141 $ & $ 2.053  \pm  0.044 $ & $ 0.361 $ & $ 0.026  \pm  0.009 $ \\

  &  \koral  & $ 1.221 $ & $ 0.199  \pm  0.062 $ & $ 2.351 $ & $ 0.517  \pm  0.097 $ & $ 2.084 $ & $ 2.017  \pm  0.025 $ & $ 0.332 $ & $ 0.056  \pm  0.008 $ \\

  & max/min  &  2.525  &  8.604  &  11.126  &  12.24  &  1.16  &  1.201  &  1.552  &  4.631  \\

   \hline

192  &  \athenapp  & $ 0.875 $ & $ 0.134  \pm  0.071 $ & $ 2.739 $ & $ 1.405  \pm  0.39 $ & $ 2.109 $ & $ 2.025  \pm  0.047 $ & $ 0.332 $ & $ 0.065  \pm  0.008 $ \\
  
  &  \bhac & $ 0.825 $ & $ 0.238  \pm  0.042 $ & $ 2.754 $ & $ 0.773  \pm  0.107 $ & $ 2.121 $ & $ 2.072  \pm  0.021 $ & $ 0.331 $ & $ 0.059  \pm  0.009 $ \\

  &  \echo  & $ 0.843 $ & $ 0.167  \pm  0.047 $ & $ 1.698 $ & $ 0.506  \pm  0.057 $ & $ 2.139 $ & $ 2.047  \pm  0.033 $ & $ 0.332 $ & $ 0.05  \pm  0.008 $ \\

  &  \hamr  & $ 0.671 $ & $ 0.237  \pm  0.104 $ & $ 2.798 $ & $ 0.691  \pm  0.049 $ & $ 2.086 $ & $ 2.019  \pm  0.026 $ & $ 0.265 $ & $ 0.048  \pm  0.009 $ \\
  
  &  \nobleharm  & $ 0.994 $ & $ 0.201  \pm  0.116 $ & $ 3.593 $ & $ 0.972  \pm  0.216 $ & $ 2.224 $ & $ 2.097  \pm  0.044 $ & $ 0.33 $ & $ 0.057  \pm  0.013 $ \\
    
  &  \harm  & $ 1.106 $ & $ 0.18  \pm  0.091 $ & $ 3.161 $ & $ 1.292  \pm  0.19 $ & $ 2.066 $ & $ 1.871  \pm  0.046 $ & $ 0.32 $ & $ 0.057  \pm  0.01 $ \\
  
  &  \koral  & $ 1.01 $ & $ 0.183  \pm  0.081 $ & $ 2.256 $ & $ 1.217  \pm  0.338 $ & $ 2.082 $ & $ 1.987  \pm  0.041 $ & $ 0.331 $ & $ 0.046  \pm  0.007 $ \\
  
  & max/min  &  1.649  &  1.772  &  2.116  &  2.777  &  1.077  &  1.121  &  1.25  &  1.405  \\
\hline

$256^*$  &  \bhac  & $ 0.697 $ & $ 0.203  \pm  0.065 $ & $ 2.552 $ & $ 1.021  \pm  0.091 $ & $ 2.097 $ & $ 2.034  \pm  0.019 $ & $ 0.332 $ & $ 0.064  \pm  0.007 $ \\
\hline 

384  &  \bhac  & $ 0.91 $ & $ 0.221  \pm  0.067 $ & $ 3.424 $ & $ 0.946  \pm  0.128 $ & $ 2.12 $ & $ 2.043  \pm  0.021 $ & $ 0.331 $ & $ 0.049  \pm  0.006 $ \\
\hline

1056  &  \hamr  & $ 1.143 $ & $ 0.152  \pm  0.056 $ & $ 2.324 $ & $ 1.341  \pm  0.068 $ & $ 1.975 $ & $ 1.901  \pm  0.028 $ & $ 0.336 $ & $ 0.05  \pm  0.006 $ 
\end{tabular}
\end{center}
\end{table}

\FloatBarrier 
\subsection{Disk profiles}\label{sec:disk}

\begin{figure}[htbp]
\begin{center}
\includegraphics[width=15cm]{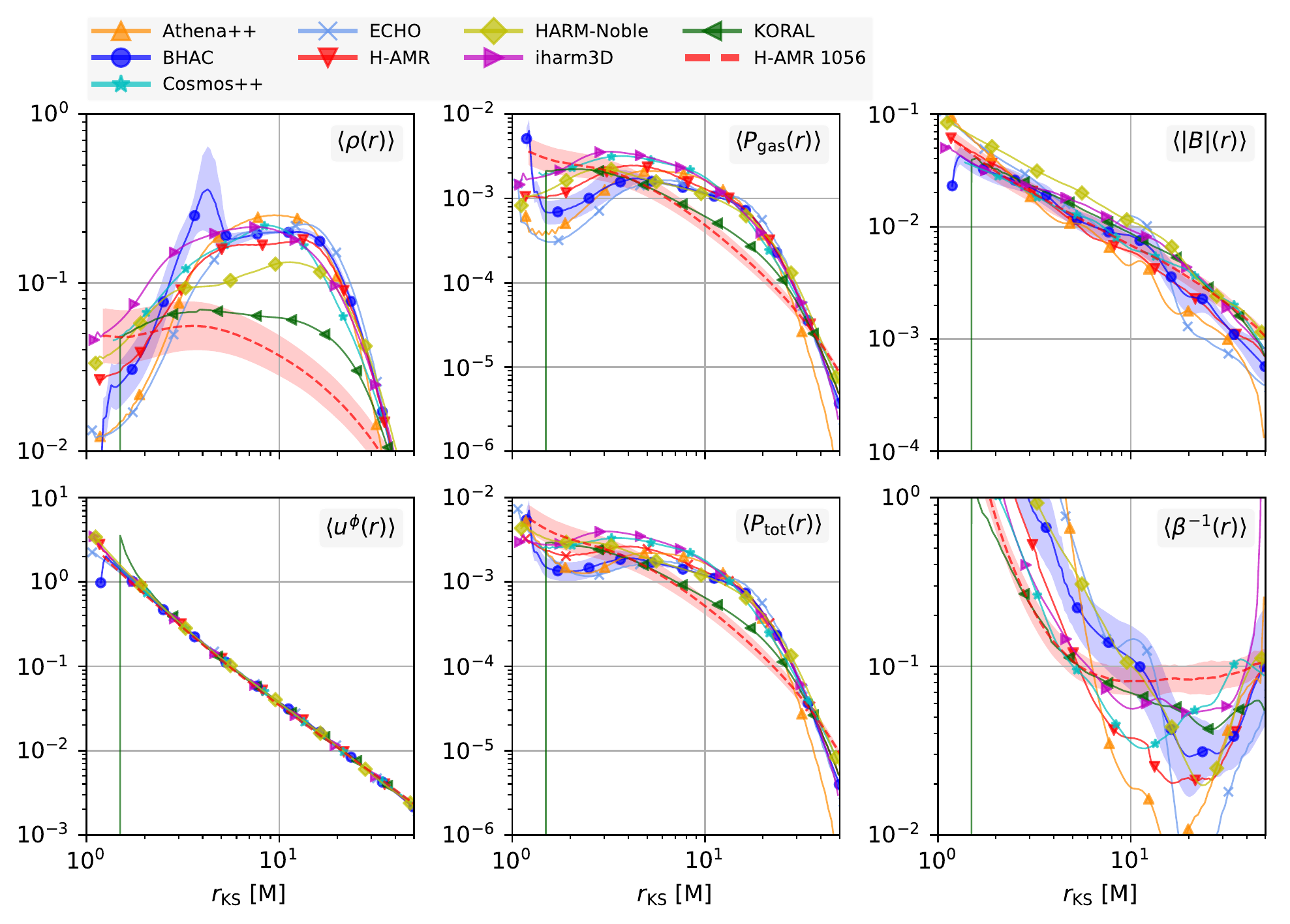}
\caption{Disk-averaged quantities at resolution of $96^3$ in the domain of interest for all codes.  Data has been averaged over the $\theta$ and $\phi$ angles as well as over the time-interval $t\in[5000M,10000M]$.  Standard deviation over time is marked as shaded region (where data was available) to judge the secular evolution within the averaging window.  For reference, the \hamr $N_{\theta}=1056$ solution was added.  See text for details. 
}
\label{fig:pr96}
\end{center}
\end{figure}

\begin{figure}[!htbp]
\begin{center}
\includegraphics[width=15cm]{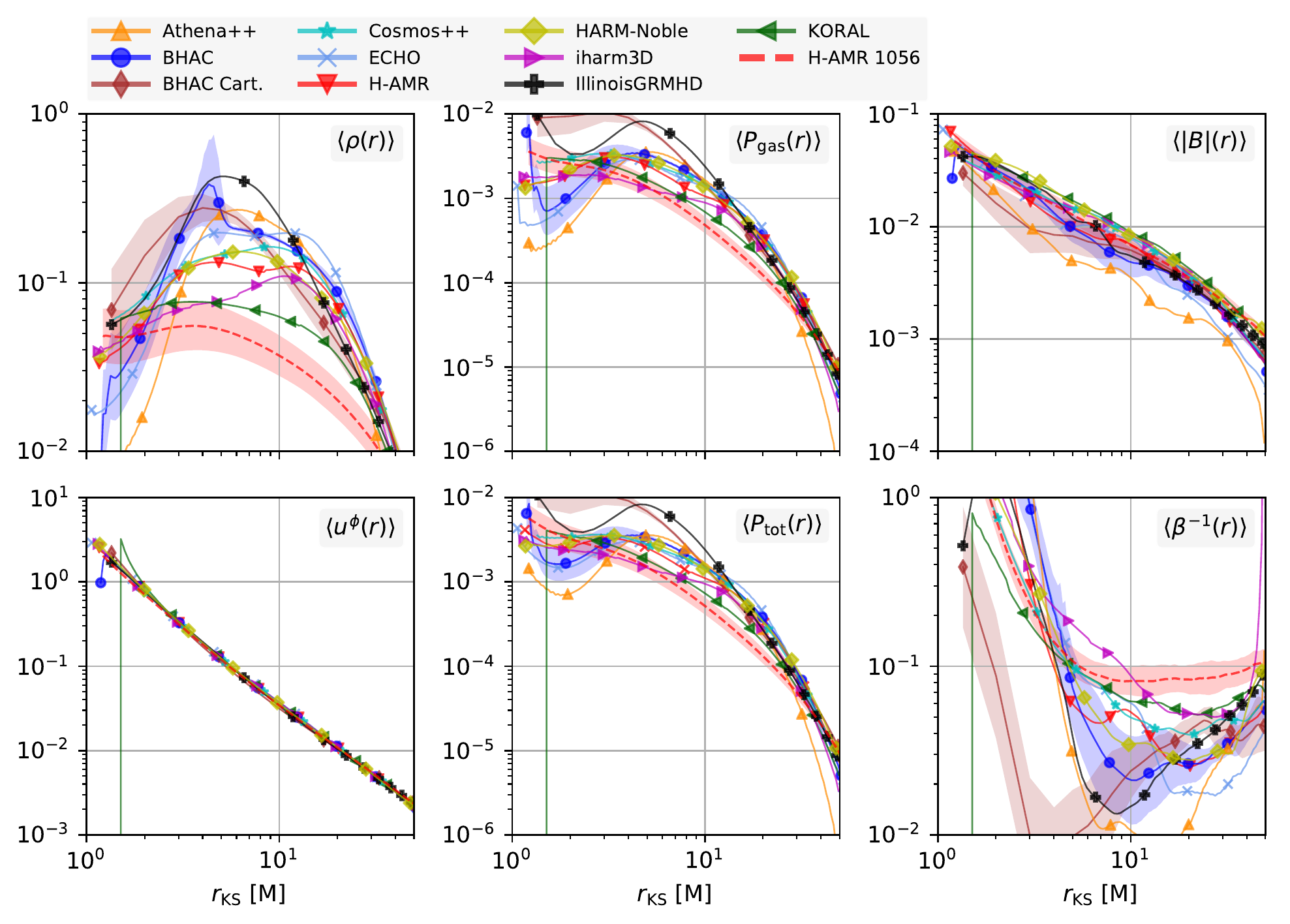}
\caption{As figure \ref{fig:pr96} for the $128^3$ data.}
\label{fig:pr128}
\end{center}
\end{figure}

\begin{figure}[htbp]
\begin{center}
\includegraphics[width=15cm]{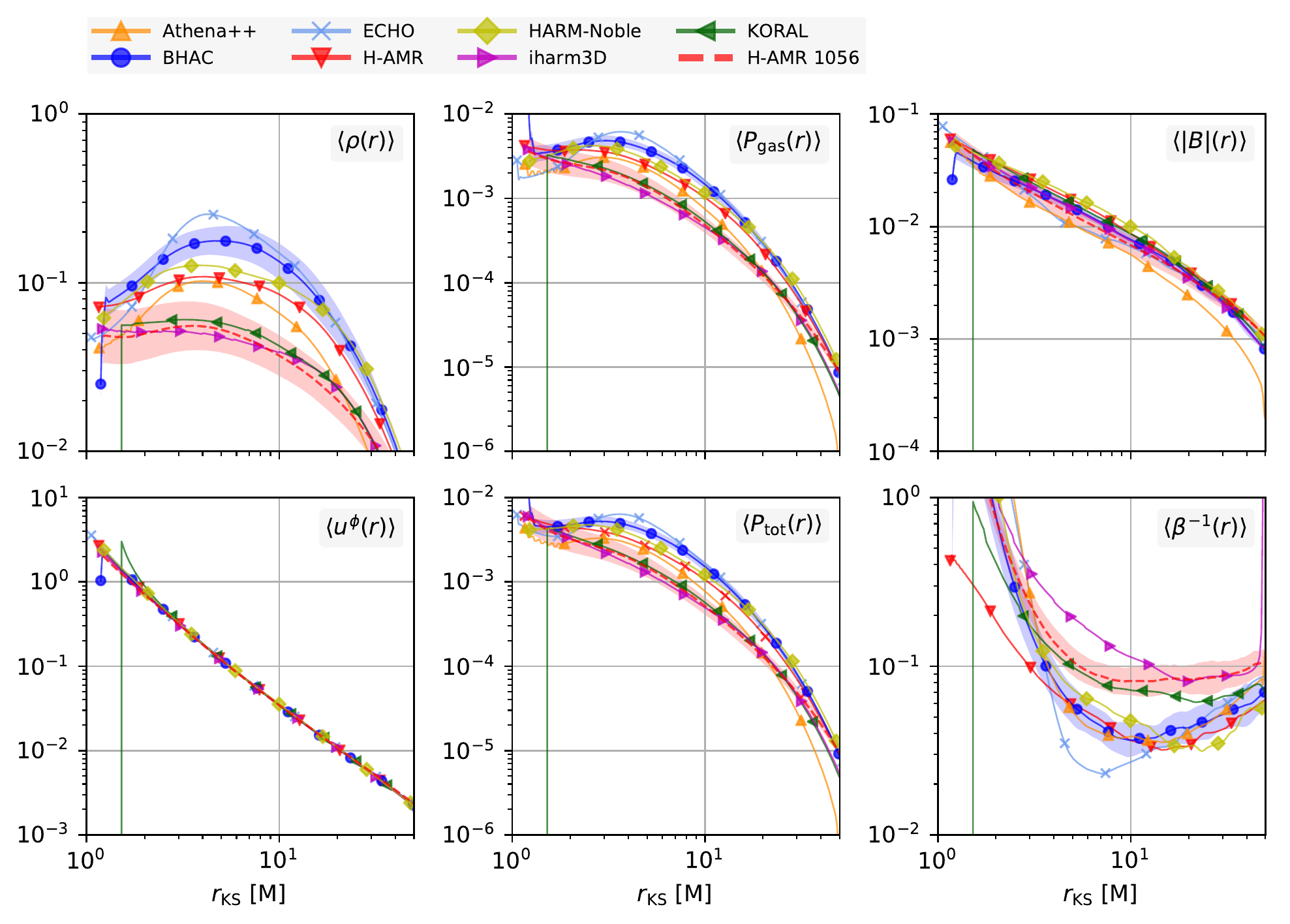}
\caption{As figure \ref{fig:pr96} for the $192^3$ data.}
\label{fig:pr192}
\end{center}
\end{figure}

The disk-averaged profiles of relevant quantities are presented in Figures~\ref{fig:pr96} to \ref{fig:pr192}.
At the late times under consideration, viscous spreading and accretion has significantly transformed the initial distributions; for example the peak densities are now an order of magnitude below the initial state.  
There is a fair spread in some quantities between codes, most notably in the profiles of density and (inverse) plasma beta while all codes capture very well the rotation law of the disk which can be approximated by a powerlaw with slope $r^{-1.75}$, somewhat steeper than the Keplerian case.  Interestingly, also the profiles of magnetic field are captured quite accurately and tend to agree very well with a very high resolution reference case computed with the \hamr code (dashed red curves).  
The approximate scaling as $r^{-1}$ of the disk magnetic field indicates a dominance of the toroidal field component \cite[e.g.][]{HiroseKrolikEtAl2004}.

At low and medium resolution, the density profile in \bhac (blue curves) demonstrates an inner peak or ``mini torus'' which goes away in the high resolution.  As the additional analysis of the MRI quality reveals (see Section \ref{sec:further}), the low- and mid- resolution \bhac runs do not properly resolve the fastest growing wavelengths $\lambda_r, \lambda_\theta$ and hence we consider the stable mini torus a numerical artefact due to lack of angular momentum transport in the under-resolved simulations.  

Taking the inverse plasma-$\beta$ (bottom right panel) as a proxy for the Maxwell stresses, the decrease in torus density is consistent with an increase in turbulent transport of angular momentum due to the increase in magnetization.  Inspection of the torus density profiles shows another trend: We obtain systematically smaller tori in setups where the outer boundary is near the torus edge ({\athenapp}, {\harm}, {\koral}).  This is expected, as the outflow boundary prohibits matter to fall back from large radii as would otherwise occur and hence mass is effectively removed from the simulation.  

Boundary effects are also visible in the run of magnetic field and pressures: again the three small domain models have steepest pressure profiles, with the {\athenapp} showing clear kinks in several quantities at the boundary and the \harm~ measurement for $\langle \beta^{-1}\rangle$ rising sharply at the boundary (due to a few high $\beta^{-1}$ cells in the equatorial region close to the boundary, see for example also Section \ref{sec:t-phi-averages} and Figure \ref{fig:t-phi3}).  

Turning to the high resolution data, the spread in all quantities becomes dramatically reduced.  
Starting at a few gravitational radii, the magnetization of the disk is nearly constant, with values in the range $0.04-0.1$.  Here the Cartesian runs are the exception; their magnetization decreases within $r\sim10\rm M$ and reaches a minimum of $\sim0.01$.  We suppose that this stems from increased numerical dissipation due to the comparatively low resolution in the inner regions of the Cartesian grids.  In fact, we checked how many resolution elements capture the fastest growing MRI mode (quality factor) for \bhac Cart. and confirm that these fall below a critical value of six for $r\sim10\rm M$.  

Taking advantage of the scale-freedom in the ideal MHD approximation, we have also performed a comparison with density, pressure and magnetic field variables re-scaled to the \hamr $N_{\theta}=1056$ solution.  This is exemplified in Appendix \ref{sec:rescaled} which shows the expected better match for the density and pressure profiles.  However, as can already be anticipated from the difference in plasma-$\beta$, the variance in magnetic field is bound to increase.

\FloatBarrier 
\subsection{Axisymmetrized data}\label{sec:t-phi-averages}

\begin{figure}[htbp]
\begin{center}
\includegraphics[height=\textheight]{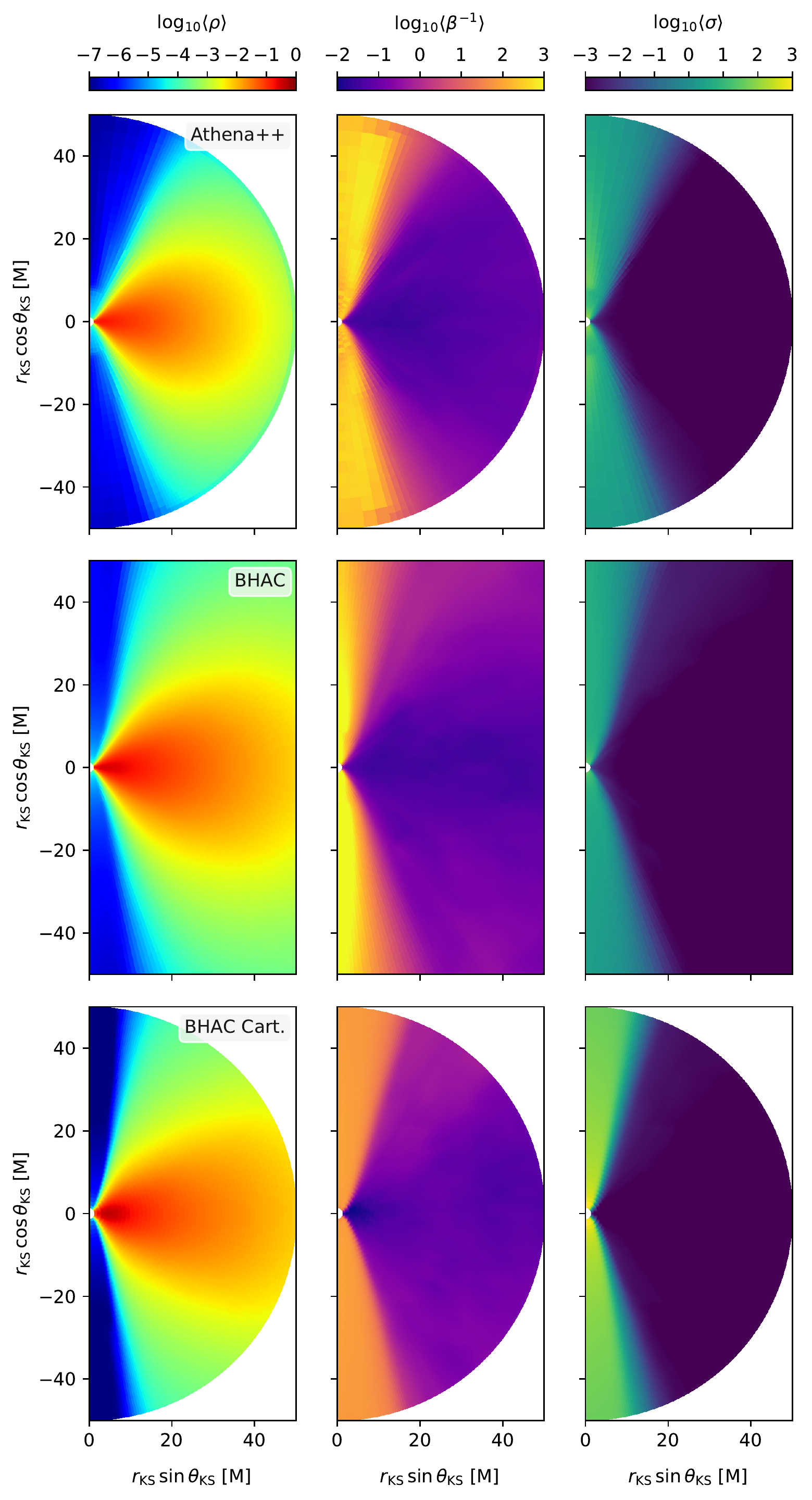}
\caption{$t-$ and $\phi-$averaged data.  We show averages of rest-frame density, inverse plasma-$\beta$ and the magnetization $\sigma$.  From top to bottom: \athenapp{} ($N_{\theta}=192$), \bhac ($N_{\theta}=192$) and \bhac Cart. (cf. Table \ref{tab:settings}).}
\label{fig:t-phi1}
\end{center}
\end{figure}

\begin{figure}[htbp]
\begin{center}
\includegraphics[height=\textheight]{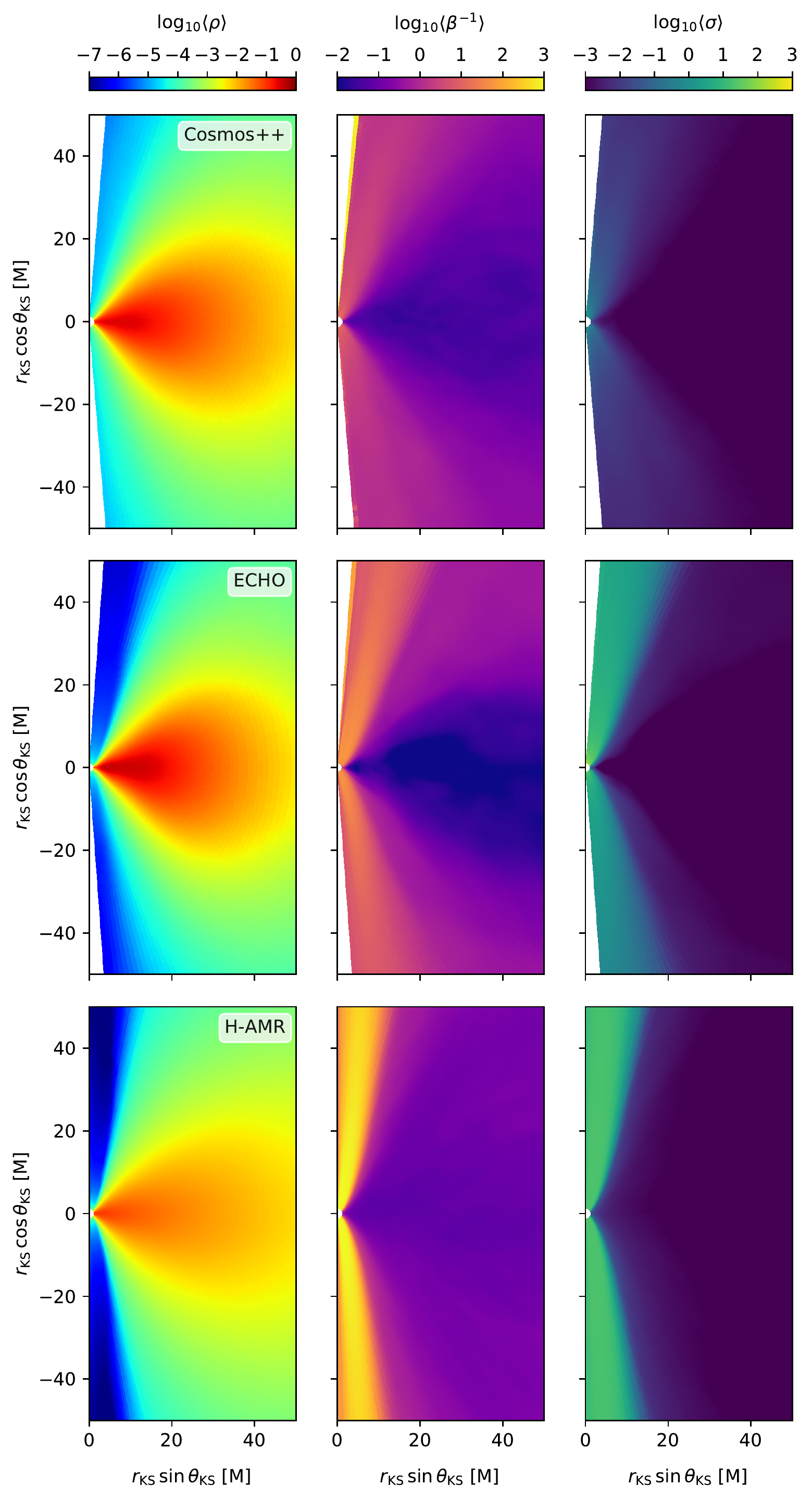}
\caption{As Figure \ref{fig:t-phi1} for the codes: \cosmos ($N_{\theta}=128$), \echo ($N_{\theta}=192$), and \hamr ($N_{\theta}=1056$). }
\label{fig:t-phi2}
\end{center}
\end{figure}

\begin{figure}[htbp]
\begin{center}
\includegraphics[height=\textheight]{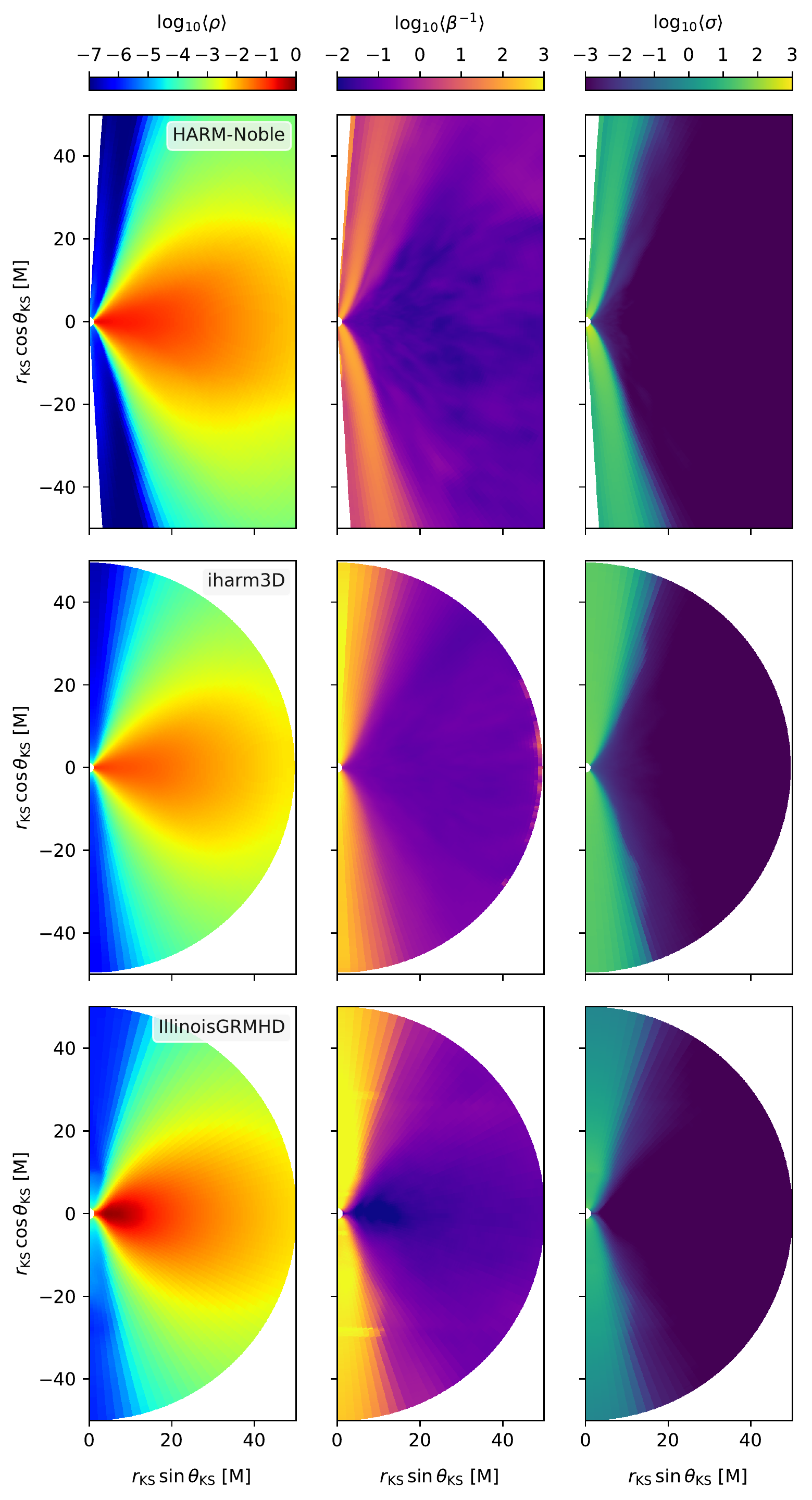}
\caption{As Figure \ref{fig:t-phi1} for the codes: \nobleharm, \harm (both $N_{\theta}=192$) and \IGM (cf. Table \ref{tab:settings}).  }
\label{fig:t-phi3}
\end{center}
\end{figure}

\begin{figure}[htbp]
\begin{center}
\includegraphics[height=0.42\textheight]{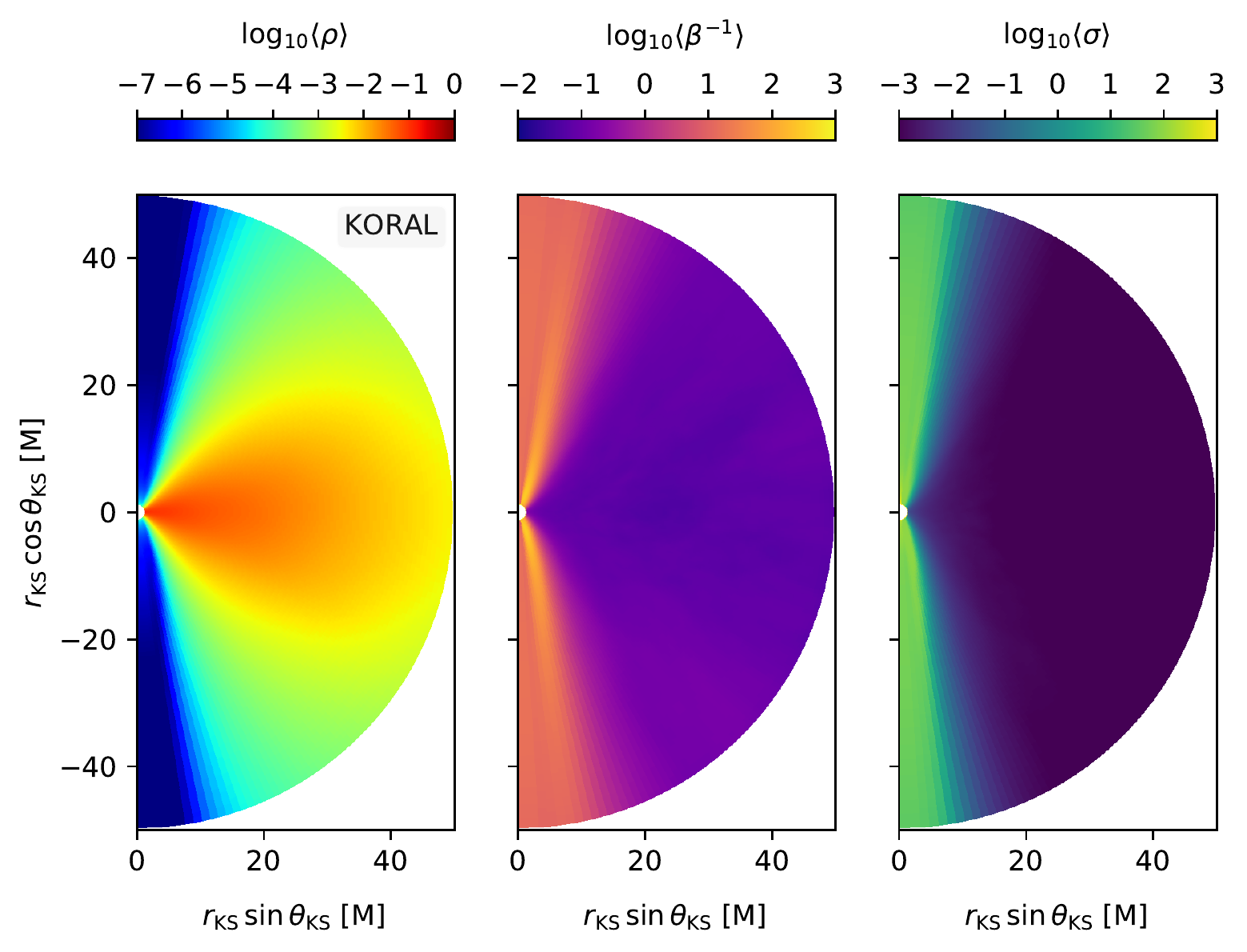}

\caption{As Figure \ref{fig:t-phi1} for \koral with $N_\theta=192$.}
\label{fig:t-phi4}
\end{center}
\end{figure}

To gain an overall impression of the solutions in the quasi-stationary state, we compare averages over the $t-$ and $\phi-$ coordinates in Figures \ref{fig:t-phi1}-\ref{fig:t-phi4}.  The different panels depict rest-frame density, inverse plasma-$\beta$ and the magnetization obtained for each code at the highest resolution available.  

Though qualitatively very similar between codes, the density maps portray correlations between torus size and position of the outflow boundary, noted already in Section \ref{sec:disk}.  Visually the boundary is most pronounced in the \athenapp~ run where the low density region is spread out over a large polar angle.  

As a large variety of floor models is employed, maps of plasma-$\beta$ and magnetization exhibit substantial spread in the funnel region.  
Besides the difference in absolute values reached in the funnel (which are merely related to the floor level), there are also qualitative differences concerning the cells near the axis:  Depending on the pole treatment, the inverse plasma-$\beta$ decreases towards the axis for some codes (\cosmos, \koral, \nobleharm, \athenapp, \hamr) while others see a increase of $\beta^{-1}$ (\bhac, \harm) or a near constant behavior (\bhac Cart.).  
We should note that the cartesian runs do not require any pole treatment at all and as such give perhaps the most reliable answer in this region.  However, also cartesian grids are far from perfect: in the \IGM run we observe some artefacts in the form of horizontal features in $\langle\beta^{-1}\rangle$ which are associated with resolution jumps in the computational grid.  

Significant leakage of magnetic flux leads to a complete loss of the magnetically dominated region in the medium resolution \cosmos run.  A similar behaviour is observed with the medium resolution \nobleharm case (not shown).   In both cases, the polar axis was excised which necessitates the use of boundary conditions at the polar cutout.  If no special precautions are taken, magnetic field simply leaks through the boundary, leaving less flux on the black hole (cf. Section \ref{sec:ts}).  The open outflow boundaries adopted in the \cosmos runs are particularly prone to this effect.  

\begin{figure}[htbp]
\begin{center}
\includegraphics[height=7cm]{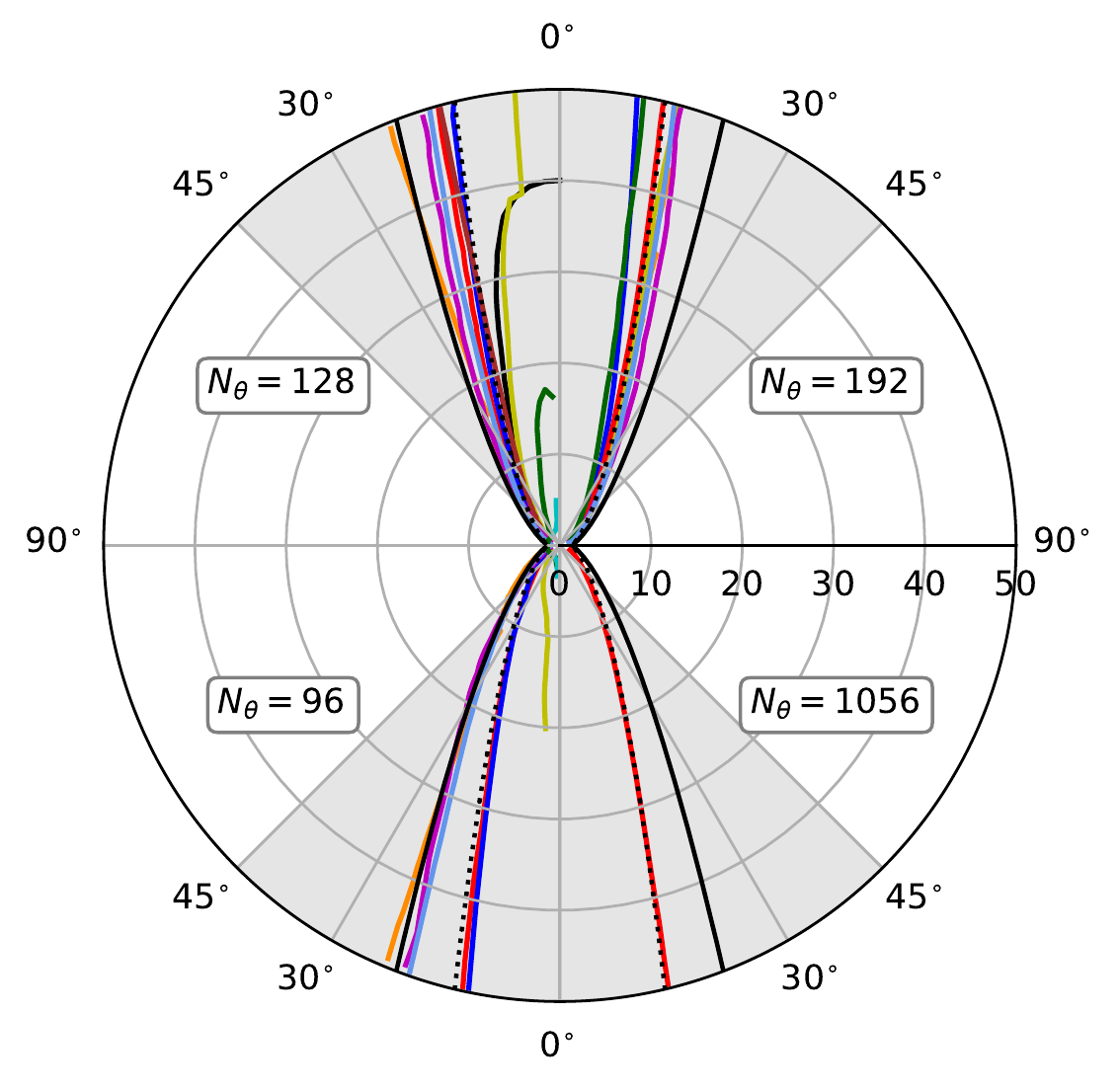}
\includegraphics[height=7cm]{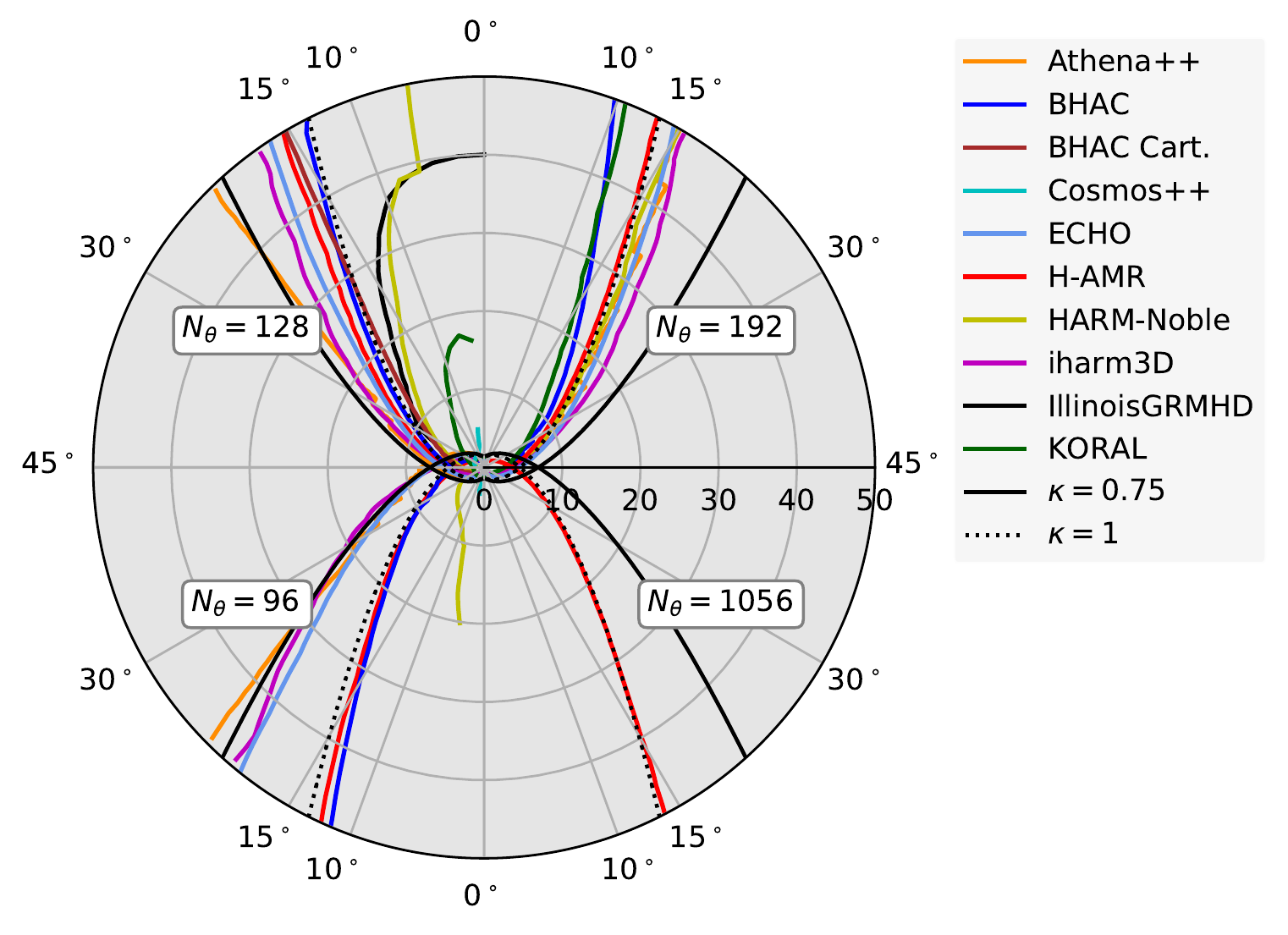}
\caption{Contours of the jet-disk boundary $\sigma=1$ for increasing resolution \textit{(left)} and a zoom into the region $\theta \in [0,45^\circ]$ \textit{(right)}.  Only in the high-resolution case the jet is recovered with all codes.  To guide the eye, we also show paraboloidal curves according to Eq. (\ref{eq:paraboloidal}), corresponding to $z\propto R^2$ ($\kappa=1$, thin dotted black) and $z\propto R^{1.6}$ ($\kappa=0.75$, thin solid black).  The overall spread of the jet angle (given by its theta-coordinate at $r=50\rm M$) is on the $5^\circ$ level.)}
\label{fig:t-phi_sigmacont}
\end{center}
\end{figure}

Although not the focus of this work (reflected e.g. in the choice of gridding by the various groups) it is interesting to examine the jet-disk boundary defined as $\sigma=1$ \citep[e.g.][]{McKinneyGammie2004,Nakamura2018} in more detail.  Figure \ref{fig:t-phi_sigmacont} illustrates that as resolution is increased, the contour is more faithfully recovered across codes:  whereas some low resolution runs do not capture the highly magnetised funnel, at high resolution all runs show a clearly defined jet-disk boundary.  Despite the large variances in floor treatment, at $N_\theta=192$ cells, the difference is reduced to five degrees and the polar angle of the disk-jet boundary ranges between $10^\circ$ and $15^\circ$ at $r=50\rm M$.  
For illustrative purposes, we also overplot flux-functions of the approximate force-free solutions discussed in \cite{Tchekhovskoy2008,Nakamura2018}:
\begin{equation}
    \Psi(r,\theta) = \left(\frac{r}{r_{\rm h}}\right)^{\kappa} (1-\cos(\theta)) \label{eq:paraboloidal}
\end{equation}
In the recent 2D simulations of \cite{Nakamura2018} the jet boundary given by $\sigma=1$ was accurately described by the choice $\kappa=0.75$ for a wide range of initial conditions, black hole spins and horizon penetrating flux $\Phi_{\rm BH}/\sqrt{\dot{M}}\in[5,10]$.  
This results in a field line shape $z\propto R^{1.6}$ which matches well the shape derived from VLBI observations on the scale of $10-10^5\rm M$ \citep[e.g.][]{AsadaNakamura2012,HadaKino2013}.  
In our 3D SANE models, the line $\kappa=0.75$ is recovered at low resolution by \echo and \harm and the more collimated genuinely paraboloidal shape $\kappa=1$ resembling the solution of \cite{BlandfordZnajek1977} seems to be a better match for the funnel-wall shape at higher resolutions.  Incidentally, the match with the $N_\theta=1056$ \hamr run with the paraboloidal shape is particularly good.  
We suspect that the narrower jet profile compared to \cite{Nakamura2018} is due to the lower $\phi\simeq2-3$ of our benchmark configuration (at high resolution).  

\begin{figure}[htbp]
\begin{center}
\includegraphics[width=5cm]{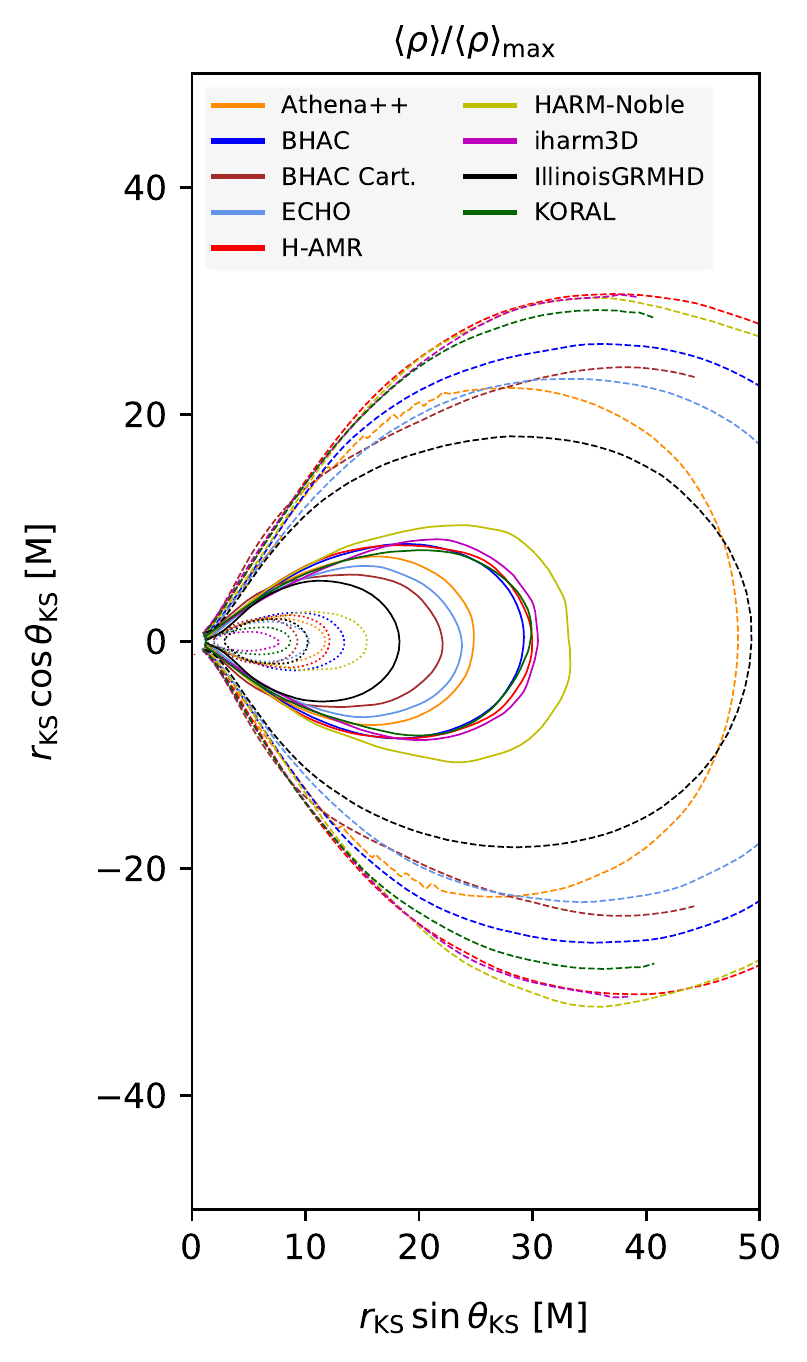}
\caption{Comparison of the density-contours at fixed ratios of the peak value for the high-resolution models.  Contours are placed at $\langle \rho\rangle/\langle \rho \rangle_{\rm max}\in[0.0078125,0.125,0.5]$.}
\label{fig:t-phi_rhocont}
\end{center}
\end{figure}

Comparing the torus density in more detail, in Figure \ref{fig:t-phi_rhocont} we illustrate contours of $\langle\rho\rangle/\langle\rho\rangle_{\rm max}$ for the runs with $N_\theta=192$ cells.  
Normalised in this way, the agreement in torus extent generally improves as it compensates the spread in peak densities (cf. Fig. \ref{fig:pr192}).  Nonetheless there remains a large variance in the given contours.  

\begin{figure}[htbp]
\begin{center}
\includegraphics[width=7cm]{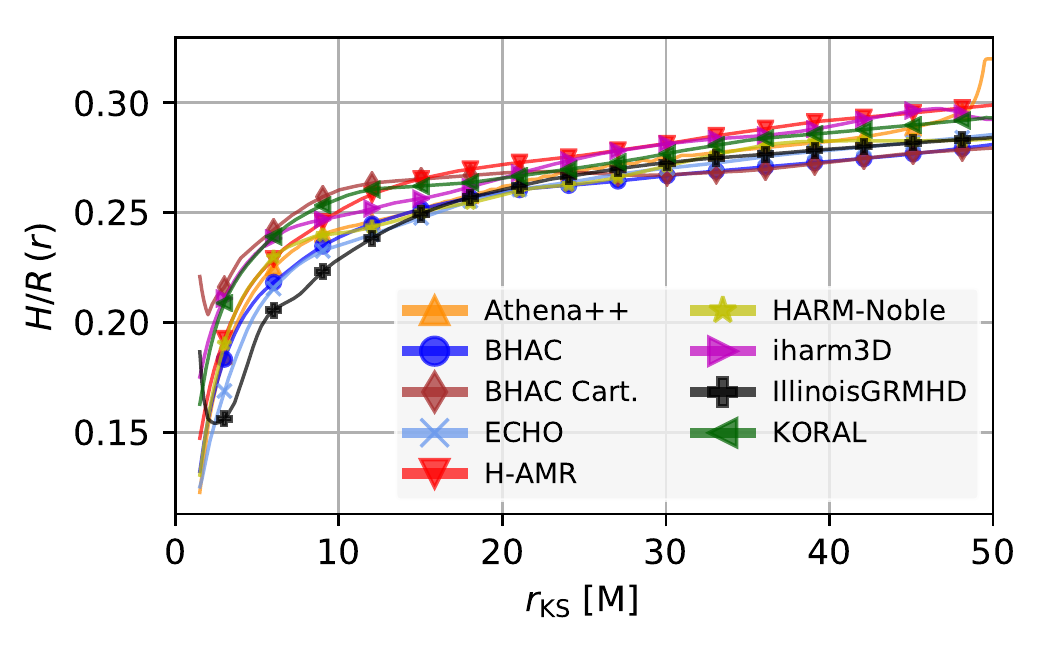}
\caption{Scale height profiles of the high-resolution models ($N_\theta=192$) according to Eq. (\ref{eq:scale-height}).  As expected, all codes are consistent in this quantity, with largest departures seen in the inner regions for the cartesian runs.  }
\label{fig:scaleheight}
\end{center}
\end{figure}

For a more quantitative view, we compute the density-scale height 
\begin{align}
  H/R\, (r) := \frac{\int_{t_{\rm beg}}^{t_{\rm end}}\int_0^{2\pi}\int_{0}^{\pi} \ |\pi/2-\theta_{\rm KS}|\ \rho(r,\theta,\phi,t) \sqrt{-g} \, dt\, d\phi \, d\theta}
  {\int_{t_{\rm beg}}^{t_{\rm end}}\int_0^{2\pi}\int_{0}^{\pi} \rho(r,\theta,\phi,t) \sqrt{-g}\, dt\, d\phi \, d\theta}\, , \label{eq:scale-height}
\end{align}
where the averaging-time is again taken between $t_{\rm beg}=5000\rm M$ and $t_{\rm end}=10\, 000 \rm M$.  As shown in Figure \ref{fig:scaleheight}, a scale height of $H/R=0.25 - 0.3$ between $r_{\rm KS}\in [10,50]$ is consistently recovered with all codes.  The largest departures from the general trend are seen in the inner regions for the cartesian runs which show a slight ``flaring up'' which can also be observed in Figures \ref{fig:t-phi1} and \ref{fig:t-phi3}.  Furthermore, the presence of the outflow boundary in \athenapp leads to a decrease of density in the outer equatorial region (cf. Figure \ref{fig:t-phi1}) which shows up as kink in Figure \ref{fig:scaleheight}.  

\FloatBarrier 
\subsection{Variability analysis}\label{sec:lightcurves}

The analysis of lightcurves from accreting compact objects is a powerful diagnostic of the dynamics of the inner accretion flow.  To get a feeling for the predictive power of variability from GRMHD simulations in a turbulent regime, we here compare the salient features between codes.  

In Sgr A*, episodic flares are detected in X-ray and near-infrared (NIR) on a roughly daily basis \citep{BaganoffBautz2001,NowakNeilsen2012,MossouxGrosso2017} and the recent detection of orbital motion by \cite{GravityCollaborationAbuterEtAl2018}  places the origin of an IR flare within 10 gravitational radii of the black hole.  
In addition to large flares, a low-level continuous variability is present in NIR and radio \citep[e.g.][]{Dodds-EdenGillessen2011}.  
While the nature of the strong coincident X-ray and IR flares is most likely associated with discrete events of particle heating or acceleration \citep{MarkoffFalcke2001,Dodds-EdenPorquet2009}, weaker IR flares could also be caused by lensed thermal plasma in the turbulent accretion flow \citep{ChanPsaltis2015a} and flares in the sub-mm band can be attributed solely to turbulent fluctuations in the accretion flow \citep{DexterAgolEtAl2009,DexterAgol2010}.  
On short timescales (below the characteristic timescales $\tau\sim4$ hrs in IR and $\tau\sim8$ hrs in sub-mm) the continuous variability is well described by a red-noise spectrum with powerlaw slope of $\approx2$  with no indication for a further break down to $\sim 2-10$ min, delimited by the typical sampling \citep{DoGhez2009,DexterKelly2014,WitzelMartinez2018}.  As pointed out by \cite{WitzelMartinez2018}, the IR- and sub-mm characteristic timescales are in fact statistically compatible indicating a direct relation of the emitting regions.  
For low-luminosity AGN, \cite{BowerMarkoff2015} have found that the sub-mm characteristic timescale is consistent with a linear dependence on the black hole mass as would be expected for optically thin emission in radiatively inefficient general relativistic models.  Hence in M87, the variability with its characteristic timescale of $45^{+61}_{-24}$ days \citep{BowerMarkoff2015} can be obtained by scaling of the Sgr A* result.  

Due to the potential impact of the space-time on the variability of the emission\footnote{Numerical simulations by \cite{DolenceGammie2012} have observed a steepening of the IR and X-ray spectral index to $\sim3$ accompanied by quasi periodic oscillations (QPOs) at the frequency of the innermost stable circular orbit (ISCO) (corresponding to $f_0=0.041/M$ in our case and thus periods of $8$ minutes if scaled to Sgr A*).}, it is clear that the characterisation of the spectrum can hold great merit.  
From theoretical grounds, a break in the power spectrum at the orbital frequency of the innermost emitting annulus (often identified with the ISCO) is expected (see also the discussion in \citealt{ArmitageReynolds2003}).  
This high-frequency break is however close to the current sampling frequency of a few minutes for Sgr A*.  

To get a feeling for the mm- variability without actually subjecting each code's output to a rigorous general-relativistic radiative transfer calculation (GRRT), an approximate optically thin lightcurve was calculated as a volume integral over the domain of interest according to Eq. (\ref{eq:j}).  
Before we discuss the variability from this pseudo-emissivity, let us briefly compare its properties to an actual GRRT calculation of an exemplary simulation output.  For this purpose, we employ the \bhoss~ code \citep{YounsiEtAl2019} to compute the 1.3 mm emission corresponding to the EHT observing frequency from the high resolution \bhac data scaled to Sgr A*.  

The lightcurves are computed assuming a resolution of $1024\times 1024$ pixels and a field-of-view of 100 M in the horizontal and vertical directions, i.e., $[-50~{\rm M},50~{\rm M}]$.  We here neglect the effect of the finite light travel time through the domain (also known as the ``fast light'' approximation, see e.g.  \cite{DexterAgol2010} for a discussion).  
The chosen synchrotron emissivity is the photon pitch angle-dependent approximate formula from Leung et al.~(2011) [eqn.~(72) therein] and the corresponding self-absorption is included. The ion-to-electron temperature ratio used to obtain the local electron temperature as a function of the local ion temperature is fixed to $3$ as in the intermediate model of \cite{MoscibrodzkaGammieEtAl2009}.
The domain within which the GRRT calculations are performed is set identical to that specified in the emission proxy, i.e., eqn.~\eqref{eq:j}.
The mass and distance of Sgr A* adopt the standard fiducial values \citep{GenzelEisenhauer2010}, and the 1.3 mm flux is normalised to $3.4$~Jy 
\citep{MarroneMoran2007} for an observer inclination of $60^{\circ}$ using the fiducial snapshot at 10\,000~M.

\begin{figure}[htbp]
\begin{center}
\includegraphics[width=\textwidth]{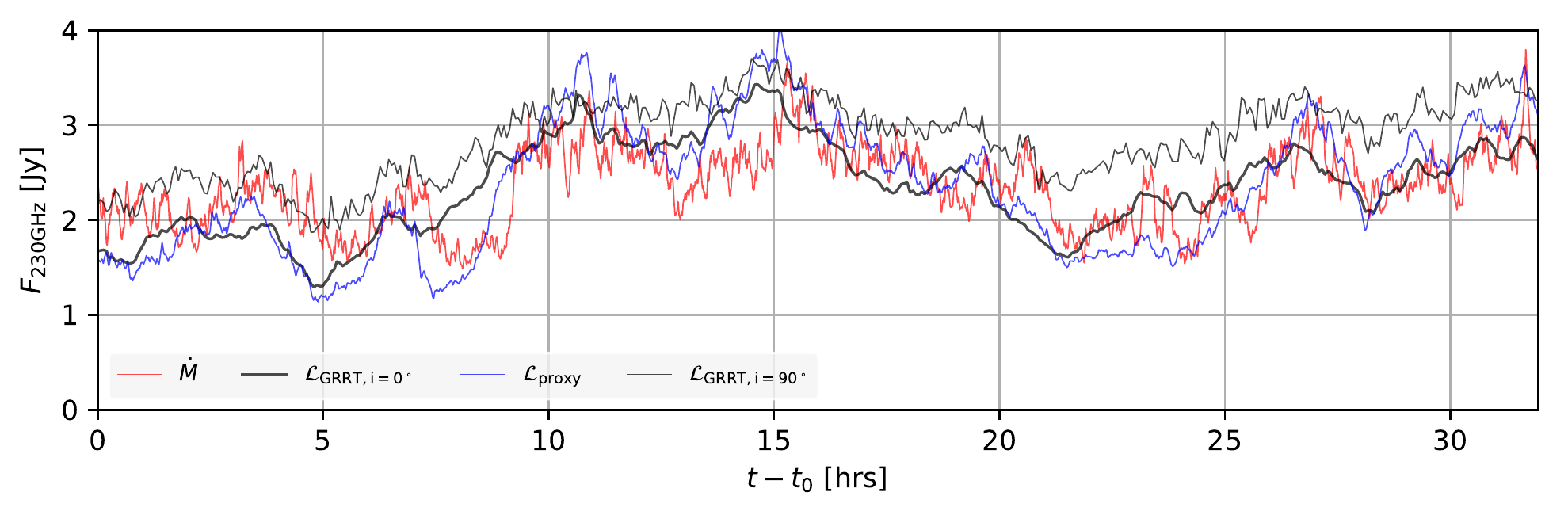}
\caption{Comparison of the lightcurves scaled to $\rm Sgr~A^*$ from the \bhac high-resolution run showing the typical short-term variability and long-term stability of low-luminosity AGN: emission proxy $\mathcal{L}_{\rm proxy}$ (blue), accretion rate $\dot{M}$ (red) and two synthetic lightcurves obtained with the GRRT \bhoss code: inclination $i=0^\circ$ ($\mathcal{L}_{\rm proxy,\ i=0^\circ}$; black, thick) and $i=90^\circ$ ($\mathcal{L}_{\rm proxy,\ i=90^\circ}$; black, thin).  The accretion rate and emission proxy have been scaled to match the mean flux of the $i=0^\circ$ calculation and time has been reset at $t_0=5000 M$ of the simulation.  Both follow the overall trend of the GRRT lightcurve, though the accretion rate exhibits substantially more small-scale variability.  One hour corresponds to $\sim 180\rm \, M$ simulation time.}  
\label{fig:lightcurve}
\end{center}
\end{figure}

A comparison of the lightcurves for $i=0^\circ$, $i=90^\circ$, the (scaled) accretion rate and the emission proxy is given in Figure \ref{fig:lightcurve}.  
It demonstrates that both the accretion rate and the emission proxy follow reasonably well the trend of the GRRT lightcurve on the scale of a few hours.  However, the accretion rate shows more pronounced variability on small scales than the proxy or the GRRT calculations which can be understood as a consequence of the extended size of the emission region in the latter two prescriptions.
Doppler boosting adds additional variability to the lightcurve which leads to larger fluctuations in the edge-on lightcurve compared  to the face on ($i=0^\circ$) curve.  

Having validated the emission proxy as a reasonable descriptor for the flux in the mm-band, we now compute the power-spectrum-densities (PSDs) of the emission proxy $\mathcal{L}$.  The PSD is obtained by dividing the data into three non-overlapping segments between $5000$ M and $10\, 000$ M and fast Fourier transform of each Hamming-windowed segment, followed by Fibonnacci binning in frequency space.  The three binned PSDs are then averaged and we also take note of the standard deviation between them.  To validate this procedure, also synthetic pure red-noise data with a PSD of $f^{-\beta}$ with $\beta\in \{2,3\}$ was created and is faithfully recovered by the algorithm.\footnote{Note that due to the spectral leakage of the finite time lightcurves, using a boxcar window resulted in artificially flattening of the $\beta=3$ spectrum to $\beta\sim2.5$}  It is clear that as each window only contains $10\, \rm hrs$, the observed break at the characteristic frequency $\tau\sim8\rm hrs$ (for Sgr A*) is not measurable by this procedure.  

\begin{figure}[htbp]
\begin{center}
\includegraphics[width=\textwidth]{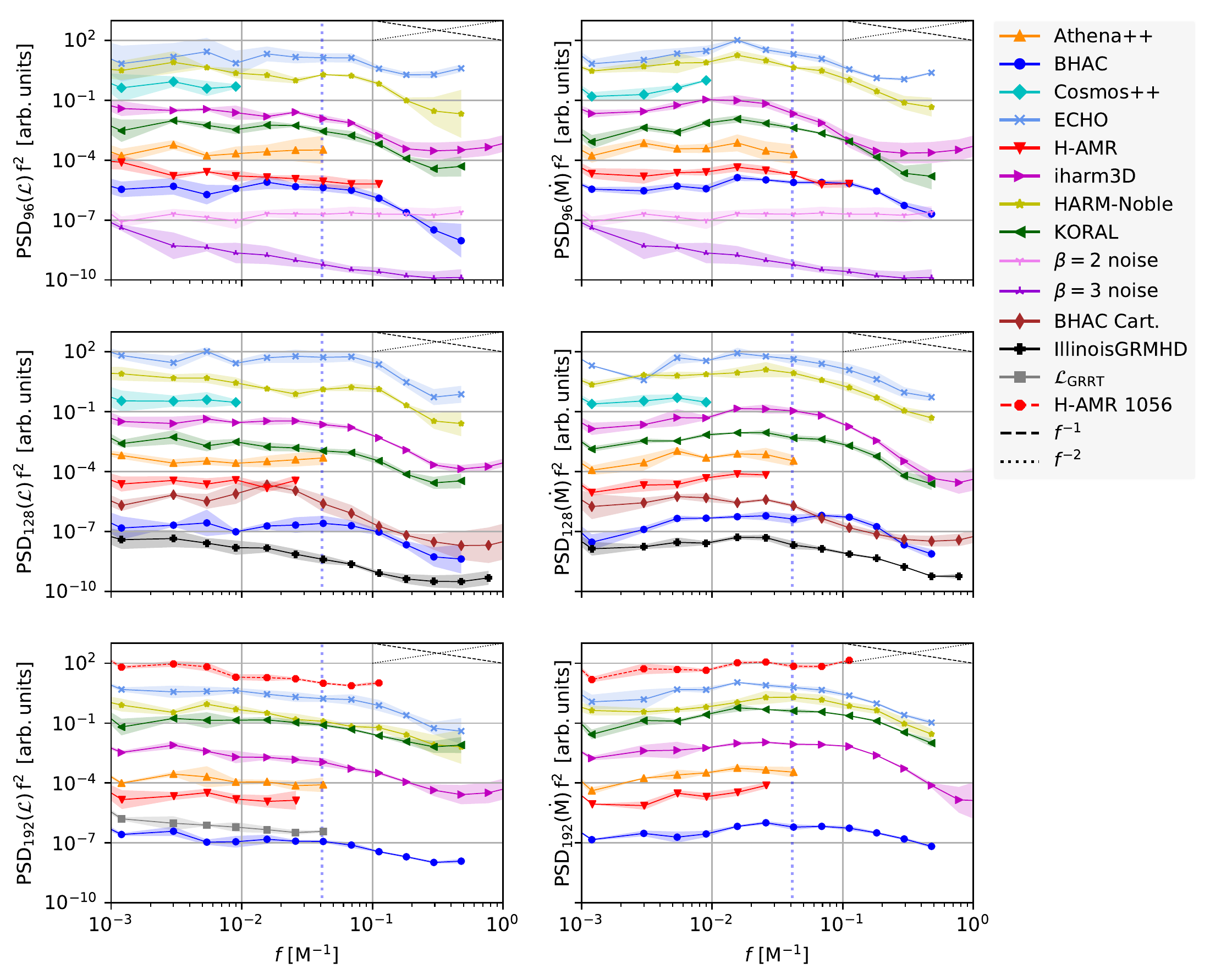}
\caption{Periodograms, compensated with $f^2$, of the emission proxy $\mathcal{L}$ (left) and for the accretion rate $\dot{M}$ (right) for increasing resolutions (top to bottom).  Also the PSD of a synthetic lightcurve at $230\,\rm GHz$ and inclination $i=0^\circ$ is shown in the lower panel (light gray curve).  To better visually separate the curves, they have been progressively shifted by factors of ten. The blue dotted vertical line denotes the orbital frequency of the ISCO.  
In the data for $\dot{M}$, we obtain a flatter low-frequency spectrum and a peak in the vicinity of the ISCO frequency of $f_0=0.04\rm M^{-1}$. 
}
\label{fig:periodograms}
\end{center}
\end{figure}

The data is presented in Figure \ref{fig:periodograms} for all available resolutions.  
At low- and mid-resolution, all lightcurves are compatible with a red-noise spectrum with $\beta=2$ and we obtain a steepening near $f=0.1 M^{-1}$.  
For high resolution, this break is less apparent, however there is a slight trend for an overall steepening to $\beta=3$ e.g. in \nobleharm, \harm and \IGM.   
It is reassuring that the lightcurve obtained by GRRT at $i=0^\circ$ (light gray) shown in the lower left panel displays a very similar behavior as the corresponding data obtained via the proxy (\bhac code, blue).  

Since the accretion rates have been computed with all codes and generally are an easily accessible and relevant diagnostic in GRMHD simulations which can give some guidance on the overall variability features \citep[see e.g.][]{ReynoldsMiller2009,DexterAgol2010}, we perform the same analysis with the time-series of $\dot{M}$.  
The right panels of Figure \ref{fig:periodograms} indicates that the low-frequency powerlaw seen in $\mathcal{L}$ is not recovered in $\dot{M}$.  For frequencies $f\lesssim 0.01 \rm M^{-1}$, the power in all codes is definitely shallower than $f^{-2}$, approaching a flicker-type noise, $f^{-1}$ and we find indications for a spectral break in the vicinity of the ISCO frequency in several codes. 
The $\beta=1$ slope for $f<10^{-2}\rm M$ is consistent with e.g. \cite{hoggReynolds2016} who extracted the accretion rate at the ISCO itself.  The latter authors also note that the PSD of an ``emission proxy'' is steeper than the one obtained from $\dot{M}$, due to the fact that additional low-frequency power is added from larger radii in the extended emission proxy.  

\begin{figure}[htbp]
\begin{center}
\includegraphics[width=0.45\textwidth]{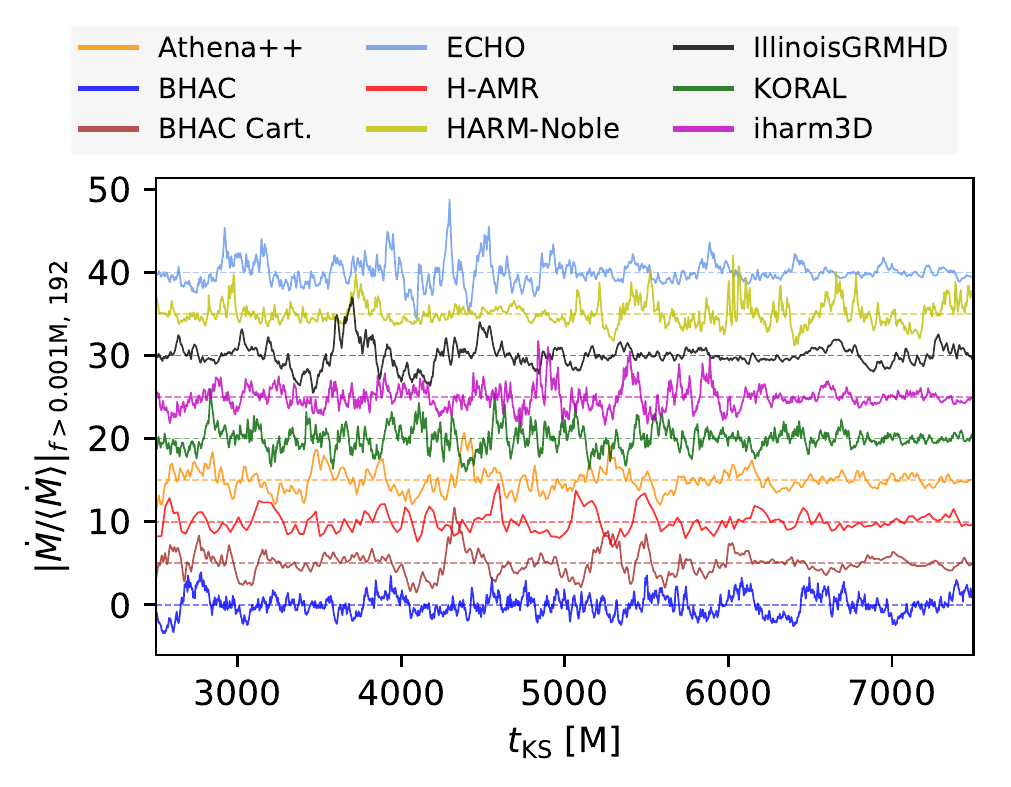}
\includegraphics[width=0.45\textwidth]{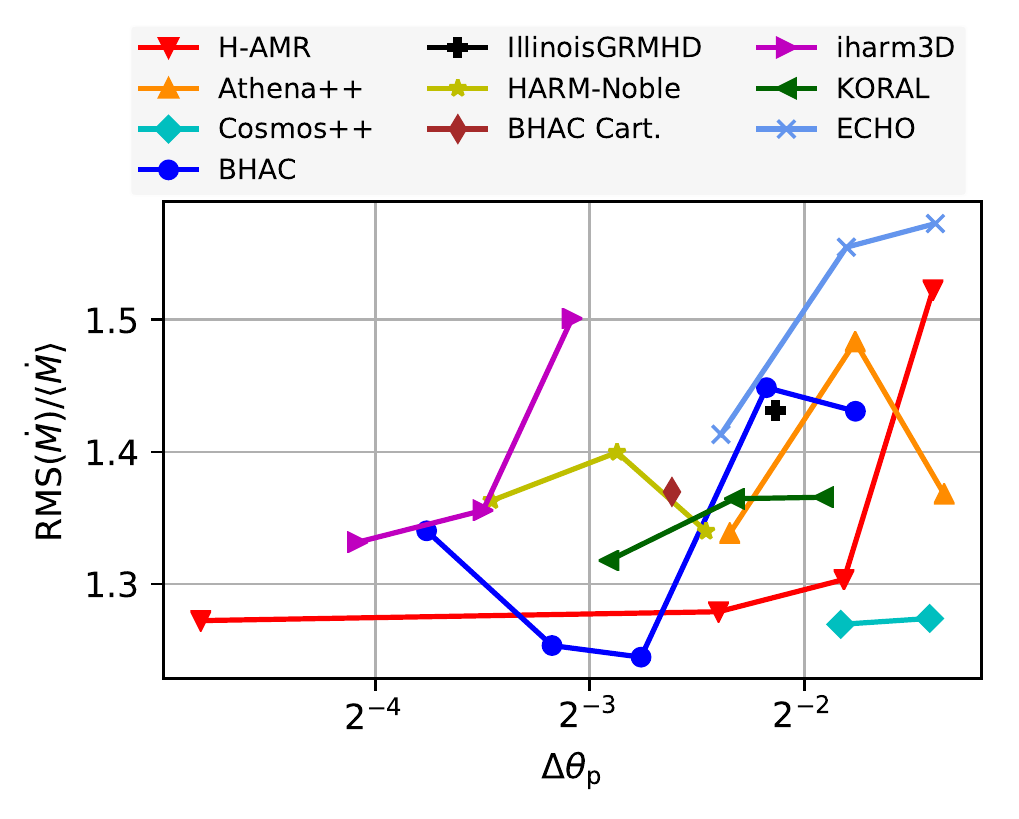}
\caption{Variability of the detrended accretion rates.  The long time secular evolution has been removed by zeroing out all Fourier amplitudes below $1/1000\rm M$.  An example of the high-resolution data is shown in the left panel.  In the right, we compare the resolution dependence in the root-mean-square accretion rate normalized by its mean value.  We obtain a clear ``RMS-flux'' relation in all codes with quotient converging to $k\simeq1.3$.}
\label{fig:RMS}
\end{center}
\end{figure}

To also compare the variability magnitude, we have computed the root-mean-square (RMS) of the accretion rate after zeroing out all Fourier amplitudes with frequency below $1/1000\rm M$.  This serves as an effective detrending of the low-frequency secular evolution.  In order to avoid edge-effects, we compute the RMS in the region $t\in[2500,7500]\rm M$.  Exemplary data of the remaining high-frequency variability is shown for the high-resolution runs in the left panel of Figure \ref{fig:RMS} where we have normalized by the average accretion rate in the region of interest.  One can visually make out differences in variability amplitude with the \bhac data on the low end and \nobleharm, \harm on the high end.  
The (normalized) RMS values shown against mid-plane resolution (right panel) quantify this further and indicate quite a universal behaviour with values in the range $1.2-1.6$ across all codes and resolutions.\footnote{Using the non-normalized (raw) values of the RMS yielded far less consistent results with differing resolution dependent trends between codes.  Furthermore, exceptional low-number statistics 'flare' events dominate over raw lightcurves, as for example seen in the  low-resolution \harm data in Figure \ref{fig:ts96}.}  
  With increased resolution, all codes tend to be attracted to the point $k=$RMS$(\dot{M})/\langle \dot{M} \rangle\simeq 1.3$.  Repeating the same analysis with the emission proxy yielded essentially the same outcome.  
This quite striking result (after all, there is a scatter in mean accretion rates itself by a factor of $\sim 8$ in the sample) is a re-statement of the RMS-flux relationship which is a ubiquitous feature of black hole accretion across all mass ranges (\cite{UttleyMcHardy2001,HeilVaughanEtAl2012}). The quotient $k\sim 1.3$ is consistent with the recent simulations of  \cite{hoggReynolds2016} who find an RMS-flux relationship of the accretion rate with $k=1.4\pm 0.4$.  

Summarising the comparison of variability, although it remains a challenge to extract accurate power spectra from the finite time series of the data, all codes agree quite well on the salient features: 
\begin{enumerate}
    \item A $\beta\simeq2$ powerlaw slope is recovered for the emission proxy $\mathcal{L}$ with all codes and resolutions.
    \item The PSD for $\mathcal{L}$ and $\dot{M}$ steepens at $f\simeq0.1\rm M^{-1}$.
    \item The PSD of the accretion rate is shallower than the one for the proxy and is more accurately described by an index of $\beta\simeq1$.
    \item The ``RMS-flux'' relationship is quite universally recovered for $\dot{M}$ and $\mathcal{L}$ for all codes and resolutions.
\end{enumerate}

\FloatBarrier 
\subsection{Further analysis}\label{sec:further}

To get a deeper understanding of the dynamical effect that increasing the grid resolution has, some further analysis was carried out with results of \bhac and \hamr.  

\subsubsection{Maxwell stress and $\alpha$}\label{sec:maxwell}

We first compute the disks $\alpha$-parameter due to Maxwell stresses, hence we are interested only in the term $-b^r b^\phi$ of the Energy-Momentum tensor (\ref{TMUNUMHD}).  
As noted however by \cite{KrolikHawley2005, Beckwith2008, PennaScadowskiEtAl2013}, the stress depends on the coordinate system and cannot unambiguously be defined in general relativity.  
The best one can do is to measure the stress in a locally flat frame co-moving with the fluid.  Even so, further ambiguities arise when defining the orientation of the co-moving tetrad and hence the $b^{(r)}$ and $b^{(\phi)}$ directions.  
With the aim of merely comparing codes and convergence properties in global disk simulations, we here settle for the simpler diagnostic in the Kerr-Schild coordinate frame.  As shown in Appendix \ref{sec:tetrads}, for our case, this is a fair approximation to the fluid-frame stress in the most commonly used basis. 
Hence using the same operation as for the disk profiles (see Equation (\ref{eq:averaged})), we compute the shell average of 
\begin{align}
    \langle w^{r\phi}\rangle(r,t) := \langle -b^r \sqrt{\gamma_{rr}} b^\phi \sqrt{\gamma_{\phi\phi}} \rangle
\end{align}
and define the $\alpha(r,t)$-parameter due to Maxwell stresses as 
\begin{align}
    \alpha(r,t) := \frac{\langle w^{r\phi}\rangle(r,t)}{\langle P_{\rm tot}\rangle(r,t)}
\end{align}
This $\alpha(r,t)$ is further averaged in time to yield $\alpha(r)$ or volume averaged to yield $\bar{\alpha}(t)$.  

\begin{figure}[htbp]
\begin{center}
\includegraphics[width=\textwidth]{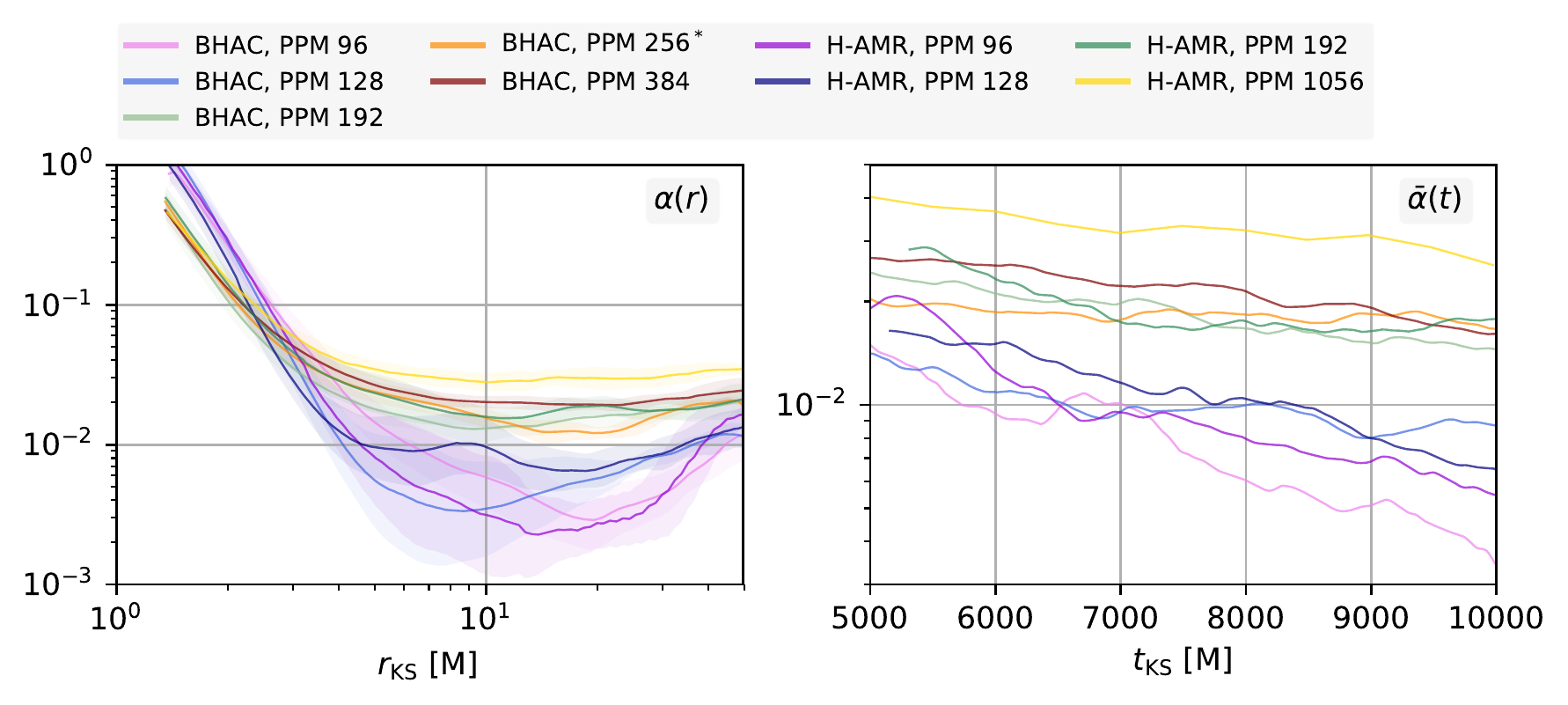}
\caption{Maxwell stress analysis.  \textit{Left:} profiles of disk $\alpha(r)$ parameter.  Shaded regions indicate the standard deviation over time.  \textit{Right:} Time evolution of the spatially averaged $\bar{\alpha}(t)$.  The $\alpha(r)$ parameter is constant throughout most of the disk and behaves similar to $\langle \beta^{-1}\rangle$ as shown in Figures \ref{fig:pr96} - \ref{fig:pr192}.}
\label{fig:stress}
\end{center}
\end{figure}

Figure \ref{fig:stress} illustrates the resolution dependence of the $\alpha$-parameter for the \hamr and \bhac runs.  Overall, there is good agreement between \bhac and \hamr for $N_\theta>128$.  Not surprisingly, the stress profiles resemble closely the run of $\langle\beta^{-1}\rangle$ already shown in Figures \ref{fig:pr96}-\ref{fig:pr192}, but $\alpha(r)$ is lower by a factor of $\sim 3$.  
In particular, there is no ``stress-edge'' at or within the ISCO (located at $r_{\rm KS}=2.04\rm M$) where $\alpha$ drops to zero.  This is consistent with results of \cite{KrolikHawley2005} for the highly spinning case and marginally consistent with \cite{PennaScadowskiEtAl2013} who find that $\alpha(r)$ peaks at $\sim 2-3\rm M$.  The latter difference can be explained in part through the coordinate choice and in part through the higher spin adopted here (see also Appendix \ref{sec:tetrads}).  
With increased resolution, $\alpha$ in our global simulations increases, as opposed to what is observed in local shearing boxes.  For example, \cite{BodoCattaneoEtAl2014,RyanGammie2017} report $\bar{\alpha}\propto N^{-1/3}$ for stratified local simulations and \cite{GuanGammieEtAl2009} found $\bar{\alpha}\propto N^{-1}$ for the unstratified case ($N$ is the number grid cells per scale-height).  Over time, $\bar{\alpha}$ decreases, but less so as the grid resolution is increased.  This secular trend and the large gap in resolution between the $N_\theta=384$ \bhac run and the $N_\theta=1056$ \hamr run make a strict quantification of the convergence behavior difficult. At a comparatively high resolution of $N_\theta=384$ though, it seems that $\bar{\alpha}$ has not converged yet.  

\subsubsection{Viscous spreading}\label{sec:viscous}
\begin{figure}[htbp]
\begin{center}
\includegraphics[width=0.7\textwidth]{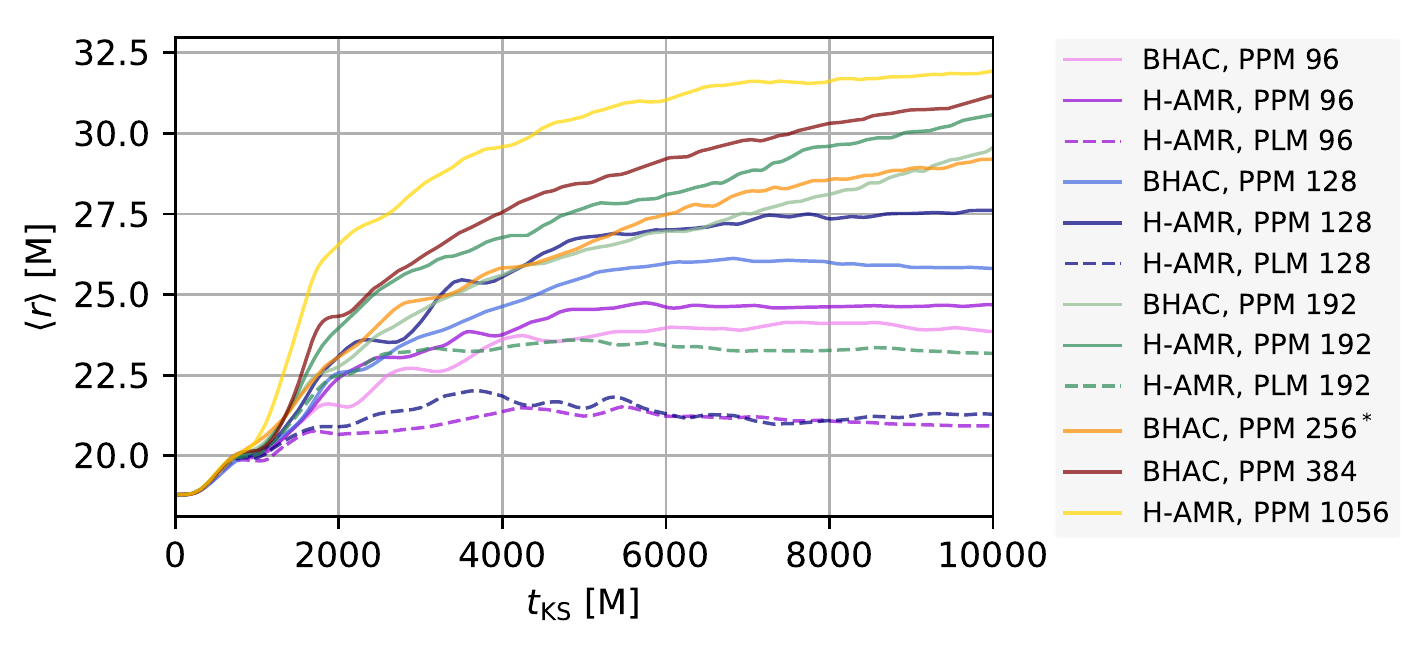}
\caption{Barycentric radii of the disk according to Eq. \ref{eq:rbarycenter}.  The data has been computed with \bhac and \hamr both using linear (PLM, dashed curves) and piecewise parabolic reconstruction (PPM, solid curves).  The disk stalls its spreading early for low- and medium resolution and even shrinks when linear reconstruction is employed.
}
\label{fig:rbary}
\end{center}
\end{figure}

The viscosity induced by the MRI leads to accretion of material and viscous spreading of the disk as a whole.  Hence with increased resolution and thus higher disk stresses, a global effect on the viscous spreading is expected.  We quantify this by calculating the rest-frame density weighted radius $\langle r \rangle$ according to 
\begin{align}
  \langle r \rangle(t) := \frac{\int_0^{2\pi}\int_{0}^{\pi}\int_{r_{\rm h}}^{r_{\rm max}} \ r_{\rm KS}\ \rho(r,\theta,\phi,t) \sqrt{-g} \, d\phi \, d\theta\, dr}
  {\int_0^{2\pi}\int_{0}^{\pi}\int_{r_{\rm h}}^{r_{\rm max}} \rho(r,\theta,\phi,t) \sqrt{-g} \, d\phi \, d\theta\, dr} \, \label{eq:rbarycenter}
\end{align}
where the outer radius of integration has again been set to the domain of interest $r_{\rm max}=50 \rm M$.  As more mass is expelled beyond $50\rm M$ though, a larger integration domain would yield somewhat different results.  
The corresponding temporal evolution is displayed in Figure~\ref{fig:rbary} for \bhac and \hamr where linear and piecewise parabolic reconstruction have been employed.  It is directly apparent that the disk size increases with resolution.  Furthermore, the saturation of the low- and medium resolution PPM runs occurring at $t\sim 6000 $ to values of $\langle r \rangle \in [24, 27]$ are indications that the turbulence starts to decay at this time.  It is noteworthy that the linear reconstruction experiments with \hamr indicate a shrinking of the disk even at a high resolution of $N_\theta=192$.  

The behavior of $\langle r \rangle$ is consistent with the fact that higher resolution in the disk leads to larger Maxwell stresses, driving stronger disk-winds and thus larger disk expansion.  This lowers the overall disk density (as seen in Fig. \ref{fig:pr96} - \ref{fig:pr192}: $\langle \rho(r) \rangle$), but the disks magnetic field is affected to a lesser extent (cf. Fig. \ref{fig:pr96} - \ref{fig:pr192}: $\langle |B|(r) \rangle$), hence the disks $\alpha$ increases further in a non-linear fashion.  As shown in Section \ref{sec:highres} however, the disk radius starts to converge for $N_\theta>192$,  indicating that the global effect of MRI stresses is accurately captured.

\subsubsection{MRI quality factors}\label{sec:mri-Q}

These resolution dependent effects can be understood by an inspection of the ``MRI quality factors'' (in short $Q$-factors), defined as the number of cells available to resolve the fastest growing MRI mode in a given direction.  While the fastest growing mode might be well resolved in the initial condition, whether this remains to be the case throughout the simulation can only established a-posteriori.  If the simulation fails to capture at least the fastest growing mode, turbulent field amplification cannot proceed and the field will decay due to numerical dissipation, leading also to a cessation of mass accretion.  Although strictly a criterion for sufficient resolution of the linear MRI, the $Q$-factors also inform on the adequacy in the non-linear regime \citep[see e.g.][]{SanoInutsoka2004,NobleKrolik2010,HawleyRichers2013}.  
The consensus values for adequate resolution are given by $Q^{(z)} \ge 10$, $Q^{(\phi)} \ge 20-25$ of \cite{HawleyGuanKrolik2011,HawleyRichers2013}, however, lower toroidal resolution requirements have also been suggested, for example $ Q^{(z)} \ge 10-15$ for $Q^{(\phi)} \approx 10$ of \cite{SorathiaReynolds2012}.  Furthermore, the latter authors note that the toroidal and poloidal resolutions are coupled and low $Q^{(z)}$ can be compensated by a $Q^{(\phi)} \ge 25$.  Hence the product of quality factors $Q^{(z)}Q^{(\phi)} \ge 200-250$ has also been suggested \citep{NarayanSadowski2012,DhangSharma2018} as quality metric.  

\begin{figure}[htbp]
\begin{center}
\includegraphics[width=0.32\textwidth]{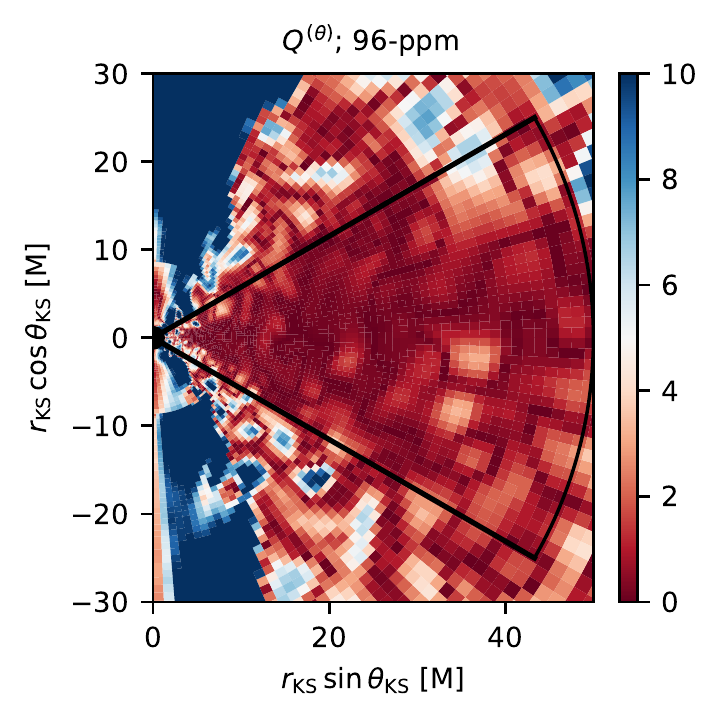}
\includegraphics[width=0.32\textwidth]{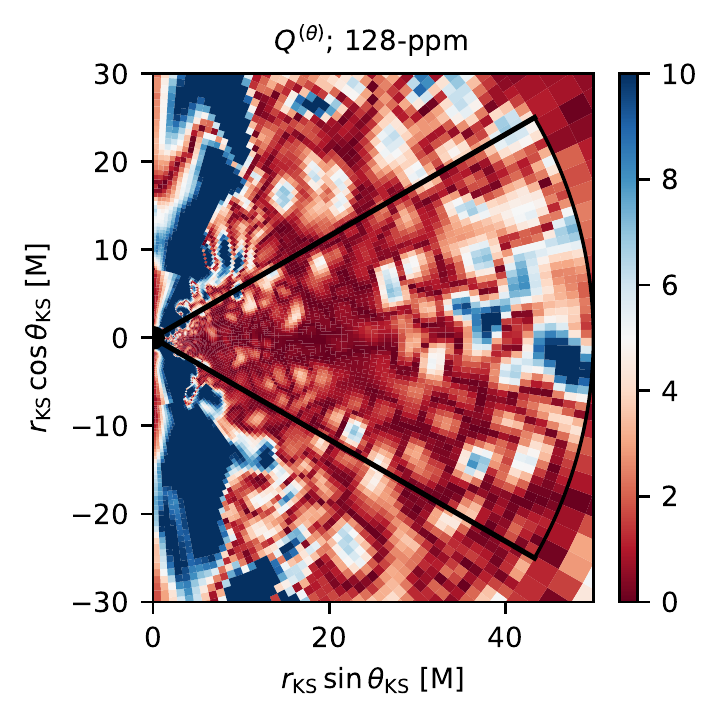}
\includegraphics[width=0.32\textwidth]{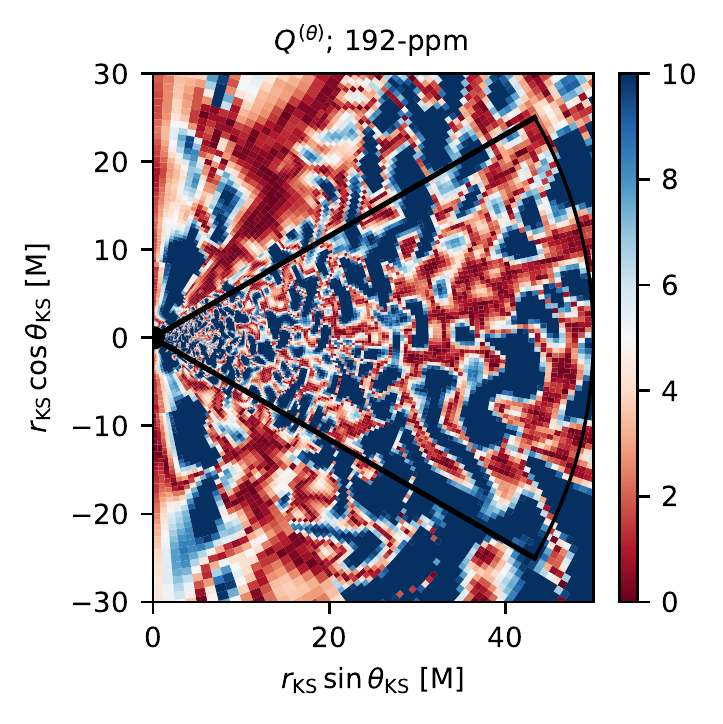}
\includegraphics[width=0.32\textwidth]{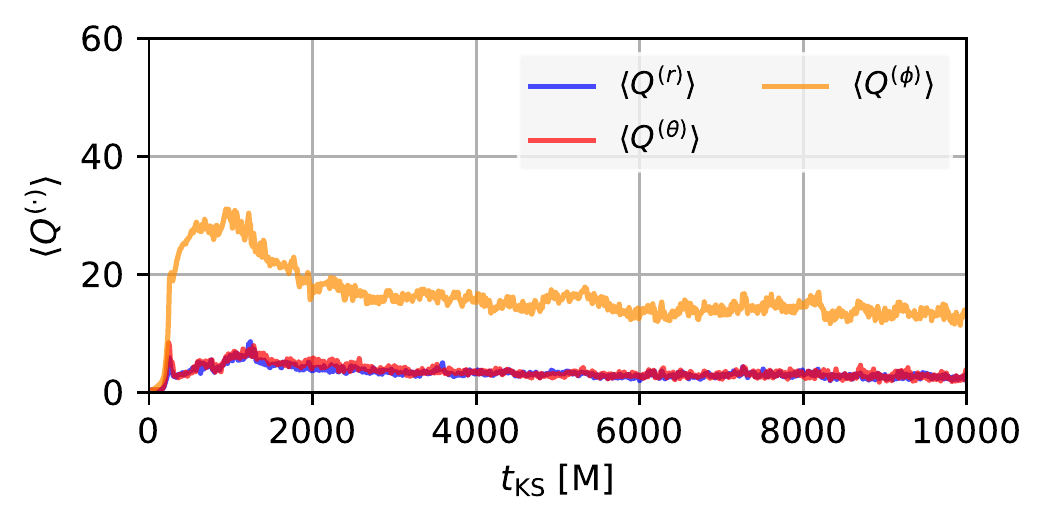}
\includegraphics[width=0.32\textwidth]{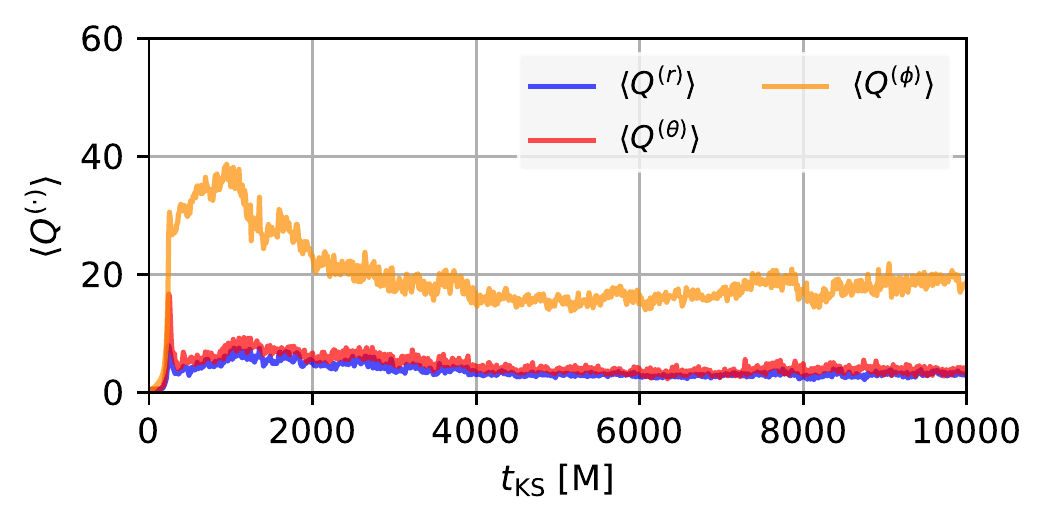}
\includegraphics[width=0.32\textwidth]{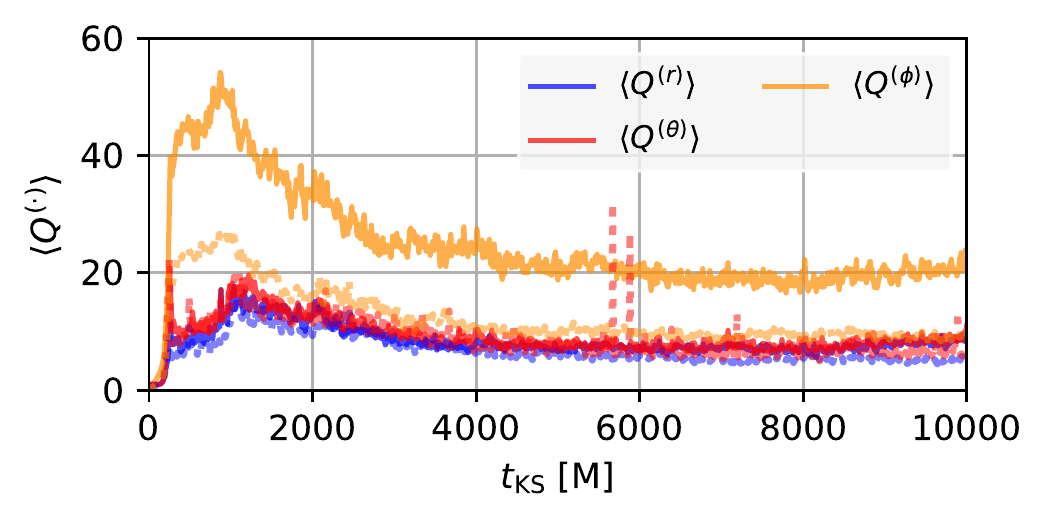}
\caption{MRI Quality factors for increasing resolution (left to right) obtained at $t=10\, 000\rm M$ with the \bhac code (top row) and corresponding temporal evolution of the average quality factor in the above highlighted region (bottom row).  In the high-res timeseries, dotted lines indicate the Q-factors for a run with half the azimuthal resolution (hence 96 cells).  At high resolution, the late-time decay of $Q^{(r)}, Q^{(\theta)}$ is halted and a mean value of $\sim 8$ is reached, indicating just about sufficient resolution to keep the MRI going. 
}
\label{fig:Qfac}
\end{center}
\end{figure}

As basis for our MRI quality factors, we compare the wavelength of the fastest growing MRI mode evaluated in a fluid-frame tetrad basis $\hat{e}_\mu^{(\cdot)}$ with the grid spacing.  In particular, we orient the tetrads along the locally non-rotating reference frame (LNRF) \citep[see e.g.][for the transformations]{Takahashi2008}.  
Take for example the $\theta$-direction:  
\begin{align}
  \lambda^{(\theta)}:= \frac{2\pi}{\sqrt{(\rho h+b^2})\Omega} b^\mu e^{(\theta)}_\mu
\end{align}
and adopt the corresponding grid resolutions as seen in the orthonormal fluid-frame
\begin{align}
  \Delta x^{(\theta)} := \Delta \theta^\mu e^{(\theta)}_\mu\,
\end{align}
where $\Delta \theta^\mu=(0,0,\Delta\theta,0)$ is the distance between two adjacent grid lines and $\Omega=u^\phi/u^t$ the angular velocity of the fluid.  
The resulting $Q^{(\theta)}:=\lambda^{(\theta)}/\Delta x^{(\theta)}$  are shown for the standard resolutions at the final simulation times in the top panel of Figure~\ref{fig:Qfac}.  It is clear that the low- and medium- resolution simulations are not particularly well resolved regarding the poloidal MRI wavelengths.  The time- and spatial average values in the domain of interest are $Q^{(r,\theta,\phi)}_{96}\simeq 2.9,3,14$, $Q^{(r,\theta,\phi)}_{128}\simeq 3,3.8,17$, $Q^{(r,\theta,\phi)}_{192}\simeq 7.3,7.9,19.7$.  
This is still somewhat below the nominal $Q^{(z,\phi)}\sim 10,20$ of \cite{HawleyGuanKrolik2011} but well above the six cells suggested by \cite{SanoInutsoka2004} to adequately capture the saturation level of the MRI.  
However, it seems that only the highest resolution case keeps MRI driven turbulence alive up to the end of the simulation.  
At $96^3$, $128^3$ or $192^3$ resolution in the domain of interest, the bottle-neck is clearly in the poloidal direction with $Q^{(r)} \simeq Q^{(\theta)}$ but the azimuthal factor $Q^{(\phi)}$ is approximately twice larger.  Hence one might surmise that the resolution in $\phi-$ direction will have little influence on the overall evolution.   This has been confirmed by additional experiments with 96 instead of 192 azimuthal cells using \bhac and by runs where also the $r$ direction was reduced using the \koral code. In both cases, the agreement was very good with horizon penetrating fluxes consistent with the full resolution case.  
As noted by \cite{HawleyGuanKrolik2011}, an increase in azimuthal resolution can also lead to an improved behavior of the poloidal quality factors.  We find the same, however the effect is not dramatic.  
The dotted curve in Figure \ref{fig:Qfac} shows that with half azimuthal resolution, indeed $Q^{(r)}$ and $Q^{(\theta)}$ slightly decrease (to $Q^{(r,\theta)}\simeq 5.3, 6.8$) and $Q^{(\phi)}\simeq 9.6$ essentially halves.  
Upon increasing the resolution to $N_{\theta}=384$, we obtain $Q^{(r,\theta,\phi)}_{384}\simeq 21.3, 24.5, 40.9$ and hence a super-linear increase in $Q-$factor compared to the $N_{\theta}=192$ case.  This behavior will be encountered again for the very high resolution run and is discussed further in the next Section.  

These results give a framework to explain the resolution dependent trends in the data.  As an example, the formation of the peculiar ``mini-torus'' in the low- and mid- resolution \bhac runs is likely a result of insufficiently resolved MRI in these runs.  Furthermore, the convergence of accretion rates and MAD parameter $\phi$ shown in Figure \ref{fig:quantify-ts} coincides with the point when $Q-$factors reach a sufficient level.   

\FloatBarrier 
\subsection{A very high resolution run}\label{sec:highres}

With \hamr, we are able to produce a global simulation with a resolution of $N_{r}\times N_{\theta}\times N_{\phi}$ = $1608\times1056\times1024$ with 5 levels of static mesh refinement in the $\phi$ direction (see Sec.~\ref{sec:setup-hamr}), bringing us to local shearing box regimes (e.g., \citealt{DavisStone2010,RyanGammie2017}). 
In terms of the density scale height $H/R\simeq 0.25-0.3$ we obtain 85-100 cells per scale height and disk-averaged quality factors $Q^{(r,\theta,\phi)}_{1056}\geq 120,100, 190$.  This is more than sufficient for capturing MRI-driven turbulence as per the criteria ($Q^{(z,\phi)} \geq 10,25$ in cylindrical coordinates).
Indeed, \cite{HawleyGuanKrolik2011} suggests a resolution of $600\times450\times200$ for a global simulation with $H/R=0.1$, which we have achieved. Sufficiently resolved global simulations are essential for reproducing micro-physical phenomena such as MRI in a macro-physical environment. \cite{HawleyRichers2013} notes that even though MRI is a local instability, turbulent structures may form on larger scales (seen in large shearing boxes by \citealt{SimonBeckwithEtAl2012}). Additionally, shearing box simulations inadvertently affect poloidal flux generation by implicitly conserving the total vertical magnetic flux, and hence, may not be as reliable as global simulations in dealing with large scale poloidal flux generation \citep{LiskaTchekhovskoy2018}, which is required for jet launching and propagation. 

It is quite striking that increase of the MRI quality factors for the very high resolution run is beyond the factor of $\sim 5.5$ which would be gained by going from $N_\theta=192$ to $N_\theta=1056$.  Instead we find an extra factor of two increase in $Q$-factors (cf. Appendix \ref{sec:linear}), consistent with the simultaneous raise in disk-magnetization by the same amount.  
Overall trends of parameters over time shown in Fig. \ref{fig:ts192} are quite similar to the \harm and \koral high-resolution data, with some other $192^3$ PPM data within range. 

\begin{figure}[htbp]
\begin{center}
\includegraphics[width=0.45\textwidth]{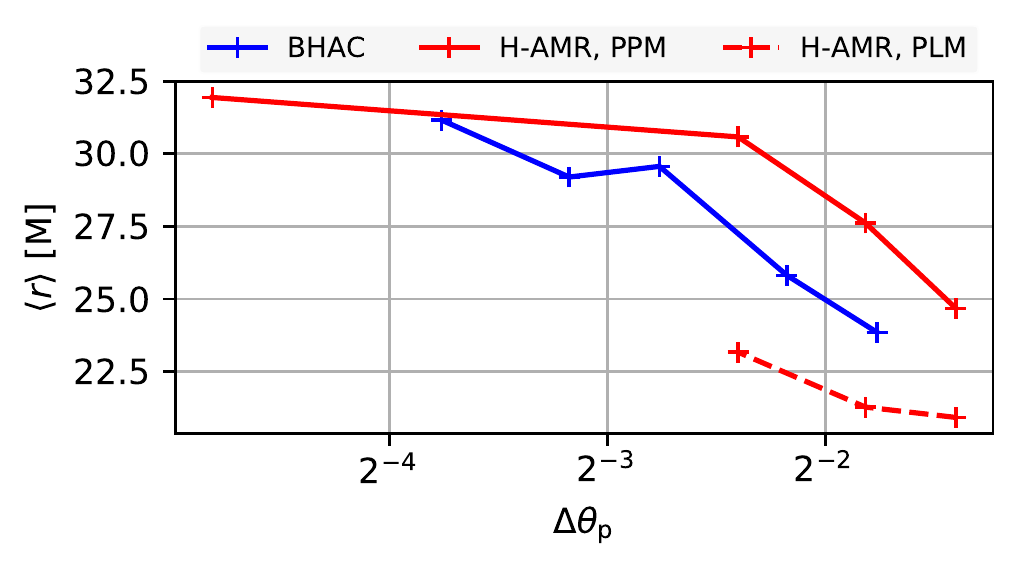}
\includegraphics[width=0.45\textwidth]{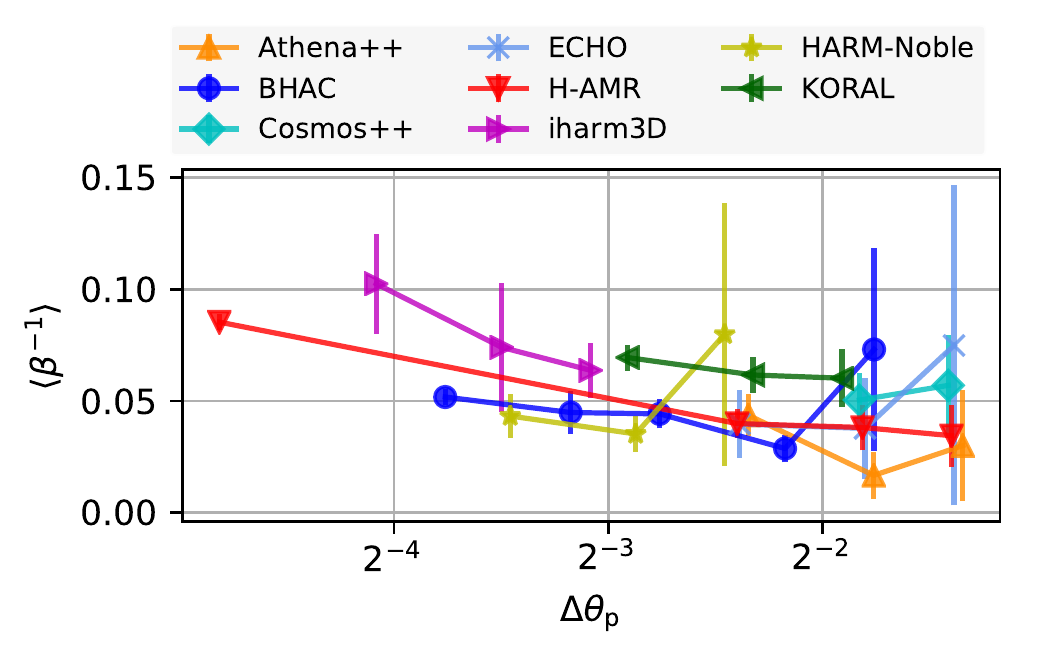}
\caption{\textit{Left:} Convergence in the disk spreading as quantified by the barycentric radius at $t=10K \rm M$.  The data has been computed with \bhac an \hamr both using linear (PLM) and piecewise parabolic reconstruction (PPM).  We note indications for convergence against the very high resolution run.  
\textit{Right:} average magnetization of the disk for $r_{\rm KS}\in[6,40]\rm M$ against physical mid-plane resolution.  Similar to the disk $\alpha$, a weak dependence on resolution is obtained in all codes with $\langle \beta^{-1}\rangle$ increasing by $\sim \times 3$ from lowest to highest resolution in the set.  
}
\label{fig:rbary_converge}
\end{center}
\end{figure}

This gives a first impression on the behavior of viscous angular momentum transport and the turbulent dissipation and generation of magnetic energy.  
As evidenced by Fig. \ref{fig:rbary_converge}, there are signs of convergence in the disk radius at $N_\theta=1056$, though the disk magnetization continues to increase slightly with increased resolution.  Due to the lack of intermediate data between $N_\theta=192$ and $N_\theta=1056$ it is however not possible to judge whether a saturation point has already been reached in between.  
The steady increase of the disk magnetization with resolution has also been noted by \cite{Shiokawa2012} who performed \harm simulations up to $N_{\theta}=384$.   Quantitatively, the values of $\beta\sim 15-20$ and the resolution dependent trend reported in their Figure 3 seem consistent with our findings.  

One notes that the magnetization reverses its trend in most codes as $\langle\beta^{-1}\rangle$ at first even decreases but then picks up at $N_{\theta}=128$ and continues to increase up to the highest resolutions tested.  This is suggestive of a ``resolution threshold'' above which turbulent amplification of the magnetic field starts to operate leading to the higher average value of the magnetization.   
At the same time, the near constant profile of $\langle\beta^{-1}(r)\rangle$ is established throughout the disk, reflected also in a decrease of the scatter in Figure \ref{fig:rbary_converge}.  

\FloatBarrier 
\section{Discussion and Conclusions} \label{sec:discussion}

Using a standardised setup of radiatively inefficient black hole accretion and a set of relevant diagnostics, we have compared the MRI-driven turbulent quasi-stationary state in nine GRMHD codes which are widely used in the community.  
Many of the codes have been developed independently, others share the same heritage but are completely re-written.  There are both differences and commonalities between the codes.  
Listing the most important common elements, we note that all apply overall 2nd order conservative schemes and preserve the divergence of the magnetic field to machine precision. All except \harm applied the piecewise-parabolic-method (PPM) for spatial reconstruction.  
Regarding the algorithmic differences, the order of time-integration, use of approximate Riemann solver (HLL and LLF) and specific integration of the induction equation are the most striking ones.  Furthermore, two configurations have been computed in cartesian coordinates, while mostly spherical coordinates are used.  
The results shed light on the systematics between codes and elucidate the resolution requirements to reach a certain level of agreement.  
Generally, we find that the level of agreement between codes improves for all diagnostics when the resolution is increased towards convergence.  
Once sufficient resolution is used, key parameters like the accretion rate of the saturated turbulent state agree within their temporal variations among all codes and the spread in the mean accretion rate is given by a factor of $\sim1.7$ between the lowest and the highest.  

\subsection{Resolution and convergence}
We have performed simulations at resolutions commonly used for black hole accretion and forward modeling of EHT observations.  In fact, early GRMHD models employed simulations that use comparable or less cells than our low- and mid-resolution setups of $96^3-128^3$ cells. For example, the fiducial runs of \cite{mckinney2009} use $256\times128\times32$ cells (though with fourth-order reconstruction) and \cite{FragileBlaes2007} used a standard effective resolution of $128^3$ for their tilted disks.  Economizing on the spatial resolution is particularly important for large parameter surveys as in \cite{MoscibrodzkaFalcke2016} who applied \texttt{HARM3D} with $96\times96\times64$ cells and for long time evolutions as in \cite{PennaMcKinney2010} who ran a large suite of simulations with resolutions of effective $256\times64\times64$ cells (accounting for the restricted $\Delta \phi=\pi$).  
Although comparing cell numbers should be done with a fair grain of salt, taking into account gridding and initial conditions, these numbers give an indication of the recent state of affairs.  
  
Balancing the desire of exploring a large parameter space (as in the generation of a simulation library for EHT observations) against numerical satisfaction, one will end up performing simulations that are ``just about'' sufficiently resolved.  Hence it is interesting to note that with the choice of our resolutions we have sampled a resolution edge in most codes.  This is best seen in the runs of accretion rates and magnetic fluxes which substantially reduce their scatter once mid-plane resolutions of $\Delta\theta_{\rm p}\simeq 0.25-0.0125 \rm M$ are adopted.  Below this threshold, the algorithmic differences still dominate leading to large scatter with sometimes opposite trends between codes.  
The resolution threshold was also confirmed by inspection of the MRI quality factors in \bhac and \hamr.  Only the high-resolution PPM runs would qualify as resolving the saturated MRI as per the criteria of \cite{HawleyGuanKrolik2011}. 

Using our sample of runs at different resolutions, we have also investigated the convergence of global quantities of interest.  While it is clear that convergence in the strict sense cannot be achieved for ideal MHD simulations of turbulence as viscous and resistive length-scales depend on the numerical resolution and numerical scheme \citep[e.g.][]{KritsukNordlundEtAl2011}, convergence in certain physical quantities such as the stress parameter $\alpha$ or the ``magnetic tilt angle'' between the average poloidal field directions could still be achieved (see also the discussions in \cite{SorathiaReynolds2012,HawleyRichers2013}).  
Hence we find that while the viscous spreading of the disk appears converged at accessible resolutions ($\sim 192^3$ cells), the disk $\alpha$ parameter and magnetization continue their weak resolution dependent rise up to $\sim1024^3$ cells.  The latter confirms and extends the findings of \cite{Shiokawa2012} by a resolution factor of $\times 2.75$.  

\subsection{Systematics}
Turning to the systematics between codes, we observe a veritable diversity in averaged density profiles. Although the spread diminishes with increasing resolution, there is still a factor of $\sim3$ difference in peak density at nominal $192^3$ resolution.  As the magnetic field profiles in the disk are quite robust against codes and resolutions, the lower density and pressure values can lead to a non-linear runnaway effect as Maxwell stresses in the simulations with lower densities will increase, leading to additional viscous spreading of the disk and hence again lower peak densities.

When modeling radiative properties of accretion systems as e.g. in the EHT workflow, the variance in absolute disk densities is somewhat mitigated by the scale-freedom of the ideal MHD simulations which allow to re-scale the value of density together with the other MHD variables.  
Not only do the absolute dissipative length-scales depend on resolution, also their ratio given through the effective magnetic Prandtl number $Pr_{\rm m}$ varies with Riemann solver, reconstruction method, magnetic field solver, grid and resolution \citep[e.g.][]{KritsukNordlundEtAl2011}.  Moreover, it is well known that transport properties of the non-linear MRI depend critically on $Pr_{\rm m}$ \citep{LesurLongaretti2007,SimonHawley2009,FromangPapaloizouEtAl2007,LongarettiLesur2010} and hence some systematic differences in the saturated state are in fact expected to prevail even for the highest resolutions.  

Another systematic is introduced by the axial boundary condition: the jet-disk boundary $\sigma=1$ is faithfully recovered only at high resolution and the jet can be particularly skimpy when a polar cutout is used.  In the low- and medium- resolution \nobleharm and \cosmos simulations we found that the axial excision significantly decreases the magnetization of the funnel region.  The reason for the loss of the magnetized funnel is well known: standard ``soft'' reflective boundaries at the polar cones allow a finite (truncation error) numerical flux of $B^r$ to leave the domain. This effect diminishes with resolution and can be counter-acted by zero-flux ``hard'' boundaries at the polar cutout as noticed by \citep{Shiokawa2012}.  In \cosmos the use of  outflow boundaries in the excised region leads to a particularly striking difference, resembling the situation described by \cite{igumenshchev2003}.  

Inspecting the trends in the jet collimation profiles, we note that there is a close relationship between the jet width and the horizon penetrating magnetic flux.  Indeed, Table \ref{tab:quantify-ts} and Figure \ref{fig:t-phi_sigmacont} for the $N_\theta=128$ data reveal that a one to one relation holds for the four models with smallest $\langle \phi\rangle$ and the one with the absolute largest value of $\langle \phi \rangle$ where the differences in opening angle are most pronounced.  
Since the jet collimation plays an important role in the acceleration of the bulk flow beyond the light cylinder \citep[e.g.][]{Camenzind1986,Komissarov2007}, this systematic will translate into an effect on the overall flow acceleration of the funnel jets.  

\subsection{The floor region}
We have also calculated 2D averaged maps of density, (inverse-) plasma-$\beta$ and the magnetization. These allow to quantify the variance of jet opening angles between codes and we find that the spread around a 'mean simulation' is reduced to $\sim 2.5^\circ$ at $r=50\rm M$ at high resolutions.  Interestingly, the jet boundaries up to $r_{\rm KS}= 50 \rm M$ follow closely the paraboloidal shape of the solution by \cite{BlandfordZnajek1977}.  It will be insightful to see how this result changes quantitatively when the magnetic flux on the black hole is increased towards the MAD case which is known to exhibit a wider jet base.  
The maps of plasma-$\beta$ and magnetization illustrate truthfully the variance in the jet region which is introduced by the different floor treatments.  Demonstrating the considerable diversity serves two purposes: \textit{i)} to underline that the plasma parameters of the Poynting dominated jet region should not be taken at face value in current MHD treatments and \textit{ii)} that despite this variance in the jet region itself, the effect on the dynamics (e.g. jet opening angle and profile) is minor.

\subsection{Time variability}
Using an emission proxy taylored to optically thin synchrotron emission from electrons distributed according to a relativistic Maxwellian, we have computed lightcurves with all codes and resolutions and have compared their power spectra.  The same analysis was performed with the accretion rates extracted at the black hole horizon.  In broad terms, all PSDs of the lightcurves are compatible with a red-noise spectrum up to $f\simeq 1/10\rm M$ where a steepening is observed.  Inspection of the PSD for the accretion rate yields flatter PSDs $\sim f^{-1}$ for $f<1/100M$ and a similar steepening at $f\simeq 1/10\rm M$.  This is quite consistent across all codes and resolutions and agrees with earlier results of \cite{ArmitageReynolds2003,hoggReynolds2016}.  The steeper low-frequency PSDs of the proxy can be explained by noting that the integration over the disk adds additional low frequency fluctuations from larger radii.    
Whether the high-frequency break occurs at the ISCO frequency of $f_0=0.041/M$ or at somewhat higher frequencies $f>0.1\rm M$ is not that clear cut in the data, however the presence of a high-frequency break indicates an inner annulus of the emission at or within the ISCO.  
With de-trended timeseries of accretion rates and emission proxy, we have computed the root-mean-square (RMS) variability and find an ubiquitous relationship between the RMS and the absolute value (RMS-flux relationship) with slope of $k\simeq1.2-1.6$ across all codes and resolutions. Upon increased resolution, $k$ tends to converge towards $\sim1.3$ for our benchmark problem.   

\subsection{Implications}
This first large GRMHD code comparison effort shows that simulation outcomes are quite robust against the numerical algorithm, implementation and choice of grid geometry in current state of the art codes.  
Once certain resolution standards are fulfilled (which might be reached at differing computational expense for different codes), we can find no preference favoring one solution against the other.  In modeling the EHT observations, we find it beneficial to use several of the codes tested here interchangeably.  In fact, a large simulation library comprising $43$ well resolved GRMHD simulations has been created for comparison to the observations, using \harm, \koral, \hamr and \bhac \citep{CollaborationAkiyamaEtAl2019d}.  Several parameter combinations have been run with multiple codes for added redundancy and the diagnostics which were calibrated here are used for cross-checks.  

To serve as benchmark for future developments, the results obtained here are freely available on the platform \cyverse. 
This also includes the unprecedented high resolution \hamr run for which a thorough analysis might provide with new general insights into rotating turbulence.  
See Section \ref{sec:data} for access instructions.  
Furthermore, animations of simulation output can be found in the supporting material at \dataset[10.7910/DVN/UCFCLK]{\doi{10.7910/DVN/UCFCLK}}.  

\subsection{Outlook}
The benchmark problem of this work, a spin $a=0.9375$, turbulent MHD torus with scale height $H/R\approx 0.3$ and normalized horizon-penetrating magnetic flux of $\phi\approx 2$ falls into the class of radiatively inefficient \citep[][]{Narayan1995} SANE disks and is perhaps the simplest case one might consider (its widespread use likely goes back to the initial conditions provided with the public \harmtwod code \citealt{Gammie2003}).  
To increase the challenge, one might consider MHD simulations of magnetically arrested disks (MAD, e.g. \citealt{NarayanIgumenshchev2003}).  The numerical problems and new dynamics introduced by the increase of magnetic flux are considerable.  New violent interchange type accretion modes and stronger magnetised disks (potentially with suppressed MRI) come into play \citep[e.g.][]{TchekhovskoyNarayan2011,McKinneyTchekhovskoy2012} presenting a physical scenario well suited to bring GRMHD codes to their limits.  
A resolution study of the MAD scenario was recently presented by \cite{WhiteStoneEtAl2019}, showing the difficulty of converging various quantities of interest, e.g. the synchrotron variability.  
How different GRMHD codes fare with the added challenge shall be studied in a future effort where also the jet dynamics will be more in the focus.  This shall include the dynamics in the funnel, e.g. the properties of the stagnation surface \citep{Nakamura2018} and the acceleration/collimation profiles.  

Another area where code comparison will prove useful is in the domain of non-ideal GRMHD modeling e.g. radiative MHD \citep[e.g.][]{FragileGillespieEtAl2012,sadowski14,McKinneyTchekhovskoy2014,RyanDolence2015,TakahashiOhsugaEtAl2016} and resistive MHD \citep[e.g.][]{BugliDel-Zanna2014,QianFendt2016,2019MNRAS.485..299R}.  This would be particularly informative as such codes are not yet widely used in the community, e.g. no public versions have been released to date.  

One of the prevailing systematics in GRMHD modeling lies in the ILES approximation which is typically applied.  Development of explicit general relativistic treatments of magnetic diffusivity and viscosity \citep[e.g.][]{FragileEtheridgeEtAl2018,FujibayashiKiuchi2018} will soon allow to perform direct numerical simulations of turbulent black hole accretion covering both the low and high $Pr_{\rm m}$ regimes which are expected to be present in such systems \citep{BalbusHenri2008}.

\section{Supplementary information}\label{sec:data}
Animations of the quantities given in Figures \ref{fig:t-phi1}-\ref{fig:t-phi4} were prepared with \bhac, \echo and \hamr for meridional slices and can be found at \dataset[10.7910/DVN/UCFCLK]{\doi{10.7910/DVN/UCFCLK}}.  

Processed data used to create all Figures in this manuscript as well as 
raw snapshot data of several high-resolution runs (including the 
$N_{\theta}=1056$ \hamr run) is available through \cyverse 
Data Commons (\url{http://datacommons.cyverse.org/}).  The data is freely accessible at \url{http://datacommons.cyverse.org/browse/iplant/home/shared/eht/2019/GRMHDCodeComparisonProject}.  Please consult \texttt{README.txt} and \texttt{LICENSE.txt} for usage instructions.  

\acknowledgments
RN thanks the National Science Foundation (NSF, grants OISE-1743747, AST-1816420) and acknowledges computational support from the NSF via XSEDE resources (grant TG-AST080026N).  LDZ acknowledges support from the PRIN-MIUR project \emph{Multi-scale Simulations of High-Energy Astrophysical Plasmas} (Prot.~2015L5EE2Y) and from the INFN - TEONGRAV initiative. CJW made use of the Extreme Science and Engineering Discovery Environment (XSEDE) Comet at the San Diego Supercomputer Center through allocation AST170012. The \hamr high resolution simulation was made possible by NSF PRAC awards no. 1615281 and OAC-1811605 at the Blue Waters sustained-petascale computing project and supported in part under grant no. NSF PHY-1125915 (PI A. Tchekhovskoy). KC and SM are supported by the Netherlands Organization for Scientific Research (NWO) VICI grant (no. 639.043.513), ML is supported by the NWO Spinoza Prize (PI M.B.M. van der Klis).
The HARM-Noble simulations were made possible by NSF PRAC award no. NSF OAC-1515969, OAC-1811228 at the Blue Waters sustained-petascale computing project and supported in part under grant no. NSF PHY-1125915.
The BHAC CKS-GRMHD simulations were performed on the Dutch National Supercomputing cluster Cartesius and are funded by the NWO computing grant 16431. 
S. C. N. was supported by an appointment to the NASA Postdoctoral Program at the Goddard Space Flight Center administered by USRA through a contract with NASA.
Y.M., H.O. O.P. and L.R. acknowledge support from the ERC synergy grant ``BlackHoleCam: Imaging the Event Horizon of Black Holes'' (Grant No. 610058). MB acknowledges support from the European Research Council (grant no. 715368 -- MagBURST) and from the Gauss Centre for Supercomputing e.V. ({\tt www.gauss-centre.eu}) for funding this project by providing computing time on the GCS Supercomputer SuperMUC at Leibniz Supercomputing Centre ({\tt www.lrz.de}). P.C.F. was supported by NSF grant AST-1616185 and used resources from the Extreme Science and Engineering Discovery Environment (XSEDE), which is supported by NSF grant number ACI-1548562. Work by P.A. was performed in part under the auspices of the U. S. Department of Energy by Lawrence Livermore National Laboratory under contract DE-AC52-07NA27344.

The authors of the present paper further thank the following organizations and programs: 
the Academy of Finland (projects 274477, 284495, 312496);
the Advanced European Network of E-infrastructures for Astronomy with the SKA (AENEAS) project, supported by the European Commission Framework Programme Horizon 2020 Research and Innovation action under grant agreement 731016;
the Alexander von Humboldt Stiftung; 
the Black Hole Initiative at Harvard University, through a grant (60477) from the John Templeton Foundation; 
the China Scholarship Council;
Comisi\'{o}n Nacional de Investigaci\'{o}n Cient\'{\i}fica y Tecnol\'{o}gica (CONICYT, Chile, via PIA ACT172033, Fondecyt 1171506, BASAL AFB-170002, ALMA-conicyt 31140007);
Consejo Nacional de Ciencia y Tecnolog\'{\i}a (CONACYT, Mexico, projects 104497, 275201, 279006, 281692);
the Delaney Family via the Delaney Family John A.\ Wheeler Chair at Perimeter Institute; 
Direcci\'{o}n General de Asuntos del Personal Acad\'{e}mico-—Universidad Nacional Aut\'{o}noma de M\'{e}xico (DGAPA-—UNAM, project IN112417); 
the European Research Council Synergy Grant "BlackHoleCam: Imaging the Event Horizon of Black Holes" (grant 610058); 
the Generalitat Valenciana postdoctoral grant APOSTD/2018/177; 
the Gordon and Betty Moore Foundation (grants GBMF-3561, GBMF-5278); 
the Istituto Nazionale di Fisica Nucleare (INFN) sezione di Napoli, iniziative specifiche TEONGRAV;
the International Max Planck Research School for Astronomy and Astrophysics at the Universities of Bonn and Cologne; 
the Jansky Fellowship program of the National Radio Astronomy Observatory (NRAO);
the Japanese Government (Monbukagakusho: MEXT) Scholarship; 
the Japan Society for the Promotion of Science (JSPS) Grant-in-Aid for JSPS Research Fellowship (JP17J08829);
the Key Research Program of Frontier Sciences, Chinese Academy of Sciences (CAS, grants QYZDJ-SSW-SLH057, QYZDJ-SSW-SYS008);
the Leverhulme Trust Early Career Research Fellowship;
the Max-Planck-Gesellschaft (MPG);
the Max Planck Partner Group of the MPG and the CAS;
the MEXT/JSPS KAKENHI (grants 18KK0090, JP18K13594, JP18K03656, JP18H03721, 18K03709, 18H01245, 25120007);
the MIT International Science and Technology Initiatives (MISTI) Funds; 
the Ministry of Science and Technology (MOST) of Taiwan (105-2112-M-001-025-MY3, 106-2112-M-001-011, 106-2119-M-001-027, 107-2119-M-001-017, 107-2119-M-001-020, and 107-2119-M-110-005);
the National Aeronautics and Space Administration (NASA, Fermi Guest Investigator grant 80NSSC17K0649); 
the National Institute of Natural Sciences (NINS) of Japan;
the National Key Research and Development Program of China (grant 2016YFA0400704, 2016YFA0400702); 
the National Science Foundation (NSF, grants AST-0096454, AST-0352953, AST-0521233, AST-0705062, AST-0905844, AST-0922984, AST-1126433, AST-1140030, DGE-1144085, AST-1207704, AST-1207730, AST-1207752, MRI-1228509, OPP-1248097, AST-1310896, AST-1312651, AST-1337663, AST-1440254, AST-1555365, AST-1715061, AST-1615796, AST-1716327, OISE-1743747, AST-1816420); 
the Natural Science Foundation of China (grants 11573051, 11633006, 11650110427, 10625314, 11721303, 11725312); 
the Natural Sciences and Engineering Research Council of Canada (NSERC, including a Discovery Grant and the NSERC Alexander Graham Bell Canada Graduate Scholarships-Doctoral Program);
the National Youth Thousand Talents Program of China;
the National Research Foundation of Korea (the Global PhD Fellowship Grant: grants NRF-2015H1A2A1033752, 2015-R1D1A1A01056807, the Korea Research Fellowship Program: NRF-2015H1D3A1066561); 
the Netherlands Organization for Scientific Research (NWO) VICI award (grant 639.043.513) and Spinoza Prize SPI 78-409; the New Scientific Frontiers with Precision Radio Interferometry Fellowship awarded by the South African Radio Astronomy Observatory (SARAO), which is a facility of the National Research Foundation (NRF), an agency of the Department of Science and Technology (DST) of South Africa;
the Onsala Space Observatory (OSO) national infrastructure, for the provisioning of its facilities/observational support (OSO receives funding through the Swedish Research Council under grant 2017-00648)
the Perimeter Institute for Theoretical Physics (research at Perimeter Institute is supported by the Government of Canada through the Department of Innovation, Science and Economic Development and by the Province of Ontario through the Ministry of Research, Innovation and Science);
the Russian Science Foundation (grant 17-12-01029); 
the Spanish Ministerio de Econom\'{\i}a y Competitividad (grants AYA2015-63939-C2-1-P,  AYA2016-80889-P); 
the State Agency for Research of the Spanish MCIU through the "Center of Excellence Severo Ochoa" award for the Instituto de Astrof\'{\i}sica de Andaluc\'{\i}a (SEV-2017-0709); 
the Toray Science Foundation; 
the US Department of Energy (USDOE) through the Los Alamos National Laboratory (operated by Triad National Security, LLC, for the National Nuclear Security Administration of the USDOE (Contract 89233218CNA000001));
the Italian Ministero dell'Istruzione Universit\`{a} e Ricerca through the grant Progetti Premiali 2012-iALMA (CUP C52I13000140001);
the European Union’s Horizon 2020 research and innovation programme under grant agreement No 730562 RadioNet;
ALMA North America Development Fund;
the Academia Sinica;
Chandra TM6-17006X.
This work used the Extreme Science and Engineering Discovery Environment (XSEDE), supported by NSF grant ACI-1548562, and CyVerse, supported by NSF grants DBI-0735191, DBI-1265383, and DBI-1743442.  XSEDE Stampede2 resource at TACC was allocated through TG-AST170024 and TG-AST080026N.  XSEDE JetStream resource at PTI and TACC was allocated through AST170028.  The simulations were performed in part on the SuperMUC cluster at the LRZ in Garching, on the LOEWE cluster in CSC in Frankfurt, and on the HazelHen cluster at the HLRS in Stuttgart.  This research was enabled in part by support provided by Compute Ontario (http://computeontario.ca), Calcul Quebec (http://www.calculquebec.ca) and Compute Canada (http://www.computecanada.ca).

\vspace{5mm}

\software{
\athenapp, \cite{White2016}; 
\bhac, \cite{PorthOlivares2017};
\cosmos, \cite{Anninos2005,Fragile2012,Fragile2014};
\echo, \cite{DelZanna2007};
\hamr, \cite{LiskaHesp2018,ChatterjeeEtAl2019};
\nobleharm, \citep{Gammie2003,Noble2006,Noble2009};
\harm, \cite{Gammie2003,Noble2006,Noble2009};
\IGM, \cite{Etienne:2015cea};
\koral, \citep{sadowski13,sadowski14}
}

\appendix

\FloatBarrier 
\section{Comparison with linear reconstruction}\label{sec:linear}

Using \hamr, \koral and \athenapp, additional simulations with piecewise-linear reconstruction (PLM) were performed.  To be more specific, in order of increasing numerical diffusivity, \athenapp used the modified ``van Leer'' limiter of \cite{Mignone2014}, \hamr the $MC_{\beta}$ limiter \citep{Leer1977} with $\beta=1.5$ and \koral $MC_{\beta}$ with $\beta=1.9$ where $MC_{\beta}$ reduces to the most diffusive ``minmod-limiter'' in case of $\beta=1$ and to ``MC'' for $\beta=2$.  
It has been observed before that the higher order reconstruction of PPM compensates for the large dissipation in the LLF and HLL approximate Riemann solvers which are applied throughout this work.  For example, \cite{FlockDzyurkevichEtAl2010} note that with PPM reconstruction, the HLL and LLF methods properly resolve the growth of the linear MRI with 10 cells per mode as opposed to 16 cells for the PLM reconstruction.  

For completeness, we list the values of horizon penetrating fluxes from these runs in table \ref{tab:quantify-ts-plm}.  It is striking that even with a resolution as high as $192^3$, the PLM runs can give a very different answer:  the time-averaged accretion rate in the \hamr and \athenapp{} PLM runs falls short of their PPM counterpart by a factor of two to almost six.  For \koral on the other hand, we obtain a more consistent behavior between the reconstruction techniques.  

\begin{table}[htp]
\caption{Quantifications -- Time series data, linear reconstruction.   Quantities in angular brackets $\langle\cdot\rangle$ denote time averages in the interval $t\in[5000,10000]\rm M$ with error given by the standard deviation.  
\label{tab:quantify-ts-plm}}
\begin{center}
\begin{tabular}{clllllllll}
$N_{\theta}$ & 
Code & 
$\dot{M}_{\rm Peak}$ & 
$\langle\dot{M}\rangle$ & 
$\left.\frac{\Phi_{\rm BH}}{\sqrt{\dot{M}}}\right|_{\rm Peak}$ & 
$\langle\Phi_{\rm BH}/\sqrt{\dot{M}}\rangle$ & 
$\left.\dot{L}/\dot{M}\right|_{\rm Peak}$ & 
$\langle\dot{L}/\dot{M}\rangle$ & 
$\left.\frac{\dot{E}-\dot{M}}{\dot{M}}\right|_{\rm Peak}$ & 
$\langle\frac{\dot{E}-\dot{M}}{\dot{M}}\rangle$ \\
  \hline \hline
96  &  \athenapp  & $ 0.209 $ & $ 0.049  \pm  0.026 $ & $ 4.762 $ & $ 3.308  \pm  0.732 $ & $ 1.929 $ & $ 1.512  \pm  0.225 $ & $ 0.332 $ & $ 0.145  \pm  0.024 $ \\
  &  \hamr  & $ 0.201 $ & $ 0.044  \pm  0.015 $ & $ 1.341 $ & $ 0.914  \pm  0.15 $ & $ 1.931 $ & $ 1.473  \pm  0.187 $ & $ 0.446 $ & $ 0.241  \pm  0.049 $ \\
  &  \koral  & $ 0.863 $ & $ 0.09  \pm  0.077 $ & $ 5.247 $ & $ 3.226  \pm  1.022 $ & $ 2.146 $ & $ 1.734  \pm  0.286 $ & $ 0.328 $ & $ 0.033  \pm  0.015 $ \\
  & max/min  &  4.296  &  2.062  &  3.913  &  3.62  &  1.112  &  1.177  &  1.358  &  7.258  \\
\hline
128  &  \athenapp{}  & $ 0.266 $ & $ 0.035  \pm  0.011 $ & $ 6.553 $ & $ 3.845  \pm  0.664 $ & $ 1.918 $ & $ 1.387  \pm  0.272 $ & $ 0.371 $ & $ 0.167  \pm  0.041 $ \\
  &  \hamr  & $ 0.244 $ & $ 0.089  \pm  0.023 $ & $ 1.211 $ & $ 0.707  \pm  0.105 $ & $ 1.907 $ & $ 1.738  \pm  0.099 $ & $ 0.336 $ & $ 0.098  \pm  0.012 $ \\
  &  \koral  & $ 0.65 $ & $ 0.179  \pm  0.066 $ & $ 4.671 $ & $ 0.881  \pm  0.143 $ & $ 2.245 $ & $ 2.118  \pm  0.046 $ & $ 0.332 $ & $ 0.041  \pm  0.013 $ \\
  & max/min  &  2.665  &  5.165  &  5.41  &  5.439  &  1.177  &  1.527  &  1.117  &  4.049  \\
\hline
192  &  \athenapp  & $ 0.42 $ & $ 0.069  \pm  0.038 $ & $ 3.108 $ & $ 1.73  \pm  0.311 $ & $ 2.064 $ & $ 1.947  \pm  0.047 $ & $ 0.332 $ & $ 0.094  \pm  0.007 $ \\
  &  \hamr  & $ 0.612 $ & $ 0.04  \pm  0.018 $ & $ 1.349 $ & $ 0.916  \pm  0.192 $ & $ 2.022 $ & $ 1.487  \pm  0.22 $ & $ 0.336 $ & $ 0.194  \pm  0.045 $ \\
&  \koral  & $ 0.789 $ & $ 0.202  \pm  0.036 $ & $ 3.357 $ & $ 0.776  \pm  0.059 $ & $ 2.084 $ & $ 2.025  \pm  0.02 $ & $ 0.337 $ & $ 0.065  \pm  0.008 $ \\
  & max/min  &  1.88  &  5.027  &  2.489  &  2.23  &  1.031  &  1.362  &  1.016  &  2.967  \\
\end{tabular}
\end{center}
\end{table}

Apart from a slightly higher mid-plane resolution in the \koral case, there are significant algorithmic differences between the codes.  Whereas the implementation of \koral follows closely the one by \cite{Gammie2003} where a cell-centered representation of the magnetic field along with arithmetic averaging of the electric fields, (ACT) is chosen, \hamr and \athenapp{} both implement a staggered upwind constrained transport scheme (UCT) following \cite{2005JCoPh.205..509G}. In contrast to ACT methods, the UCT scheme reproduces the correct solution for plane-parallel grid-aligned fields, however the amount of dissipation of the scheme is effectively doubled.  

\begin{figure}[htbp]
\begin{center}
\includegraphics[width=0.7\textwidth]{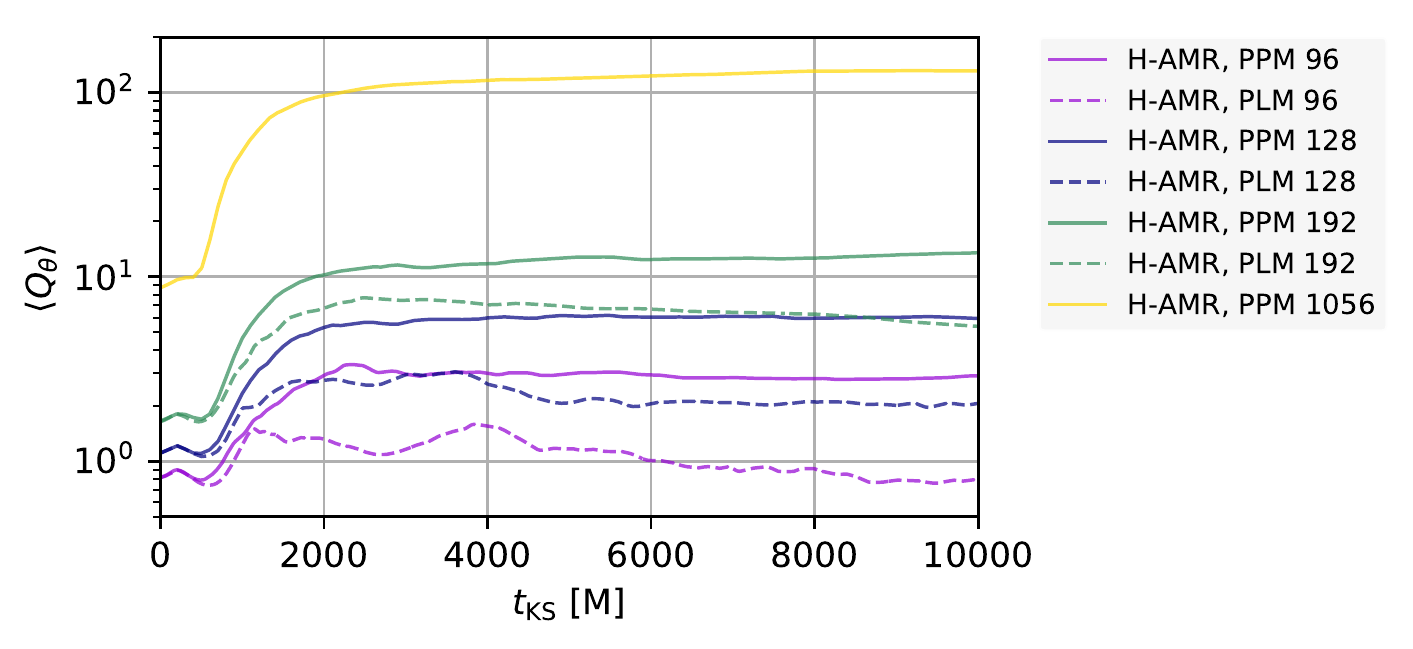}
\caption{Domain averaged and density weighted quality factors with the \hamr code using linear (PLM) and parabolic reconstruction (PPM).  The values are consistent with the ones reported in section \ref{sec:further} with the \bhac code.  
All PLM runs remain below $\sim 8$ and decrease further in the late time evolution.  }
\label{fig:Qfacs-hamr}
\end{center}
\end{figure}

Inspection of the MRI quality factors for \hamr as shown in Figure \ref{fig:Qfacs-hamr} reveals that the turbulence is not sufficiently resolved in the $96$-$192$ PLM runs.  Not only does $\langle Q_\theta\rangle$ stay below 10 during the entire run, it is also decreasing over time indicating net dissipation of magnetic energy. 

These results show that even when the overall order of the scheme is fixed to second order, the PPM method can significantly reduce the dissipation of the numerical integration and improve the results for a given resolution.  

\section{Run-to-run variation}\label{sec:run-to-run}

Due to the chaotic nature of the turbulence at late times, different initial random perturbations can accumulate to large differences in the realization of the dynamics.  
Likewise, since the compiler version and optimization (e.g. order of execution) influences the round-off error, a similar effect can be observed if the same physical scenario is run on two machines with differing compiler and run-time configurations.  
In order to judge the impact this makes compared to the inter-code differences, the \koral runs were repeated on two clusters: Harvard's \texttt{Odyssey} machine and \texttt{Stampede2} of the Texas Advanced Computing Center.  Here, both the initial random perturbations and the machine architecture differed.  

Again we list the standard quantifications of the time-series data in Table \ref{tab:quantify-ts-r2r}.  It shows that with a few exceptions of the peak values (and normalized magnetic flux), the differences are typically in the low percent regime.  
Hence the differences promoted by the various adopted algorithms are larger than the run to run variation of the \koral code.

\begin{table}[htp]
\caption{Quantifications -- Time series data, run to run variation.   Quantities in angular brackets $\langle\cdot\rangle$ denote time averages in the interval $t\in[5000,10000]\rm M$ with error given by the standard deviation.  
\label{tab:quantify-ts-r2r}}
\begin{center}
\begin{tabular}{clllllllll}
$N_{\theta}$ & 
Code & 
$\dot{M}_{\rm Peak}$ & 
$\langle\dot{M}\rangle$ & 
$\left.\frac{\Phi_{\rm BH}}{\sqrt{\dot{M}}}\right|_{\rm Peak}$ & 
$\langle\Phi_{\rm BH}/\sqrt{\dot{M}}\rangle$ & 
$\left.\dot{L}/\dot{M}\right|_{\rm Peak}$ & 
$\langle\dot{L}/\dot{M}\rangle$ & 
$\left.\frac{\dot{E}-\dot{M}}{\dot{M}}\right|_{\rm Peak}$ & 
$\langle\frac{\dot{E}-\dot{M}}{\dot{M}}\rangle$ \\
  \hline \hline
96  &  \koral Odyssey  & $ 0.821 $ & $ 0.408  \pm  0.129 $ & $ 3.029 $ & $ 0.791  \pm  0.362 $ & $ 2.126 $ & $ 2.031  \pm  0.038 $ & $ 0.33 $ & $ 0.061  \pm  0.008 $ \\
  &  \koral Stampede2  & $ 0.742 $ & $ 0.418  \pm  0.112 $ & $ 2.818 $ & $ 1.068  \pm  0.16 $ & $ 2.105 $ & $ 2.014  \pm  0.043 $ & $ 0.33 $ & $ 0.061  \pm  0.01 $ \\
  & max/min  &  1.106  &  1.024  &  1.075  &  1.349  &  1.01  &  1.008  &  1.0  &  1.009  \\
\hline
128  &  \koral Odyssey  & $ 0.859 $ & $ 0.318  \pm  0.102 $ & $ 2.899 $ & $ 1.279  \pm  0.306 $ & $ 2.073 $ & $ 1.978  \pm  0.049 $ & $ 0.331 $ & $ 0.066  \pm  0.007 $ \\
  &  \koral Stampede2  & $ 1.217 $ & $ 0.361  \pm  0.099 $ & $ 2.988 $ & $ 1.171  \pm  0.17 $ & $ 2.051 $ & $ 1.963  \pm  0.032 $ & $ 0.331 $ & $ 0.062  \pm  0.008 $ \\
  & max/min  &  1.417  &  1.134  &  1.031  &  1.093  &  1.011  &  1.008  &  1.0  &  1.053  \\
\hline
192  &  \koral Odyssey  & $ 1.067 $ & $ 0.29  \pm  0.132 $ & $ 2.459 $ & $ 1.254  \pm  0.231 $ & $ 2.036 $ & $ 1.953  \pm  0.023 $ & $ 0.331 $ & $ 0.056  \pm  0.007 $ \\
  &  \koral Stampede2  & $ 0.933 $ & $ 0.327  \pm  0.116 $ & $ 2.58 $ & $ 1.013  \pm  0.1 $ & $ 2.047 $ & $ 1.981  \pm  0.025 $ & $ 0.331 $ & $ 0.056  \pm  0.008 $ \\
  & max/min  &  1.144  &  1.128  &  1.049  &  1.237  &  1.006  &  1.014  &  1.0  &  1.014
\end{tabular}
\end{center}
\end{table}

\section{Rescaled disk profiles}\label{sec:rescaled}

\begin{figure}[htbp]
\begin{center}
\includegraphics[width=15cm]{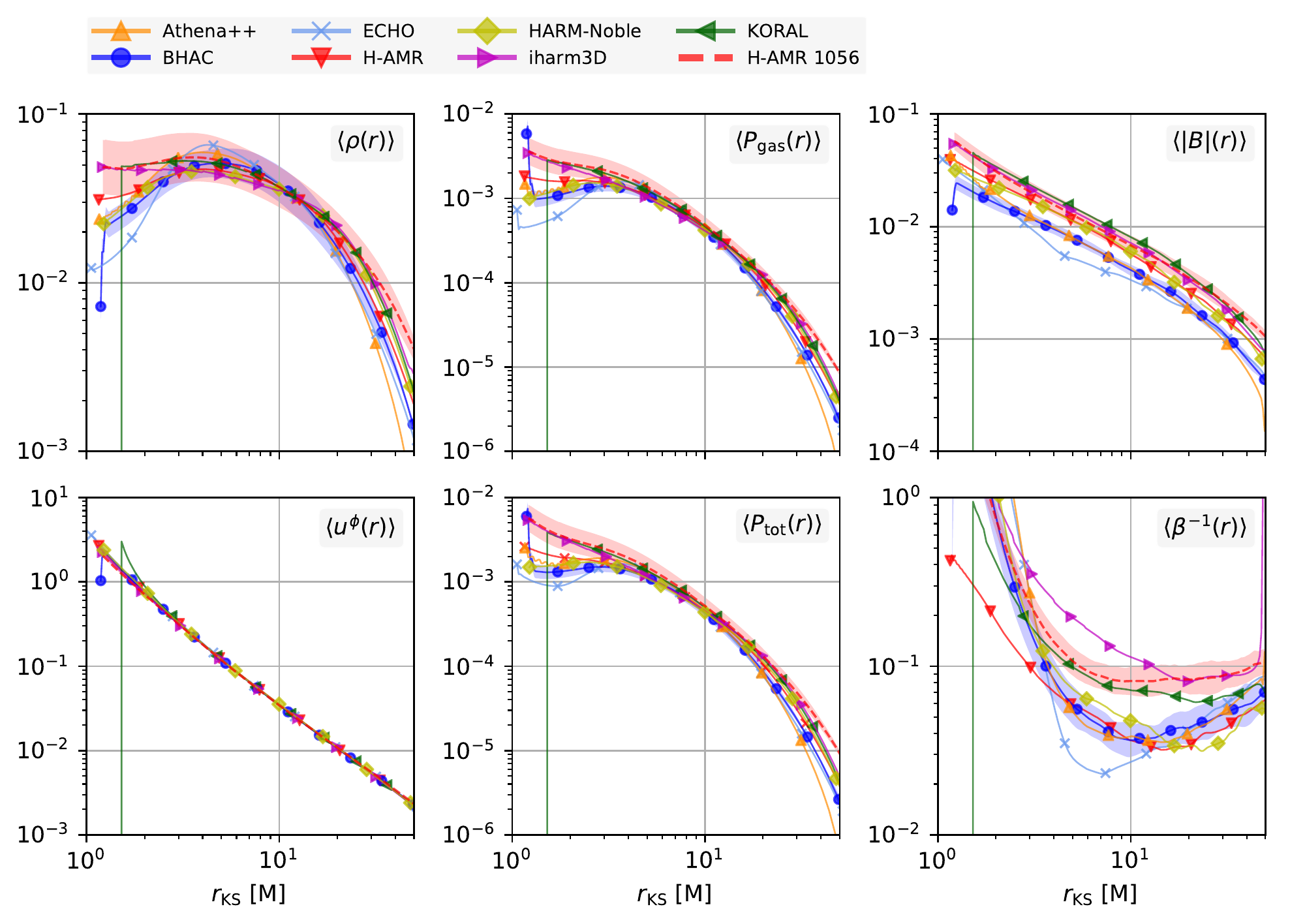}
\caption{Disk-profiles rescaled to match a 'reference solution' at the point $r=12\rm M$.  We take the \hamr very high resolution run for reference.  }
\label{fig:pr-rescaled-192}
\end{center}
\end{figure}

With the scale-freedom allowed by the test-fluid assumption, density, pressure and the magnetic fields of a given simulation can in principle be re-scaled by a constant factor (respecitvely its square root for the magnetic fields), for example to perform spectral fitting of observations.  
We here exploit this freedom and re-scale the simulations to a reference-case for which we use the $N_\theta=1056$ \hamr simulation.  In particular, we match the density at the initial density maximum of the disk, $r=12\rm M$.  
The result of this procedure is exemplified in Figure \ref{fig:pr-rescaled-192} for the high-resolution data.  
It is obvious that this improves the overlap in the inner regions $r<\rm 12 M$ in densities and pressure, but the spread in magnetic fields is worsened.   
This is merely a consequence of the non-convergence of magnetization, for if the density is made to match, the spread in magnetic field increases with the reference run having roughly a twice larger magnetization than most $192^3$ runs.

\section{Measuring the Maxwell Stress}\label{sec:tetrads}

Maxwell stresses play an important role in disk accretion, however their definition is ambiguous in general relativity due to the difficulty of identifying spatial directions in one frame with the ones in another.  In particular, due to the local nature of the magneto-rotational instability, the physical interpretation is best guided by stresses measured in a frame co-rotating with the disk.  
Hence to relate to the classical disk model of \cite{NovikovThorne1973}, \cite{KrolikHawley2005} devised a set of co-moving tetrads most closely preserving the Boyer-Lindquist azimuthal and radial directions.  The expressions for the basis are written explicitly in Appendix A of \cite{Beckwith2008} and need not be reproduced here.  Note that in order to use them, the four-vectors $b^{\mu}$ (here assumed to be in Kerr-Schild coordinates) first need to be converted to $b'_{\mu}$ in Boyer-Lindquist coordinates.  

To validate the simplified approach taken in Section \ref{sec:maxwell}, we here compare Maxwell stresses obtained with the following two prescriptions:

\begin{align}
  \langle w^{r\phi}\rangle(r,t) &:= - \frac{\int_{0}^{2\pi}d\phi \int_{\theta_{\rm min}}^{\theta_{\rm max}}d\theta\, \sqrt{-g} \sqrt{g_{rr}} b^r \sqrt{g_{\phi\phi}} b^\phi }{ \int_{0}^{2\pi}d\phi \int_{\theta_{\rm min}}^{\theta_{\rm max}}d\theta\, \sqrt{-g}} \label{eq:alphar_coord}\\
  \langle w^{(\bar{r})(\bar{\phi})}\rangle(r,t) &:= - \frac{\int_{\bar{\phi}_{\rm min}}^{\bar{\phi}_{\rm max}}dx^{(\bar{\phi})} \int_{\bar{\theta}_{\rm min}}^{\bar{\theta}_{\rm max}}dx^{(\bar{\theta})}\, b^{(\bar{r})} b^{(\bar{\phi})} }{\int_{\bar{\phi}_{\rm min}}^{\bar{\phi}_{\rm max}}dx^{(\bar{\phi})} \int_{\bar{\theta}_{\rm min}}^{\bar{\theta}}dx^{(\bar{\theta})}} \label{eq:alphar_fluid} 
\end{align}

where $\langle w^{r\phi}\rangle(r,t)$ is the Kerr-Schild frame measurement used in Section \ref{sec:maxwell} and $\langle w^{(\bar{r})(\bar{\phi})}\rangle(r,t)$ is the stress in a frame co-moving with the local fluid velocity as in \cite{KrolikHawley2005}. Integration is carried out over the equatorial wedge $\theta_{\rm KS}\in[\pi/3,2\pi/3]$ and the full azimuthal range.  
Note that we have taken the fluid frame integration over the co-moving coordinate increments $dx^{(\bar{\mu})}$ which result from the transformation of 
$dx^\mu:=(0,\Delta r, \Delta \theta, \Delta \phi)$ as appropriate (see discussion in \citealt{NobleKrolik2010}).  

\begin{figure}[htbp]
\begin{center}
\includegraphics[width=15cm]{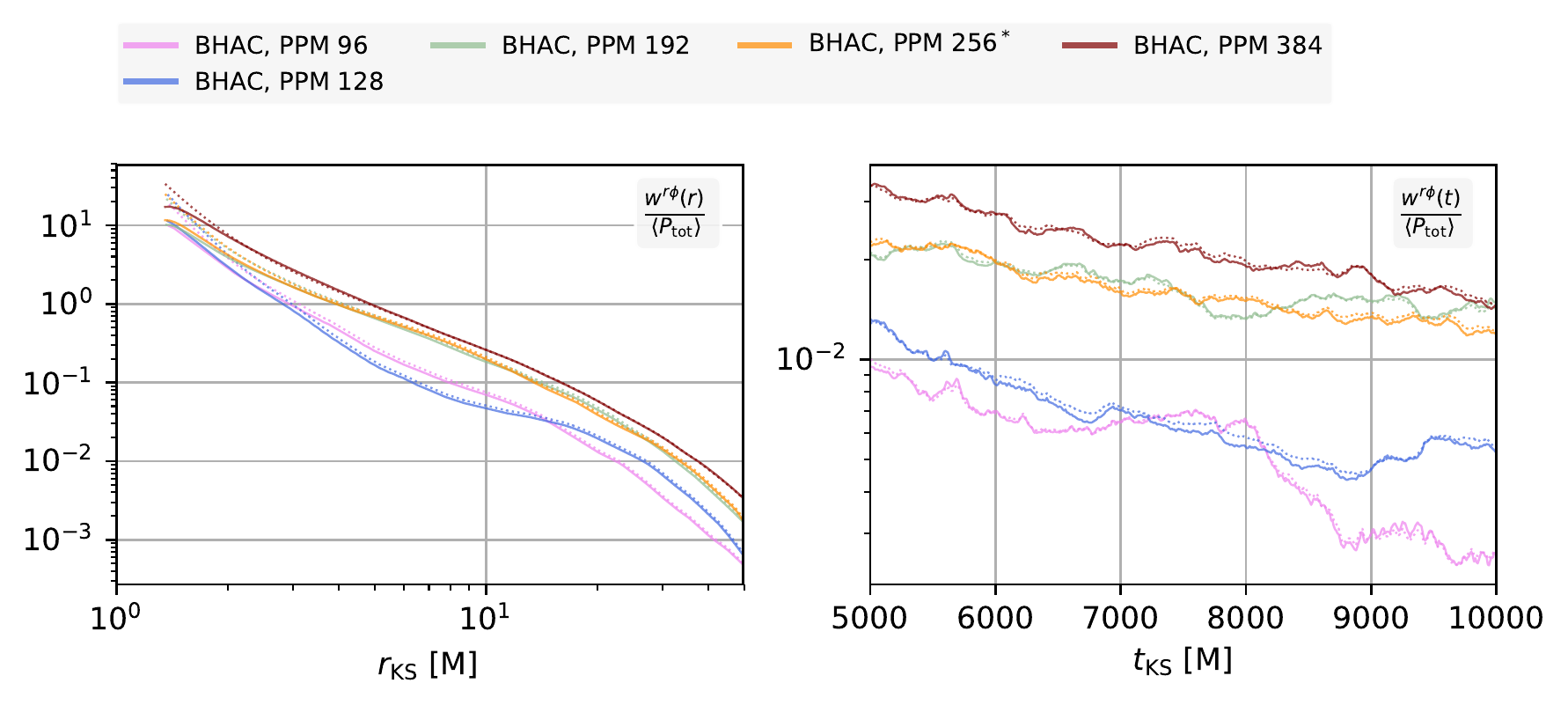}
\caption{Comparison of the Maxwell-Stress in different frames.  Solid curves show the fluid-frame measurement and dotted curves the Kerr-Schild coordinate-frame stress.  There is very good correspondence between the profiles up to $\sim 2\rm M$.  Consequently, the time evolution of the volume-averaged stresses agrees within a few percent.   }
\label{fig:BLstress}
\end{center}
\end{figure}

The resulting stress profiles and volume integrated time-series (non-dimensionalised by time- and volume- averaged total pressure $\langle P_{\rm tot}\rangle$) are shown in Figure \ref{fig:BLstress} for the five runs with the \bhac code.  
Significant departures of the two measurements only occur within $r_{\rm KS}\simeq 2\rm M$ where the fluid-frame stress flattens out and shows indications for a maximum for the two highest resolution runs.  For reference, the ISCO is located at $2.04 \rm M$ which roughly coincides with the change of slope.  
The overall stress differs only by a few percent between diagnostics (\ref{eq:alphar_coord}) and (\ref{eq:alphar_fluid}) as demonstrated in the time-series.  This is much smaller than differences between resolutions.  
We hence conclude that the simple coordinate frame measurement is appropriate for our purpose of studying the convergence and robustness of $\alpha$ as carried out in Section \ref{sec:maxwell}.  
It is important however to stress that the good correspondence between fluid- and coordinate- frame stresses does not necessarily hold for all values of the black hole spin.  In the Schwarzschild case for example, while the fluid frame stress drops to zero at the horizon \citep{KrolikHawley2005,NobleKrolik2010}, the coordinate frame stress remains monotonous.

\bibliography{astro.bib,more-ref.bib,aeireferences.bib}{}

\begin{thebibliography}{}
\expandafter\ifx\csname natexlab\endcsname\relax\def\natexlab#1{#1}\fi
\providecommand{\url}[1]{\href{#1}{#1}}
\providecommand{\dodoi}[1]{doi:~\href{http://doi.org/#1}{\nolinkurl{#1}}}
\providecommand{\doeprint}[1]{\href{http://ascl.net/#1}{\nolinkurl{http://ascl.net/#1}}}
\providecommand{\doarXiv}[1]{\href{https://arxiv.org/abs/#1}{\nolinkurl{https://arxiv.org/abs/#1}}}

\bibitem[{{Abbott} {et~al.}(2016){Abbott}, {Abbott}, {Abbott}, {Abernathy},
  {Acernese}, {Ackley}, {Adams}, {Adams}, {Addesso}, {Adhikari}, \&
  et~al.}]{AbbottAbbott2016a}
{Abbott}, B.~P., {Abbott}, R., {Abbott}, T.~D., {et~al.} 2016, Physical Review
  Letters, 116, 241103, \dodoi{10.1103/PhysRevLett.116.241103}

\bibitem[{Abbott {et~al.}(2017)}]{Abbott2017_etal}
Abbott, B.~P., {et~al.} 2017, Phys. Rev. Lett., 119, 161101,
  \dodoi{10.1103/PhysRevLett.119.161101}

\bibitem[{Akiyama {et~al.}(2015)Akiyama, Lu, Fish, Doeleman, Broderick, Dexter,
  Hada, Kino, Nagai, Honma, Johnson, Algaba, Asada, Brinkerink, Blundell,
  Bower, Cappallo, Crew, Dexter, Dzib, Freund, Friberg, Gurwell, Ho, Inoue,
  Krichbaum, Loinard, MacMahon, Marrone, Moran, Nakamura, Nagar, Ortiz-Leon,
  Plambeck, Pradel, Primiani, Rogers, Roy, SooHoo, Tavares, Tilanus, Titus,
  Wagner, Weintroub, Yamaguchi, Young, Zensus, \& Ziurys}]{AkiyamaLuEtAl2015}
Akiyama, K., Lu, R.-S., Fish, V.~L., {et~al.} 2015, \apj, 807, 150,
  \dodoi{10.1088/0004-637X/807/2/150}

\bibitem[{{Anninos} {et~al.}(2017){Anninos}, {Bryant}, {Fragile}, {Holgado},
  {Lau}, \& {Nemergut}}]{AnninosBryant2017}
{Anninos}, P., {Bryant}, C., {Fragile}, P.~C., {et~al.} 2017, \apjs, 231, 17,
  \dodoi{10.3847/1538-4365/aa7ff5}

\bibitem[{{Anninos} {et~al.}(2005){Anninos}, {Fragile}, \&
  {Salmonson}}]{Anninos2005}
{Anninos}, P., {Fragile}, P.~C., \& {Salmonson}, J.~D. 2005, \apj, 635, 723,
  \dodoi{10.1086/497294}

\bibitem[{{Anninos} {et~al.}(2012){Anninos}, {Fragile}, {Wilson}, \&
  {Murray}}]{Anninos2012}
{Anninos}, P., {Fragile}, P.~C., {Wilson}, J., \& {Murray}, S.~D. 2012, \apj,
  759, 132, \dodoi{10.1088/0004-637X/759/2/132}

\bibitem[{Ant{\'o}n {et~al.}(2006)Ant{\'o}n, Zanotti, Miralles, Mart{\'\i},
  Ib{\'a}{\~n}ez, Font, \& Pons}]{Anton05}
Ant{\'o}n, L., Zanotti, O., Miralles, J.~A., {et~al.} 2006, Astrophys. J., 637,
  296

\bibitem[{{Armitage} \& {Reynolds}(2003)}]{ArmitageReynolds2003}
{Armitage}, P.~J., \& {Reynolds}, C.~S. 2003, \mnras, 341, 1041,
  \dodoi{10.1046/j.1365-8711.2003.06491.x}

\bibitem[{{Asada} \& {Nakamura}(2012)}]{AsadaNakamura2012}
{Asada}, K., \& {Nakamura}, M. 2012, \apjl, 745, L28,
  \dodoi{10.1088/2041-8205/745/2/L28}

\bibitem[{{Bacchini} {et~al.}(2018){Bacchini}, {Ripperda}, {Chen}, \&
  {Sironi}}]{BacchiniRipperda2018a}
{Bacchini}, F., {Ripperda}, B., {Chen}, A.~Y., \& {Sironi}, L. 2018, \apjs,
  237, 6, \dodoi{10.3847/1538-4365/aac9ca}

\bibitem[{{Bacchini} {et~al.}(2019){Bacchini}, {Ripperda}, {Porth}, \&
  {Sironi}}]{2019ApJS..240...40B}
{Bacchini}, F., {Ripperda}, B., {Porth}, O., \& {Sironi}, L. 2019, \apjs, 240,
  40, \dodoi{10.3847/1538-4365/aafcb3}

\bibitem[{{Baganoff} {et~al.}(2001){Baganoff}, {Bautz}, {Brandt}, {Chartas},
  {Feigelson}, {Garmire}, {Maeda}, {Morris}, {Ricker}, {Townsley}, \&
  {Walter}}]{BaganoffBautz2001}
{Baganoff}, F.~K., {Bautz}, M.~W., {Brandt}, W.~N., {et~al.} 2001, \nat, 413,
  45, \dodoi{10.1038/35092510}

\bibitem[{{Baiotti} {et~al.}(2005){Baiotti}, {Hawke}, {Montero}, {L{\"o}ffler},
  {Rezzolla}, {Stergioulas}, {Font}, \& {Seidel}}]{Baiotti04}
{Baiotti}, L., {Hawke}, I., {Montero}, P.~J., {et~al.} 2005, Phys. Rev. D, 71,
  024035, \dodoi{10.1103/PhysRevD.71.024035}

\bibitem[{{Balbus} \& {Hawley}(1991)}]{Balbus1991}
{Balbus}, S.~A., \& {Hawley}, J.~F. 1991, Astrophys. J., 376, 214,
  \dodoi{10.1086/170270}

\bibitem[{{Balbus} \& {Hawley}(1998)}]{BalbusHawley1998}
---. 1998, Rev. Mod. Phys., 70, 1, \dodoi{10.1103/RevModPhys.70.1}

\bibitem[{Balbus \& Henri(2008)}]{BalbusHenri2008}
Balbus, S.~A., \& Henri, P. 2008, \apj, 674, 408, \dodoi{10.1086/524838}

\bibitem[{{Balsara} \& {Spicer}(1999)}]{Balsara1999b}
{Balsara}, D.~S., \& {Spicer}, D. 1999, Journal of Computational Physics, 148,
  133, \dodoi{10.1006/jcph.1998.6108}

\bibitem[{{Beckwith} {et~al.}(2008){Beckwith}, {Hawley}, \&
  {Krolik}}]{Beckwith2008}
{Beckwith}, K., {Hawley}, J.~F., \& {Krolik}, J.~H. 2008, \mnras, 390, 21,
  \dodoi{10.1111/j.1365-2966.2008.13710.x}

\bibitem[{{Beskin} {et~al.}(1992){Beskin}, {Istomin}, \&
  {Parev}}]{BeskinIstomin1992}
{Beskin}, V.~S., {Istomin}, Y.~N., \& {Parev}, V.~I. 1992, \sovast, 36, 642

\bibitem[{Bisnovatyi-Kogan \& Ruzmaikin(1976)}]{Bisnovatyi-KoganRuzmaikin1976}
Bisnovatyi-Kogan, G.~S., \& Ruzmaikin, A.~A. 1976, \apss, 42, 401,
  \dodoi{10.1007/BF01225967}

\bibitem[{{Blandford} \& {Znajek}(1977)}]{BlandfordZnajek1977}
{Blandford}, R.~D., \& {Znajek}, R.~L. 1977, \mnras, 179, 433

\bibitem[{Bodo {et~al.}(2014)Bodo, Cattaneo, Mignone, \&
  Rossi}]{BodoCattaneoEtAl2014}
Bodo, G., Cattaneo, F., Mignone, A., \& Rossi, P. 2014, \apjl, 787, L13,
  \dodoi{10.1088/2041-8205/787/1/L13}

\bibitem[{{Bowen} {et~al.}(2018){Bowen}, {Mewes}, {Campanelli}, {Noble},
  {Krolik}, \& {Zilh{\~a}o}}]{Bowen2018}
{Bowen}, D.~B., {Mewes}, V., {Campanelli}, M., {et~al.} 2018, \apjl, 853, L17,
  \dodoi{10.3847/2041-8213/aaa756}

\bibitem[{{Bower} {et~al.}(2015){Bower}, {Markoff}, {Dexter}, {Gurwell},
  {Moran}, {Brunthaler}, {Falcke}, {Fragile}, {Maitra}, {Marrone}, {Peck},
  {Rushton}, \& {Wright}}]{BowerMarkoff2015}
{Bower}, G.~C., {Markoff}, S., {Dexter}, J., {et~al.} 2015, \apj, 802, 69,
  \dodoi{10.1088/0004-637X/802/1/69}

\bibitem[{{Broderick} \& {Tchekhovskoy}(2015)}]{BroderickTchekhovskoy2015}
{Broderick}, A.~E., \& {Tchekhovskoy}, A. 2015, \apj, 809, 97,
  \dodoi{10.1088/0004-637X/809/1/97}

\bibitem[{{Bucciantini} \& {Del Zanna}(2011)}]{BucciantiniDel-Zanna2011}
{Bucciantini}, N., \& {Del Zanna}, L. 2011, \aap, 528, A101,
  \dodoi{10.1051/0004-6361/201015945}

\bibitem[{{Bucciantini} \& {Del Zanna}(2013)}]{BucciantiniDel-Zanna2013}
---. 2013, \mnras, 428, 71, \dodoi{10.1093/mnras/sts005}

\bibitem[{{Bugli} {et~al.}(2014){Bugli}, {Del Zanna}, \&
  {Bucciantini}}]{BugliDel-Zanna2014}
{Bugli}, M., {Del Zanna}, L., \& {Bucciantini}, N. 2014, \mnras, 440, L41,
  \dodoi{10.1093/mnrasl/slu017}

\bibitem[{{Bugli} {et~al.}(2018){Bugli}, {Guilet}, {M{\"u}ller}, {Del Zanna},
  {Bucciantini}, \& {Montero}}]{Bugli2018}
{Bugli}, M., {Guilet}, J., {M{\"u}ller}, E., {et~al.} 2018, \mnras, 475, 108,
  \dodoi{10.1093/mnras/stx3158}

\bibitem[{{Camenzind}(1986)}]{Camenzind1986}
{Camenzind}, M. 1986, \aap, 162, 32

\bibitem[{{Chael} {et~al.}(2018){Chael}, {Narayan}, \&
  {Johnson}}]{ChaelNarayan2018}
{Chael}, A., {Narayan}, R., \& {Johnson}, M.~D. 2018, ArXiv e-prints.
\newblock \doarXiv{1810.01983}

\bibitem[{{Chael} {et~al.}(2017){Chael}, {Narayan}, \& {Sa{\c
  d}owski}}]{chael17}
{Chael}, A.~A., {Narayan}, R., \& {Sa{\c d}owski}, A. 2017, \mnras, 470, 2367,
  \dodoi{10.1093/mnras/stx1345}

\bibitem[{{Chan} {et~al.}(2015){Chan}, {Psaltis}, {{\"O}zel}, {Medeiros},
  {Marrone}, {Sa{\c d}owski}, \& {Narayan}}]{ChanPsaltis2015a}
{Chan}, C.-k., {Psaltis}, D., {{\"O}zel}, F., {et~al.} 2015, \apj, 812, 103,
  \dodoi{10.1088/0004-637X/812/2/103}

\bibitem[{{Chandra} {et~al.}(2017){Chandra}, {Foucart}, \&
  {Gammie}}]{ChandraFoucart2017}
{Chandra}, M., {Foucart}, F., \& {Gammie}, C.~F. 2017, \apj, 837, 92,
  \dodoi{10.3847/1538-4357/aa5f55}

\bibitem[{{Chandra} {et~al.}(2015){Chandra}, {Gammie}, {Foucart}, \&
  {Quataert}}]{Chandra2015}
{Chandra}, M., {Gammie}, C.~F., {Foucart}, F., \& {Quataert}, E. 2015, \apj,
  810, 162, \dodoi{10.1088/0004-637X/810/2/162}

\bibitem[{Chatterjee {et~al.}(2019)Chatterjee, Liska, Tchekhovskoy, \&
  Markoff}]{ChatterjeeEtAl2019}
Chatterjee, K., Liska, M., Tchekhovskoy, A., \& Markoff, S.~B. 2019, arXiv
  e-prints.
\newblock \doarXiv{1904.03243}

\bibitem[{{Colella} \& {Woodward}(1984)}]{Colella1984}
{Colella}, P., \& {Woodward}, P.~R. 1984, Journal of Computational Physics, 54,
  174, \dodoi{10.1016/0021-9991(84)90143-8}

\bibitem[{{Davelaar} {et~al.}(2018){Davelaar}, {Mo{\'s}cibrodzka}, {Bronzwaer},
  \& {Falcke}}]{DavelaarMoscibrodzkaEtAl2018}
{Davelaar}, J., {Mo{\'s}cibrodzka}, M., {Bronzwaer}, T., \& {Falcke}, H. 2018,
  \aap, 612, A34, \dodoi{10.1051/0004-6361/201732025}

\bibitem[{{Davelaar} {et~al.}(2019){Davelaar}, {Olivares}, {Porth},
  {Bronzwaer}, {Janssen}, {Roelofs}, {Mizuno}, {Fromm}, {Falcke}, \&
  {Rezzolla}}]{2019arXiv190610065D}
{Davelaar}, J., {Olivares}, H., {Porth}, O., {et~al.} 2019, arXiv e-prints,
  arXiv:1906.10065.
\newblock \doarXiv{1906.10065}

\bibitem[{{Davis} {et~al.}(2010){Davis}, {Stone}, \& {Pessah}}]{DavisStone2010}
{Davis}, S.~W., {Stone}, J.~M., \& {Pessah}, M.~E. 2010, \apj, 713, 52,
  \dodoi{10.1088/0004-637X/713/1/52}

\bibitem[{{De Villiers} \& {Hawley}(2003)}]{DeVilliers03a}
{De Villiers}, J.-P., \& {Hawley}, J.~F. 2003, Astrophys. J., 589, 458,
  \dodoi{10.1086/373949}

\bibitem[{{Del Zanna} \& {Bucciantini}(2018)}]{Del-Zanna2018}
{Del Zanna}, L., \& {Bucciantini}, N. 2018, \mnras, 479, 657,
  \dodoi{10.1093/mnras/sty1633}

\bibitem[{{Del Zanna} {et~al.}(2016){Del Zanna}, {Papini}, {Landi}, {Bugli}, \&
  {Bucciantini}}]{2016MNRAS.460.3753D}
{Del Zanna}, L., {Papini}, E., {Landi}, S., {Bugli}, M., \& {Bucciantini}, N.
  2016, \mnras, 460, 3753, \dodoi{10.1093/mnras/stw1242}

\bibitem[{{Del Zanna} {et~al.}(2007){Del Zanna}, {Zanotti}, {Bucciantini}, \&
  {Londrillo}}]{DelZanna2007}
{Del Zanna}, L., {Zanotti}, O., {Bucciantini}, N., \& {Londrillo}, P. 2007,
  Astron. Astrophys., 473, 11, \dodoi{10.1051/0004-6361:20077093}

\bibitem[{Dexter {et~al.}(2009)Dexter, Agol, \& Fragile}]{DexterAgolEtAl2009}
Dexter, J., Agol, E., \& Fragile, P.~C. 2009, \apjl, 703, L142,
  \dodoi{10.1088/0004-637X/703/2/L142}

\bibitem[{{Dexter} {et~al.}(2010){Dexter}, {Agol}, {Fragile}, \&
  {McKinney}}]{DexterAgol2010}
{Dexter}, J., {Agol}, E., {Fragile}, P.~C., \& {McKinney}, J.~C. 2010, \apj,
  717, 1092, \dodoi{10.1088/0004-637X/717/2/1092}

\bibitem[{{Dexter} {et~al.}(2014){Dexter}, {Kelly}, {Bower}, {Marrone},
  {Stone}, \& {Plambeck}}]{DexterKelly2014}
{Dexter}, J., {Kelly}, B., {Bower}, G.~C., {et~al.} 2014, \mnras, 442, 2797,
  \dodoi{10.1093/mnras/stu1039}

\bibitem[{Dexter {et~al.}(2012)Dexter, McKinney, \&
  Agol}]{DexterMcKinneyEtAl2012}
Dexter, J., McKinney, J.~C., \& Agol, E. 2012, \mnras, 421, 1517,
  \dodoi{10.1111/j.1365-2966.2012.20409.x}

\bibitem[{{Dhang} \& {Sharma}(2018)}]{DhangSharma2018}
{Dhang}, P., \& {Sharma}, P. 2018, \mnras, \dodoi{10.1093/mnras/sty2692}

\bibitem[{{Dibi} {et~al.}(2012){Dibi}, {Drappeau}, {Fragile}, {Markoff}, \&
  {Dexter}}]{Dibi+2012}
{Dibi}, S., {Drappeau}, S., {Fragile}, P.~C., {Markoff}, S., \& {Dexter}, J.
  2012, \mnras, 426, 1928, \dodoi{10.1111/j.1365-2966.2012.21857.x}

\bibitem[{{Do} {et~al.}(2009){Do}, {Ghez}, {Morris}, {Yelda}, {Meyer}, {Lu},
  {Hornstein}, \& {Matthews}}]{DoGhez2009}
{Do}, T., {Ghez}, A.~M., {Morris}, M.~R., {et~al.} 2009, \apj, 691, 1021,
  \dodoi{10.1088/0004-637X/691/2/1021}

\bibitem[{{Dodds-Eden} {et~al.}(2009){Dodds-Eden}, {Porquet}, {Trap},
  {Quataert}, {Haubois}, {Gillessen}, {Grosso}, {Pantin}, {Falcke}, {Rouan},
  {Genzel}, {Hasinger}, {Goldwurm}, {Yusef-Zadeh}, {Clenet}, {Trippe},
  {Lagage}, {Bartko}, {Eisenhauer}, {Ott}, {Paumard}, {Perrin}, {Yuan},
  {Fritz}, \& {Mascetti}}]{Dodds-EdenPorquet2009}
{Dodds-Eden}, K., {Porquet}, D., {Trap}, G., {et~al.} 2009, \apj, 698, 676,
  \dodoi{10.1088/0004-637X/698/1/676}

\bibitem[{{Dodds-Eden} {et~al.}(2011){Dodds-Eden}, {Gillessen}, {Fritz},
  {Eisenhauer}, {Trippe}, {Genzel}, {Ott}, {Bartko}, {Pfuhl}, {Bower},
  {Goldwurm}, {Porquet}, {Trap}, \& {Yusef-Zadeh}}]{Dodds-EdenGillessen2011}
{Dodds-Eden}, K., {Gillessen}, S., {Fritz}, T.~K., {et~al.} 2011, \apj, 728,
  37, \dodoi{10.1088/0004-637X/728/1/37}

\bibitem[{{Doeleman} {et~al.}(2008){Doeleman}, {Weintroub}, {Rogers},
  {Plambeck}, {Freund}, {Tilanus}, {Friberg}, {Ziurys}, {Moran}, {Corey},
  {Young}, {Smythe}, {Titus}, {Marrone}, {Cappallo}, {Bock}, {Bower},
  {Chamberlin}, {Davis}, {Krichbaum}, {Lamb}, {Maness}, {Niell}, {Roy},
  {Strittmatter}, {Werthimer}, {Whitney}, \& {Woody}}]{DoelemanWeintroub2008}
{Doeleman}, S.~S., {Weintroub}, J., {Rogers}, A.~E.~E., {et~al.} 2008, \nat,
  455, 78, \dodoi{10.1038/nature07245}

\bibitem[{{Doeleman} {et~al.}(2012){Doeleman}, {Fish}, {Schenck}, {Beaudoin},
  {Blundell}, {Bower}, {Broderick}, {Chamberlin}, {Freund}, {Friberg},
  {Gurwell}, {Ho}, {Honma}, {Inoue}, {Krichbaum}, {Lamb}, {Loeb}, {Lonsdale},
  {Marrone}, {Moran}, {Oyama}, {Plambeck}, {Primiani}, {Rogers}, {Smythe},
  {SooHoo}, {Strittmatter}, {Tilanus}, {Titus}, {Weintroub}, {Wright}, {Young},
  \& {Ziurys}}]{doeleman2012}
{Doeleman}, S.~S., {Fish}, V.~L., {Schenck}, D.~E., {et~al.} 2012, ArXiv
  e-prints.
\newblock \doarXiv{1210.6132}

\bibitem[{{Dolence} {et~al.}(2012){Dolence}, {Gammie}, {Shiokawa}, \&
  {Noble}}]{DolenceGammie2012}
{Dolence}, J.~C., {Gammie}, C.~F., {Shiokawa}, H., \& {Noble}, S.~C. 2012,
  \apjl, 746, L10, \dodoi{10.1088/2041-8205/746/1/L10}

\bibitem[{{Duez} {et~al.}(2005){Duez}, {Liu}, {Shapiro}, \&
  {Stephens}}]{Duez05MHD0}
{Duez}, M.~D., {Liu}, Y.~T., {Shapiro}, S.~L., \& {Stephens}, B.~C. 2005, Phys.
  Rev. D, 72, 024028, \dodoi{10.1103/PhysRevD.72.024028}

\bibitem[{Etienne {et~al.}(2012{\natexlab{a}})Etienne, Liu, Paschalidis, \&
  Shapiro}]{Etienne:2011ea}
Etienne, Z.~B., Liu, Y.~T., Paschalidis, V., \& Shapiro, S.~L.
  2012{\natexlab{a}}, Phys. Rev. D, 85, 064029

\bibitem[{Etienne {et~al.}(2006)Etienne, Liu, \& Shapiro}]{Etienne:2006am}
Etienne, Z.~B., Liu, Y.~T., \& Shapiro, S.~L. 2006, Phys. Rev. D, 74, 044030

\bibitem[{{Etienne} {et~al.}(2010){Etienne}, {Liu}, \&
  {Shapiro}}]{Etienne:2010ui}
{Etienne}, Z.~B., {Liu}, Y.~T., \& {Shapiro}, S.~L. 2010, Phys. Rev. D, 82,
  084031, \dodoi{10.1103/PhysRevD.82.084031}

\bibitem[{Etienne {et~al.}(2015)Etienne, Paschalidis, Haas, Moesta, \&
  Shapiro}]{Etienne:2015cea}
Etienne, Z.~B., Paschalidis, V., Haas, R., Moesta, P., \& Shapiro, S.~L. 2015,
  Class. Quantum Grav., 32, 175009

\bibitem[{Etienne {et~al.}(2012{\natexlab{b}})Etienne, Paschalidis, Liu, \&
  Shapiro}]{Etienne:2011re}
Etienne, Z.~B., Paschalidis, V., Liu, Y.~T., \& Shapiro, S.~L.
  2012{\natexlab{b}}, Phys. Rev. D, 85, 024013

\bibitem[{{Evans} \& {Hawley}(1988)}]{evans1988}
{Evans}, C.~R., \& {Hawley}, J.~F. 1988, \apj, 332, 659, \dodoi{10.1086/166684}

\bibitem[{{Event Horizon Telescope Collaboration}
  {et~al.}(2019{\natexlab{a}}){Event Horizon Telescope Collaboration}, Akiyama,
  Alberdi, Alef, Asada, Azulay, Baczko, Ball, Balokovi{\'c}, Barrett, Bintley,
  Blackburn, Boland, Bouman, Bower, Bremer, Brinkerink, Brissenden, Britzen,
  Broderick, Broguiere, Bronzwaer, Byun, Carlstrom, Chael, Chan, Chatterjee,
  Chatterjee, Chen, Chen, Cho, Christian, Conway, Cordes, Crew, Cui, Davelaar,
  De~Laurentis, Deane, Dempsey, Desvignes, Dexter, Doeleman, Eatough, Falcke,
  Fish, Fomalont, Fraga-Encinas, Freeman, Friberg, Fromm, G{\'o}mez, Galison,
  Gammie, Garc{\'{\i}}a, Gentaz, Georgiev, Goddi, Gold, Gu, Gurwell, Hada,
  Hecht, Hesper, Ho, Ho, Honma, Huang, Huang, Hughes, Ikeda, Inoue, Issaoun,
  James, Jannuzi, Janssen, Jeter, Jiang, Johnson, Jorstad, Jung, Karami,
  Karuppusamy, Kawashima, Keating, Kettenis, Kim, Kim, Kim, Kino, Koay, Koch,
  Koyama, Kramer, Kramer, Krichbaum, Kuo, Lauer, Lee, Li, Li, Lindqvist, Liu,
  Liuzzo, Lo, Lobanov, Loinard, Lonsdale, Lu, MacDonald, Mao, Markoff, Marrone,
  Marscher, Mart{\'{\i}}-Vidal, Matsushita, Matthews, Medeiros, Menten, Mizuno,
  Mizuno, Moran, Moriyama, Moscibrodzka, M{\"u}ller, Nagai, Nagar, Nakamura,
  Narayan, Narayanan, Natarajan, Neri, Ni, Noutsos, Okino, Olivares,
  Ortiz-Le{\'o}n, Oyama, {\"O}zel, Palumbo, Patel, Pen, Pesce, Pi{\'e}tu,
  Plambeck, PopStefanija, Porth, Prather, Preciado-L{\'o}pez, Psaltis, Pu,
  Ramakrishnan, Rao, Rawlings, Raymond, Rezzolla, Ripperda, Roelofs, Rogers,
  Ros, Rose, Roshanineshat, Rottmann, Roy, Ruszczyk, Ryan, Rygl, S{\'a}nchez,
  S{\'a}nchez-Arguelles, Sasada, Savolainen, Schloerb, Schuster, Shao, Shen,
  Small, Sohn, SooHoo, Tazaki, Tiede, Tilanus, Titus, Toma, Torne, Trent,
  Trippe, Tsuda, van Bemmel, van Langevelde, van Rossum, Wagner, Wardle,
  Weintroub, Wex, Wharton, Wielgus, Wong, Wu, Young, Young, Younsi, Yuan, Yuan,
  Zensus, Zhao, Zhao, Zhu, Algaba, Allardi, Amestica, Anczarski, Bach,
  Baganoff, Beaudoin, Benson, Berthold, Blanchard, Blundell, Bustamente,
  Cappallo, Castillo-Dom{\'{\i}}nguez, Chang, Chang, Chang, Chen, Chilson,
  Chuter, C{\'o}rdova~Rosado, Coulson, Crawford, Crowley, David, Derome,
  Dexter, Dornbusch, Dudevoir, Dzib, Eckart, Eckert, Erickson, Everett, Faber,
  Farah, Fath, Folkers, Forbes, Freund, G{\'o}mez-Ruiz, Gale, Gao, Geertsema,
  Graham, Greer, Grosslein, Gueth, Haggard, Halverson, Han, Han, Hao, Hasegawa,
  Henning, Hern{\'a}ndez-G{\'o}mez, Herrero-Illana, Heyminck, Hirota, Hoge,
  Huang, Impellizzeri, Jiang, Kamble, Keisler, Kimura, Kono, Kubo, Kuroda,
  Lacasse, Laing, Leitch, Li, Lin, Liu, Liu, Lu, Marson, Martin-Cocher,
  Massingill, Matulonis, McColl, McWhirter, Messias, Meyer-Zhao, Michalik,
  Monta{\~n}a, Montgomerie, Mora-Klein, Muders, Nadolski, Navarro, Neilsen,
  Nguyen, Nishioka, Norton, Nowak, Nystrom, Ogawa, Oshiro, Oyama, Parsons,
  Paine, Pe{\~n}alver, Phillips, Poirier, Pradel, Primiani, Raffin, Rahlin,
  Reiland, Risacher, Ruiz, S{\'a}ez-Mada{\'{\i}}n, Sassella, Schellart, Shaw,
  Silva, Shiokawa, Smith, Snow, Souccar, Sousa, Sridharan, Srinivasan, Stahm,
  Stark, Story, Timmer, Vertatschitsch, Walther, Wei, Whitehorn, Whitney,
  Woody, Wouterloot, Wright, Yamaguchi, Yu, Zeballos, Zhang, \&
  Ziurys}]{CollaborationAkiyamaEtAl2019}
{Event Horizon Telescope Collaboration}, Akiyama, K., Alberdi, A., {et~al.}
  2019{\natexlab{a}}, \apjl, 875, L1, \dodoi{10.3847/2041-8213/ab0ec7}

\bibitem[{{Event Horizon Telescope Collaboration}
  {et~al.}(2019{\natexlab{b}}){Event Horizon Telescope Collaboration}, Akiyama,
  Alberdi, Alef, Asada, Azulay, Baczko, Ball, Balokovi{\'c}, Barrett, Bintley,
  Blackburn, Boland, Bouman, Bower, Bremer, Brinkerink, Brissenden, Britzen,
  Broderick, Broguiere, Bronzwaer, Byun, Carlstrom, Chael, Chan, Chatterjee,
  Chatterjee, Chen, Chen, Cho, Christian, Conway, Cordes, Crew, Cui, Davelaar,
  De~Laurentis, Deane, Dempsey, Desvignes, Dexter, Doeleman, Eatough, Falcke,
  Fish, Fomalont, Fraga-Encinas, Friberg, Fromm, G{\'o}mez, Galison, Gammie,
  Garc{\'{\i}}a, Gentaz, Georgiev, Goddi, Gold, Gu, Gurwell, Hada, Hecht,
  Hesper, Ho, Ho, Honma, Huang, Huang, Hughes, Ikeda, Inoue, Issaoun, James,
  Jannuzi, Janssen, Jeter, Jiang, Johnson, Jorstad, Jung, Karami, Karuppusamy,
  Kawashima, Keating, Kettenis, Kim, Kim, Kim, Kino, Koay, Koch, Koyama,
  Kramer, Kramer, Krichbaum, Kuo, Lauer, Lee, Li, Li, Lindqvist, Liu, Liuzzo,
  Lo, Lobanov, Loinard, Lonsdale, Lu, MacDonald, Mao, Markoff, Marrone,
  Marscher, Mart{\'{\i}}-Vidal, Matsushita, Matthews, Medeiros, Menten, Mizuno,
  Mizuno, Moran, Moriyama, Moscibrodzka, M{\"u}ller, Nagai, Nagar, Nakamura,
  Narayan, Narayanan, Natarajan, Neri, Ni, Noutsos, Okino, Olivares, Oyama,
  {\"O}zel, Palumbo, Patel, Pen, Pesce, Pi{\'e}tu, Plambeck, PopStefanija,
  Porth, Prather, Preciado-L{\'o}pez, Psaltis, Pu, Ramakrishnan, Rao, Rawlings,
  Raymond, Rezzolla, Ripperda, Roelofs, Rogers, Ros, Rose, Roshanineshat,
  Rottmann, Roy, Ruszczyk, Ryan, Rygl, S{\'a}nchez, S{\'a}nchez-Arguelles,
  Sasada, Savolainen, Schloerb, Schuster, Shao, Shen, Small, Sohn, SooHoo,
  Tazaki, Tiede, Tilanus, Titus, Toma, Torne, Trent, Trippe, Tsuda, van Bemmel,
  van Langevelde, van Rossum, Wagner, Wardle, Weintroub, Wex, Wharton, Wielgus,
  Wong, Wu, Young, Young, Younsi, Yuan, Yuan, Zensus, Zhao, Zhao, Zhu,
  Anczarski, Baganoff, Eckart, Farah, Haggard, Meyer-Zhao, Michalik, Nadolski,
  Neilsen, Nishioka, Nowak, Pradel, Primiani, Souccar, Vertatschitsch,
  Yamaguchi, \& Zhang}]{CollaborationAkiyamaEtAl2019d}
---. 2019{\natexlab{b}}, \apjl, 875, L5, \dodoi{10.3847/2041-8213/ab0f43}

\bibitem[{{Fambri} {et~al.}(2018){Fambri}, {Dumbser}, {K{\"o}ppel}, {Rezzolla},
  \& {Zanotti}}]{FambriDumbser2018}
{Fambri}, F., {Dumbser}, M., {K{\"o}ppel}, S., {Rezzolla}, L., \& {Zanotti}, O.
  2018, \mnras, 477, 4543, \dodoi{10.1093/mnras/sty734}

\bibitem[{{Fishbone} \& {Moncrief}(1976)}]{Fishbone76}
{Fishbone}, L.~G., \& {Moncrief}, V. 1976, Astrophys. J., 207, 962

\bibitem[{Flock {et~al.}(2010)Flock, Dzyurkevich, Klahr, \&
  Mignone}]{FlockDzyurkevichEtAl2010}
Flock, M., Dzyurkevich, N., Klahr, H., \& Mignone, A. 2010, Astronomy and
  Astrophysics, 516, A26, \dodoi{10.1051/0004-6361/200912443}

\bibitem[{{Font}(2008)}]{Font2008}
{Font}, J.~A. 2008, Living Reviews in Relativity, 11, 7,
  \dodoi{10.12942/lrr-2008-7}

\bibitem[{Font \& Daigne(2002)}]{Font02a}
Font, J.~A., \& Daigne, F. 2002, Mon. Not. R. Astron. Soc., 334, 383

\bibitem[{{Foucart} {et~al.}(2017){Foucart}, {Chandra}, {Gammie}, {Quataert},
  \& {Tchekhovskoy}}]{FoucartChandra2017}
{Foucart}, F., {Chandra}, M., {Gammie}, C.~F., {Quataert}, E., \&
  {Tchekhovskoy}, A. 2017, \mnras, 470, 2240, \dodoi{10.1093/mnras/stx1368}

\bibitem[{{Fragile} {et~al.}(2007){Fragile}, {Blaes}, {Anninos}, \&
  {Salmonson}}]{FragileBlaes2007}
{Fragile}, P.~C., {Blaes}, O.~M., {Anninos}, P., \& {Salmonson}, J.~D. 2007,
  \apj, 668, 417, \dodoi{10.1086/521092}

\bibitem[{{Fragile} {et~al.}(2018){Fragile}, {Etheridge}, {Anninos}, {Mishra},
  \& {Klu{\'z}niak}}]{FragileEtheridgeEtAl2018}
{Fragile}, P.~C., {Etheridge}, S.~M., {Anninos}, P., {Mishra}, B., \&
  {Klu{\'z}niak}, W. 2018, \apj, 857, 1, \dodoi{10.3847/1538-4357/aab788}

\bibitem[{{Fragile} {et~al.}(2012){Fragile}, {Gillespie}, {Monahan},
  {Rodriguez}, \& {Anninos}}]{Fragile2012}
{Fragile}, P.~C., {Gillespie}, A., {Monahan}, T., {Rodriguez}, M., \&
  {Anninos}, P. 2012, Astrophys. J., Supp., 201, 9,
  \dodoi{10.1088/0067-0049/201/2/9}

\bibitem[{Fragile {et~al.}(2012)Fragile, Gillespie, Monahan, Rodriguez, \&
  Anninos}]{FragileGillespieEtAl2012}
Fragile, P.~C., Gillespie, A., Monahan, T., Rodriguez, M., \& Anninos, P. 2012,
  \apjs, 201, 9, \dodoi{10.1088/0067-0049/201/2/9}

\bibitem[{{Fragile} {et~al.}(2014){Fragile}, {Olejar}, \&
  {Anninos}}]{Fragile2014}
{Fragile}, P.~C., {Olejar}, A., \& {Anninos}, P. 2014, Astrophys. J., 796, 22,
  \dodoi{10.1088/0004-637X/796/1/22}

\bibitem[{Fromang {et~al.}(2007)Fromang, Papaloizou, Lesur, \&
  Heinemann}]{FromangPapaloizouEtAl2007}
Fromang, S., Papaloizou, J., Lesur, G., \& Heinemann, T. 2007, \aap, 476, 1123,
  \dodoi{10.1051/0004-6361:20077943}

\bibitem[{{Fujibayashi} {et~al.}(2018){Fujibayashi}, {Kiuchi}, {Nishimura},
  {Sekiguchi}, \& {Shibata}}]{FujibayashiKiuchi2018}
{Fujibayashi}, S., {Kiuchi}, K., {Nishimura}, N., {Sekiguchi}, Y., \&
  {Shibata}, M. 2018, \apj, 860, 64, \dodoi{10.3847/1538-4357/aabafd}

\bibitem[{{Gammie} \& {Guan}(2012)}]{hamcode}
{Gammie}, C.~F., \& {Guan}, X. 2012, {HAM2D: 2D Shearing Box Model},
  Astrophysics Source Code Library.
\newblock \doeprint{1210.022}

\bibitem[{{Gammie} {et~al.}(2003){Gammie}, {McKinney}, \&
  {T{\'o}th}}]{Gammie2003}
{Gammie}, C.~F., {McKinney}, J.~C., \& {T{\'o}th}, G. 2003, \apj, 589, 444,
  \dodoi{10.1086/374594}

\bibitem[{{Gardiner} \& {Stone}(2005)}]{2005JCoPh.205..509G}
{Gardiner}, T.~A., \& {Stone}, J.~M. 2005, Journal of Computational Physics,
  205, 509, \dodoi{10.1016/j.jcp.2004.11.016}

\bibitem[{{Genzel} {et~al.}(2010){Genzel}, {Eisenhauer}, \&
  {Gillessen}}]{GenzelEisenhauer2010}
{Genzel}, R., {Eisenhauer}, F., \& {Gillessen}, S. 2010, Reviews of Modern
  Physics, 82, 3121, \dodoi{10.1103/RevModPhys.82.3121}

\bibitem[{{Giacomazzo} \& {Rezzolla}(2007)}]{Giacomazzo:2007ti}
{Giacomazzo}, B., \& {Rezzolla}, L. 2007, Class. Quantum Grav., 24, 235,
  \dodoi{10.1088/0264-9381/24/12/S16}

\bibitem[{{Goddi} {et~al.}(2017){Goddi}, {Falcke}, {Kramer}, {Rezzolla},
  {Brinkerink}, {Bronzwaer}, {Davelaar}, {Deane}, {de Laurentis}, {Desvignes},
  {Eatough}, {Eisenhauer}, {Fraga-Encinas}, {Fromm}, {Gillessen}, {Grenzebach},
  {Issaoun}, {Jan{\ss}en}, {Konoplya}, {Krichbaum}, {Laing}, {Liu}, {Lu},
  {Mizuno}, {Moscibrodzka}, {M{\"u}ller}, {Olivares}, {Pfuhl}, {Porth},
  {Roelofs}, {Ros}, {Schuster}, {Tilanus}, {Torne}, {van Bemmel}, {van
  Langevelde}, {Wex}, {Younsi}, \& {Zhidenko}}]{GoddiFalcke2017}
{Goddi}, C., {Falcke}, H., {Kramer}, M., {et~al.} 2017, International Journal
  of Modern Physics D, 26, 1730001, \dodoi{10.1142/S0218271817300014}

\bibitem[{Gold {et~al.}(2014)Gold, Paschalidis, Ruiz, Shapiro, Etienne, \&
  Pfeiffer}]{Gold:2014dta}
Gold, R., Paschalidis, V., Ruiz, M., {et~al.} 2014, Phys. Rev. D, 90, 104030

\bibitem[{{Gravity Collaboration} {et~al.}(2018){Gravity Collaboration},
  {Abuter}, {Amorim}, {Baub{\"o}ck}, {Berger}, {Bonnet}, {Brandner},
  {Cl{\'e}net}, {Coud{\'e} Du Foresto}, {de Zeeuw}, {Deen}, {Dexter}, {Duvert},
  {Eckart}, {Eisenhauer}, {F{\"o}rster Schreiber}, {Garcia}, {Gao}, {Gendron},
  {Genzel}, {Gillessen}, {Guajardo}, {Habibi}, {Haubois}, {Henning}, {Hippler},
  {Horrobin}, {Huber}, {Jim{\'e}nez-Rosales}, {Jocou}, {Kervella}, {Lacour},
  {Lapeyr{\`e}re}, {Lazareff}, {Le Bouquin}, {L{\'e}na}, {Lippa}, {Ott},
  {Panduro}, {Paumard}, {Perraut}, {Perrin}, {Pfuhl}, {Plewa}, {Rabien},
  {Rodr{\'{\i}}guez-Coira}, {Rousset}, {Sternberg}, {Straub}, {Straubmeier},
  {Sturm}, {Tacconi}, {Vincent}, {von Fellenberg}, {Waisberg}, {Widmann},
  {Wieprecht}, {Wiezorrek}, {Woillez}, \&
  {Yazici}}]{GravityCollaborationAbuterEtAl2018}
{Gravity Collaboration}, {Abuter}, R., {Amorim}, A., {et~al.} 2018, \aap, 618,
  L10, \dodoi{10.1051/0004-6361/201834294}

\bibitem[{{Guan} \& {Gammie}(2008)}]{guan2008}
{Guan}, X., \& {Gammie}, C.~F. 2008, \apjs, 174, 145, \dodoi{10.1086/521147}

\bibitem[{Guan {et~al.}(2009)Guan, Gammie, Simon, \&
  Johnson}]{GuanGammieEtAl2009}
Guan, X., Gammie, C.~F., Simon, J.~B., \& Johnson, B.~M. 2009, \apj, 694, 1010,
  \dodoi{10.1088/0004-637X/694/2/1010}

\bibitem[{{Hada} {et~al.}(2013){Hada}, {Kino}, {Doi}, {Nagai}, {Honma},
  {Hagiwara}, {Giroletti}, {Giovannini}, \& {Kawaguchi}}]{HadaKino2013}
{Hada}, K., {Kino}, M., {Doi}, A., {et~al.} 2013, \apj, 775, 70,
  \dodoi{10.1088/0004-637X/775/1/70}

\bibitem[{{Hamlin} \& {Newman}(2013)}]{HamlinNewman2013}
{Hamlin}, N.~D., \& {Newman}, W.~I. 2013, \pre, 87, 043101,
  \dodoi{10.1103/PhysRevE.87.043101}

\bibitem[{Harten {et~al.}(1983)Harten, Lax, \& Leer}]{harten1983upstream}
Harten, A., Lax, P.~D., \& Leer, B.~v. 1983, SIAM review, 25, 35

\bibitem[{{Hawley} {et~al.}(2011){Hawley}, {Guan}, \&
  {Krolik}}]{HawleyGuanKrolik2011}
{Hawley}, J.~F., {Guan}, X., \& {Krolik}, J.~H. 2011, \apj, 738, 84,
  \dodoi{10.1088/0004-637X/738/1/84}

\bibitem[{{Hawley} \& {Krolik}(2006)}]{HawleyKrolik2006}
{Hawley}, J.~F., \& {Krolik}, J.~H. 2006, \apj, 641, 103,
  \dodoi{10.1086/500385}

\bibitem[{{Hawley} {et~al.}(2013){Hawley}, {Richers}, {Guan}, \&
  {Krolik}}]{HawleyRichers2013}
{Hawley}, J.~F., {Richers}, S.~A., {Guan}, X., \& {Krolik}, J.~H. 2013, \apj,
  772, 102, \dodoi{10.1088/0004-637X/772/2/102}

\bibitem[{{Hawley} {et~al.}(1984){Hawley}, {Smarr}, \& {Wilson}}]{Hawley84a}
{Hawley}, J.~F., {Smarr}, L.~L., \& {Wilson}, J.~R. 1984, Astrophys. J., 277,
  296, \dodoi{10.1086/161696}

\bibitem[{Heil {et~al.}(2012)Heil, Vaughan, \& Uttley}]{HeilVaughanEtAl2012}
Heil, L.~M., Vaughan, S., \& Uttley, P. 2012, \mnras, 422, 2620,
  \dodoi{10.1111/j.1365-2966.2012.20824.x}

\bibitem[{Hirose {et~al.}(2004)Hirose, Krolik, De~Villiers, \&
  Hawley}]{HiroseKrolikEtAl2004}
Hirose, S., Krolik, J.~H., De~Villiers, J.-P., \& Hawley, J.~F. 2004, \apj,
  606, 1083, \dodoi{10.1086/383184}

\bibitem[{Hirotani \& Okamoto(1998)}]{HirotaniOkamoto1998}
Hirotani, K., \& Okamoto, I. 1998, \apj, 497, 563, \dodoi{10.1086/305479}

\bibitem[{{Hogg} \& {Reynolds}(2016)}]{hoggReynolds2016}
{Hogg}, J.~D., \& {Reynolds}, C.~S. 2016, \apj, 826, 40,
  \dodoi{10.3847/0004-637X/826/1/40}

\bibitem[{{Hoshino}(2015)}]{Hoshino2015}
{Hoshino}, M. 2015, Physical Review Letters, 114, 061101,
  \dodoi{10.1103/PhysRevLett.114.061101}

\bibitem[{{Hubeny} \& {Hubeny}(1998)}]{HubenyHubeny1998}
{Hubeny}, I., \& {Hubeny}, V. 1998, \apj, 505, 558, \dodoi{10.1086/306207}

\bibitem[{{Igumenshchev} {et~al.}(2003){Igumenshchev}, {Narayan}, \&
  {Abramowicz}}]{igumenshchev2003}
{Igumenshchev}, I.~V., {Narayan}, R., \& {Abramowicz}, M.~A. 2003, \apj, 592,
  1042, \dodoi{10.1086/375769}

\bibitem[{Keppens {et~al.}(2012)Keppens, Meliani, van Marle, Delmont, Vlasis,
  \& van~der Holst}]{Keppens2012718}
Keppens, R., Meliani, Z., van Marle, A., {et~al.} 2012, Journal of
  Computational Physics, 231, 718 , \dodoi{10.1016/j.jcp.2011.01.020}

\bibitem[{{Koide} {et~al.}(1999){Koide}, {Shibata}, \&
  {Kudoh}}]{KoideShibata1999}
{Koide}, S., {Shibata}, K., \& {Kudoh}, T. 1999, \apj, 522, 727,
  \dodoi{10.1086/307667}

\bibitem[{{Komissarov}(2007)}]{Komissarov2007}
{Komissarov}, S.~S. 2007, \mnras, 382, 995,
  \dodoi{10.1111/j.1365-2966.2007.12448.x}

\bibitem[{{Kozlowski} {et~al.}(1978){Kozlowski}, {Jaroszynski}, \&
  {Abramowicz}}]{Kozlowski1978}
{Kozlowski}, M., {Jaroszynski}, M., \& {Abramowicz}, M.~A. 1978, Astron. and
  Astrophys., 63, 209

\bibitem[{Kritsuk {et~al.}(2011)Kritsuk, Nordlund, Collins, Padoan, Norman,
  Abel, Banerjee, Federrath, Flock, Lee, Li, M{\"u}ller, Teyssier, Ustyugov,
  Vogel, \& Xu}]{KritsukNordlundEtAl2011}
Kritsuk, A.~G., Nordlund, {\AA}., Collins, D., {et~al.} 2011, \apj, 737, 13,
  \dodoi{10.1088/0004-637X/737/1/13}

\bibitem[{{Krolik} {et~al.}(2005){Krolik}, {Hawley}, \&
  {Hirose}}]{KrolikHawley2005}
{Krolik}, J.~H., {Hawley}, J.~F., \& {Hirose}, S. 2005, \apj, 622, 1008,
  \dodoi{10.1086/427932}

\bibitem[{{Kunz} {et~al.}(2016){Kunz}, {Stone}, \& {Quataert}}]{KunzStone2016}
{Kunz}, M.~W., {Stone}, J.~M., \& {Quataert}, E. 2016, Physical Review Letters,
  117, 235101, \dodoi{10.1103/PhysRevLett.117.235101}

\bibitem[{{Kuo} {et~al.}(2014){Kuo}, {Asada}, {Rao}, {Nakamura}, {Algaba},
  {Liu}, {Inoue}, {Koch}, {Ho}, {Matsushita}, {Pu}, {Akiyama}, {Nishioka}, \&
  {Pradel}}]{KuoAsada2014}
{Kuo}, C.~Y., {Asada}, K., {Rao}, R., {et~al.} 2014, \apjl, 783, L33,
  \dodoi{10.1088/2041-8205/783/2/L33}

\bibitem[{Lesur \& Longaretti(2007)}]{LesurLongaretti2007}
Lesur, G., \& Longaretti, P.-Y. 2007, \mnras, 378, 1471,
  \dodoi{10.1111/j.1365-2966.2007.11888.x}

\bibitem[{{Leung} {et~al.}(2011){Leung}, {Gammie}, \& {Noble}}]{Leung2011}
{Leung}, P.~K., {Gammie}, C.~F., \& {Noble}, S.~C. 2011, Astrophys. J., 737,
  21, \dodoi{10.1088/0004-637X/737/1/21}

\bibitem[{{Levermore}(1984)}]{levermore84}
{Levermore}, C.~D. 1984, \jqsrt, 31, 149, \dodoi{10.1016/0022-4073(84)90112-2}

\bibitem[{Levinson \& Rieger(2011)}]{LevinsonRieger2011}
Levinson, A., \& Rieger, F. 2011, \apj, 730, 123,
  \dodoi{10.1088/0004-637X/730/2/123}

\bibitem[{{Liska} {et~al.}(2018{\natexlab{a}}){Liska}, {Hesp}, {Tchekhovskoy},
  {Ingram}, {van der Klis}, \& {Markoff}}]{LiskaHesp2018}
{Liska}, M., {Hesp}, C., {Tchekhovskoy}, A., {et~al.} 2018{\natexlab{a}},
  \mnras, 474, L81, \dodoi{10.1093/mnrasl/slx174}

\bibitem[{{Liska} {et~al.}(2018{\natexlab{b}}){Liska}, {Tchekhovskoy}, \&
  {Quataert}}]{LiskaTchekhovskoy2018}
{Liska}, M.~T.~P., {Tchekhovskoy}, A., \& {Quataert}, E. 2018{\natexlab{b}},
  ArXiv e-prints.
\newblock \doarXiv{1809.04608}

\bibitem[{L{\"{o}}ffler {et~al.}(2012)L{\"{o}}ffler, Faber, Bentivegna, Bode,
  Diener, Haas, Hinder, Mundim, Ott, Schnetter, Allen, Campanelli, \&
  Laguna}]{Loffler:2011ay}
L{\"{o}}ffler, F., Faber, J., Bentivegna, E., {et~al.} 2012, Class. Quantum
  Grav., 29, 115001, \dodoi{doi:10.1088/0264-9381/29/11/115001}

\bibitem[{{L\"{o}hner}(1987)}]{Loehner87}
{L\"{o}hner}, R. 1987, Computer Methods in Applied Mechanics and Engineering,
  61, 323, \dodoi{10.1016/0045-7825(87)90098-3}

\bibitem[{{Londrillo} \& {Del Zanna}(2000)}]{Londrillo2000}
{Londrillo}, P., \& {Del Zanna}, L. 2000, Astrophys. J., 530, 508,
  \dodoi{10.1086/308344}

\bibitem[{Londrillo \& Del~Zanna(2004)}]{londrillo2004divergence}
Londrillo, P., \& Del~Zanna, L. 2004, Journal of Computational Physics, 195, 17

\bibitem[{{Longaretti} \& {Lesur}(2010)}]{LongarettiLesur2010}
{Longaretti}, P.-Y., \& {Lesur}, G. 2010, \aap, 516, A51,
  \dodoi{10.1051/0004-6361/201014093}

\bibitem[{Lu {et~al.}(2018)Lu, Krichbaum, Roy, Fish, Doeleman, Johnson,
  Akiyama, Psaltis, Alef, Asada, Beaudoin, Bertarini, Blackburn, Blundell,
  Bower, Brinkerink, Broderick, Cappallo, Crew, Dexter, Dexter, Falcke, Freund,
  Friberg, Greer, Gurwell, Ho, Honma, Inoue, Kim, Lamb, Lindqvist, Macmahon,
  Marrone, Mart{\'{\i}}-Vidal, Menten, Moran, Nagar, Plambeck, Primiani,
  Rogers, Ros, Rottmann, SooHoo, Spilker, Stone, Strittmatter, Tilanus, Titus,
  Vertatschitsch, Wagner, Weintroub, Wright, Young, Zensus, \&
  Ziurys}]{LuKrichbaumEtAl2018}
Lu, R.-S., Krichbaum, T.~P., Roy, A.~L., {et~al.} 2018, \apj, 859, 60,
  \dodoi{10.3847/1538-4357/aabe2e}

\bibitem[{{Mahadevan} \& {Quataert}(1997)}]{MahadevanQuataert1997}
{Mahadevan}, R., \& {Quataert}, E. 1997, \apj, 490, 605

\bibitem[{{Markoff} {et~al.}(2001){Markoff}, {Falcke}, {Yuan}, \&
  {Biermann}}]{MarkoffFalcke2001}
{Markoff}, S., {Falcke}, H., {Yuan}, F., \& {Biermann}, P.~L. 2001, \aap, 379,
  L13, \dodoi{10.1051/0004-6361:20011346}

\bibitem[{{Marrone} {et~al.}(2007){Marrone}, {Moran}, {Zhao}, \&
  {Rao}}]{MarroneMoran2007}
{Marrone}, D.~P., {Moran}, J.~M., {Zhao}, J., \& {Rao}, R. 2007, \apjl, 654,
  L57, \dodoi{10.1086/510850}

\bibitem[{{Mart{\'{\i}}} \& {M{\"u}ller}(2015)}]{2015LRCA....1....3M}
{Mart{\'{\i}}}, J.~M., \& {M{\"u}ller}, E. 2015, Living Reviews in
  Computational Astrophysics, 1, 3, \dodoi{10.1007/lrca-2015-3}

\bibitem[{{McKinney}(2006)}]{mckinney2006}
{McKinney}, J.~C. 2006, \mnras, 367, 1797,
  \dodoi{10.1111/j.1365-2966.2006.10087.x}

\bibitem[{{McKinney} \& {Blandford}(2009)}]{mckinney2009}
{McKinney}, J.~C., \& {Blandford}, R.~D. 2009, \mnras, 394, L126,
  \dodoi{10.1111/j.1745-3933.2009.00625.x}

\bibitem[{{McKinney} \& {Gammie}(2004)}]{McKinneyGammie2004}
{McKinney}, J.~C., \& {Gammie}, C.~F. 2004, \apj, 611, 977,
  \dodoi{10.1086/422244}

\bibitem[{{McKinney} {et~al.}(2012){McKinney}, {Tchekhovskoy}, \&
  {Blandford}}]{McKinneyTchekhovskoy2012}
{McKinney}, J.~C., {Tchekhovskoy}, A., \& {Blandford}, R.~D. 2012, \mnras, 423,
  3083, \dodoi{10.1111/j.1365-2966.2012.21074.x}

\bibitem[{{McKinney} {et~al.}(2014){McKinney}, {Tchekhovskoy}, {Sadowski}, \&
  {Narayan}}]{McKinneyTchekhovskoy2014}
{McKinney}, J.~C., {Tchekhovskoy}, A., {Sadowski}, A., \& {Narayan}, R. 2014,
  \mnras, 441, 3177, \dodoi{10.1093/mnras/stu762}

\bibitem[{{Meliani} {et~al.}(2016){Meliani}, {Grandcl{\'e}ment}, {Casse},
  {Vincent}, {Straub}, \& {Dauvergne}}]{Meliani2016}
{Meliani}, Z., {Grandcl{\'e}ment}, P., {Casse}, F., {et~al.} 2016, Classical
  and Quantum Gravity, 33, 155010, \dodoi{10.1088/0264-9381/33/15/155010}

\bibitem[{{Mignone}(2014)}]{Mignone2014}
{Mignone}, A. 2014, Journal of Computational Physics, 270, 784,
  \dodoi{10.1016/j.jcp.2014.04.001}

\bibitem[{{Mizuno} {et~al.}(2006){Mizuno}, {Nishikawa}, {Koide}, {Hardee}, \&
  {Fishman}}]{Mizuno06}
{Mizuno}, Y., {Nishikawa}, K.-I., {Koide}, S., {Hardee}, P., \& {Fishman},
  G.~J. 2006, ArXiv Astrophysics e-prints

\bibitem[{{Mo{\'s}cibrodzka} \& {Falcke}(2013)}]{MoscibrodzkaFalcke2013}
{Mo{\'s}cibrodzka}, M., \& {Falcke}, H. 2013, \aap, 559, L3,
  \dodoi{10.1051/0004-6361/201322692}

\bibitem[{{Mo{\'s}cibrodzka} {et~al.}(2016{\natexlab{a}}){Mo{\'s}cibrodzka},
  {Falcke}, \& {Noble}}]{MoscibrodzkaFalcke2016}
{Mo{\'s}cibrodzka}, M., {Falcke}, H., \& {Noble}, S. 2016{\natexlab{a}}, \aap,
  596, A13, \dodoi{10.1051/0004-6361/201629157}

\bibitem[{{Mo{\'s}cibrodzka} {et~al.}(2016{\natexlab{b}}){Mo{\'s}cibrodzka},
  {Falcke}, \& {Shiokawa}}]{MoscibrodzkaFalcke2016a}
{Mo{\'s}cibrodzka}, M., {Falcke}, H., \& {Shiokawa}, H. 2016{\natexlab{b}},
  \aap, 586, A38, \dodoi{10.1051/0004-6361/201526630}

\bibitem[{{Mo{\'s}cibrodzka} {et~al.}(2011){Mo{\'s}cibrodzka}, {Gammie},
  {Dolence}, \& {Shiokawa}}]{Moscibrodzka2011}
{Mo{\'s}cibrodzka}, M., {Gammie}, C.~F., {Dolence}, J.~C., \& {Shiokawa}, H.
  2011, \apj, 735, 9, \dodoi{10.1088/0004-637X/735/1/9}

\bibitem[{Mo{\'s}cibrodzka {et~al.}(2009)Mo{\'s}cibrodzka, Gammie, Dolence,
  Shiokawa, \& Leung}]{MoscibrodzkaGammieEtAl2009}
Mo{\'s}cibrodzka, M., Gammie, C.~F., Dolence, J.~C., Shiokawa, H., \& Leung,
  P.~K. 2009, \apj, 706, 497, \dodoi{10.1088/0004-637X/706/1/497}

\bibitem[{{Mossoux} \& {Grosso}(2017)}]{MossouxGrosso2017}
{Mossoux}, E., \& {Grosso}, N. 2017, \aap, 604, A85,
  \dodoi{10.1051/0004-6361/201629778}

\bibitem[{{Nakamura} {et~al.}(2018){Nakamura}, {Asada}, {Hada}, {Pu}, {Noble},
  {Tseng}, {Toma}, {Kino}, {Nagai}, {Takahashi}, {Algaba}, {Orienti},
  {Akiyama}, {Doi}, {Giovannini}, {Giroletti}, {Honma}, {Koyama}, {Lico},
  {Niinuma}, \& {Tazaki}}]{Nakamura2018}
{Nakamura}, M., {Asada}, K., {Hada}, K., {et~al.} 2018, \apj, 868, 146,
  \dodoi{10.3847/1538-4357/aaeb2d}

\bibitem[{{Narayan} {et~al.}(2003){Narayan}, {Igumenshchev}, \&
  {Abramowicz}}]{NarayanIgumenshchev2003}
{Narayan}, R., {Igumenshchev}, I.~V., \& {Abramowicz}, M.~A. 2003, \pasj, 55,
  L69, \dodoi{10.1093/pasj/55.6.L69}

\bibitem[{Narayan {et~al.}(1998)Narayan, Mahadevan, Grindlay, Popham, \&
  Gammie}]{NarayanMahadevanEtAl1998a}
Narayan, R., Mahadevan, R., Grindlay, J.~E., Popham, R.~G., \& Gammie, C. 1998,
  \apj, 492, 554, \dodoi{10.1086/305070}

\bibitem[{{Narayan} {et~al.}(2012){Narayan}, {Sadowski}, {Penna}, \&
  {Kulkarni}}]{NarayanSadowski2012}
{Narayan}, R., {Sadowski}, A., {Penna}, R.~F., \& {Kulkarni}, A.~K. 2012,
  \mnras, 426, 3241, \dodoi{10.1111/j.1365-2966.2012.22002.x}

\bibitem[{{Narayan} \& {Yi}(1995)}]{NarayanYi1995}
{Narayan}, R., \& {Yi}, I. 1995, \apj, 452, 710, \dodoi{10.1086/176343}

\bibitem[{{Narayan} {et~al.}(1995){Narayan}, {Yi}, \&
  {Mahadevan}}]{Narayan1995}
{Narayan}, R., {Yi}, I., \& {Mahadevan}, R. 1995, \nat, 374, 623,
  \dodoi{10.1038/374623a0}

\bibitem[{{Noble} {et~al.}(2006){Noble}, {Gammie}, {McKinney}, \& {Del
  Zanna}}]{Noble2006}
{Noble}, S.~C., {Gammie}, C.~F., {McKinney}, J.~C., \& {Del Zanna}, L. 2006,
  Astrophys. J., 641, 626, \dodoi{10.1086/500349}

\bibitem[{{Noble} {et~al.}(2009){Noble}, {Krolik}, \& {Hawley}}]{Noble2009}
{Noble}, S.~C., {Krolik}, J.~H., \& {Hawley}, J.~F. 2009, Astrophys. J., 692,
  411, \dodoi{10.1088/0004-637X/692/1/411}

\bibitem[{{Noble} {et~al.}(2010){Noble}, {Krolik}, \&
  {Hawley}}]{NobleKrolik2010}
---. 2010, \apj, 711, 959, \dodoi{10.1088/0004-637X/711/2/959}

\bibitem[{{Noble} {et~al.}(2012){Noble}, {Mundim}, {Nakano}, {Krolik},
  {Campanelli}, {Zlochower}, \& {Yunes}}]{Noble2012}
{Noble}, S.~C., {Mundim}, B.~C., {Nakano}, H., {et~al.} 2012, arXiv:1204.1073.
\newblock \doarXiv{1204.1073}

\bibitem[{Novikov \& Thorne(1973)}]{NovikovThorne1973}
Novikov, I.~D., \& Thorne, K.~S. 1973, in Black Holes (Les Astres Occlus), ed.
  C.~{Dewitt} \& B.~S. {Dewitt}, 343--450.
\newblock \url{http://adsabs.harvard.edu/abs/1973blho.conf..343N}

\bibitem[{{Nowak} {et~al.}(2012){Nowak}, {Neilsen}, {Markoff}, {Baganoff},
  {Porquet}, {Grosso}, {Levin}, {Houck}, {Eckart}, {Falcke}, {Ji}, {Miller}, \&
  {Wang}}]{NowakNeilsen2012}
{Nowak}, M.~A., {Neilsen}, J., {Markoff}, S.~B., {et~al.} 2012, \apj, 759, 95,
  \dodoi{10.1088/0004-637X/759/2/95}

\bibitem[{{Olivares} {et~al.}(2019){Olivares}, {Porth}, {Davelaar}, {Most},
  {Fromm}, {Mizuno}, {Younsi}, \& {Rezzolla}}]{2019arXiv190610795O}
{Olivares}, H., {Porth}, O., {Davelaar}, J., {et~al.} 2019, arXiv e-prints,
  arXiv:1906.10795.
\newblock \doarXiv{1906.10795}

\bibitem[{{Olivares} {et~al.}(2018){Olivares}, {Porth}, \&
  {Mizuno}}]{OlivaresPorth2018}
{Olivares}, H., {Porth}, O., \& {Mizuno}, Y. 2018, ArXiv e-prints.
\newblock \doarXiv{1802.00860}

\bibitem[{Pareschi \& Russo(2005)}]{Pareschi2005}
Pareschi, L., \& Russo, G. 2005, Journal of Scientific Computing, 25, 129,
  \dodoi{10.1007/BF02728986}

\bibitem[{{Parfrey} {et~al.}(2018){Parfrey}, {Philippov}, \&
  {Cerutti}}]{ParfreyPhilippovEtAl2019}
{Parfrey}, K., {Philippov}, A., \& {Cerutti}, B. 2018, arXiv e-prints.
\newblock \doarXiv{1810.03613}

\bibitem[{Paschalidis {et~al.}(2012)Paschalidis, Etienne, \&
  Shapiro}]{Paschalidis:2012ff}
Paschalidis, V., Etienne, Z.~B., \& Shapiro, S.~L. 2012, Phys. Rev. D, 86,
  064032

\bibitem[{Paschalidis {et~al.}(2011)Paschalidis, Liu, Etienne, \&
  Shapiro}]{Paschalidis:2011ez}
Paschalidis, V., Liu, Y.~T., Etienne, Z.~B., \& Shapiro, S.~L. 2011, Phys. Rev.
  D, 84, 104032

\bibitem[{Paschalidis {et~al.}(2015)Paschalidis, Ruiz, \&
  Shapiro}]{Paschalidis:2014qra}
Paschalidis, V., Ruiz, M., \& Shapiro, S.~L. 2015, Astrophys. J., 806, L14

\bibitem[{{Penna} {et~al.}(2010){Penna}, {McKinney}, {Narayan}, {Tchekhovskoy},
  {Shafee}, \& {McClintock}}]{PennaMcKinney2010}
{Penna}, R.~F., {McKinney}, J.~C., {Narayan}, R., {et~al.} 2010, \mnras, 408,
  752, \dodoi{10.1111/j.1365-2966.2010.17170.x}

\bibitem[{Penna {et~al.}(2013)Penna, S{\c a}dowski, Kulkarni, \&
  Narayan}]{PennaScadowskiEtAl2013}
Penna, R.~F., S{\c a}dowski, A., Kulkarni, A.~K., \& Narayan, R. 2013, \mnras,
  428, 2255, \dodoi{10.1093/mnras/sts185}

\bibitem[{Phinney(1995)}]{Phinney1995}
Phinney, E.~S. 1995, in Bulletin of the American Astronomical Society, Vol.~27,
  American Astronomical Society Meeting Abstracts, 1450.
\newblock \url{http://adsabs.harvard.edu/abs/1995AAS...18711801P}

\bibitem[{{Porth} {et~al.}(2017){Porth}, {Olivares}, {Mizuno}, {Younsi},
  {Rezzolla}, {Moscibrodzka}, {Falcke}, \& {Kramer}}]{PorthOlivares2017}
{Porth}, O., {Olivares}, H., {Mizuno}, Y., {et~al.} 2017, Computational
  Astrophysics and Cosmology, 4, 1, \dodoi{10.1186/s40668-017-0020-2}

\bibitem[{{Porth} {et~al.}(2014){Porth}, {Xia}, {Hendrix}, {Moschou}, \&
  {Keppens}}]{PorthXia2014}
{Porth}, O., {Xia}, C., {Hendrix}, T., {Moschou}, S.~P., \& {Keppens}, R. 2014,
  \apjs, 214, 4, \dodoi{10.1088/0067-0049/214/1/4}

\bibitem[{{Pu} {et~al.}(2015){Pu}, {Nakamura}, {Hirotani}, {Mizuno}, {Wu}, \&
  {Asada}}]{PuNakamura2015}
{Pu}, H.-Y., {Nakamura}, M., {Hirotani}, K., {et~al.} 2015, \apj, 801, 56,
  \dodoi{10.1088/0004-637X/801/1/56}

\bibitem[{{Qian} {et~al.}(2016){Qian}, {Fendt}, {Noble}, \&
  {Bugli}}]{QianFendt2016}
{Qian}, Q., {Fendt}, C., {Noble}, S., \& {Bugli}, M. 2016, ArXiv e-prints.
\newblock \doarXiv{1610.04445}

\bibitem[{{Radice} \& {Rezzolla}(2013)}]{RadiceRezzollaProceedings2013}
{Radice}, D., \& {Rezzolla}, L. 2013, in Astronomical Society of the Pacific
  Conference Series, Vol. 474, Numerical Modeling of Space Plasma Flows
  (ASTRONUM2012), ed. N.~V. {Pogorelov}, E.~{Audit}, \& G.~P. {Zank}, 25

\bibitem[{{Radice} {et~al.}(2014){Radice}, {Rezzolla}, \&
  {Galeazzi}}]{Radice2013b}
{Radice}, D., {Rezzolla}, L., \& {Galeazzi}, F. 2014, Mon. Not. R. Astron. Soc.
  L., 437, L46, \dodoi{10.1093/mnrasl/slt137}

\bibitem[{{Ressler} {et~al.}(2017){Ressler}, {Tchekhovskoy}, {Quataert}, \&
  {Gammie}}]{Ressler2017}
{Ressler}, S.~M., {Tchekhovskoy}, A., {Quataert}, E., \& {Gammie}, C.~F. 2017,
  \mnras, 467, 3604, \dodoi{10.1093/mnras/stx364}

\bibitem[{Reynolds \& Miller(2009)}]{ReynoldsMiller2009}
Reynolds, C.~S., \& Miller, M.~C. 2009, \apj, 692, 869,
  \dodoi{10.1088/0004-637X/692/1/869}

\bibitem[{{Rezzolla} \& {Zanotti}(2013)}]{Rezzolla_book:2013}
{Rezzolla}, L., \& {Zanotti}, O. 2013, Relativistic Hydrodynamics (Oxford, UK:
  Oxford University Press), \dodoi{10.1093/acprof:oso/9780198528906.001.0001}

\bibitem[{{Ripperda} {et~al.}(2019{\natexlab{a}}){Ripperda}, {Porth}, {Sironi},
  \& {Keppens}}]{2019MNRAS.485..299R}
{Ripperda}, B., {Porth}, O., {Sironi}, L., \& {Keppens}, R. 2019{\natexlab{a}},
  \mnras, 485, 299, \dodoi{10.1093/mnras/stz387}

\bibitem[{{Ripperda} {et~al.}(2019{\natexlab{b}}){Ripperda}, {Bacchini},
  {Porth}, {Most}, {Olivares}, {Nathanail}, {Rezzolla}, {Teunissen}, \&
  {Keppens}}]{2019arXiv190707197R}
{Ripperda}, B., {Bacchini}, F., {Porth}, O., {et~al.} 2019{\natexlab{b}}, arXiv
  e-prints, arXiv:1907.07197.
\newblock \doarXiv{1907.07197}

\bibitem[{{Rusanov}(1961)}]{Rusanov1961}
{Rusanov}, V.~V. 1961, Zh. Vychisl. Mat. Mat. Fiz., 1, 267

\bibitem[{{Ryan} {et~al.}(2015){Ryan}, {Dolence}, \&
  {Gammie}}]{RyanDolence2015}
{Ryan}, B.~R., {Dolence}, J.~C., \& {Gammie}, C.~F. 2015, \apj, 807, 31,
  \dodoi{10.1088/0004-637X/807/1/31}

\bibitem[{{Ryan} {et~al.}(2017){Ryan}, {Gammie}, {Fromang}, \&
  {Kestener}}]{RyanGammie2017}
{Ryan}, B.~R., {Gammie}, C.~F., {Fromang}, S., \& {Kestener}, P. 2017, \apj,
  840, 6, \dodoi{10.3847/1538-4357/aa6a52}

\bibitem[{{Ryan} {et~al.}(2018){Ryan}, {Ressler}, {Dolence}, {Gammie}, \&
  {Quataert}}]{RyanRessler2018}
{Ryan}, B.~R., {Ressler}, S.~M., {Dolence}, J.~C., {Gammie}, C., \& {Quataert},
  E. 2018, \apj, 864, 126, \dodoi{10.3847/1538-4357/aad73a}

\bibitem[{{Sano} {et~al.}(2004){Sano}, {Inutsuka}, {Turner}, \&
  {Stone}}]{SanoInutsoka2004}
{Sano}, T., {Inutsuka}, S.-i., {Turner}, N.~J., \& {Stone}, J.~M. 2004, \apj,
  605, 321, \dodoi{10.1086/382184}

\bibitem[{{S{\c a}dowski} \& {Narayan}(2015)}]{sadowski15b}
{S{\c a}dowski}, A., \& {Narayan}, R. 2015, \mnras, 454, 2372,
  \dodoi{10.1093/mnras/stv2022}

\bibitem[{{S{\c a}dowski} {et~al.}(2014){S{\c a}dowski}, {Narayan}, {McKinney},
  \& {Tchekhovskoy}}]{sadowski14}
{S{\c a}dowski}, A., {Narayan}, R., {McKinney}, J.~C., \& {Tchekhovskoy}, A.
  2014, \mnras, 439, 503, \dodoi{10.1093/mnras/stt2479}

\bibitem[{{S{\c a}dowski} {et~al.}(2015){S{\c a}dowski}, {Narayan},
  {Tchekhovskoy}, {Abarca}, {Zhu}, \& {McKinney}}]{sadowski15a}
{S{\c a}dowski}, A., {Narayan}, R., {Tchekhovskoy}, A., {et~al.} 2015, \mnras,
  447, 49, \dodoi{10.1093/mnras/stu2387}

\bibitem[{{S{\c a}dowski} {et~al.}(2013){S{\c a}dowski}, {Narayan},
  {Tchekhovskoy}, \& {Zhu}}]{sadowski13}
{S{\c a}dowski}, A., {Narayan}, R., {Tchekhovskoy}, A., \& {Zhu}, Y. 2013,
  \mnras, 429, 3533, \dodoi{10.1093/mnras/sts632}

\bibitem[{{S{\c a}dowski} {et~al.}(2017){S{\c a}dowski}, {Wielgus}, {Narayan},
  {Abarca}, {McKinney}, \& {Chael}}]{sadowski17}
{S{\c a}dowski}, A., {Wielgus}, M., {Narayan}, R., {et~al.} 2017, \mnras, 466,
  705, \dodoi{10.1093/mnras/stw3116}

\bibitem[{{Shiokawa} {et~al.}(2012){Shiokawa}, {Dolence}, {Gammie}, \&
  {Noble}}]{Shiokawa2012}
{Shiokawa}, H., {Dolence}, J.~C., {Gammie}, C.~F., \& {Noble}, S.~C. 2012,
  Astrophys. J., 744, 187, \dodoi{10.1088/0004-637X/744/2/187}

\bibitem[{Simon {et~al.}(2012)Simon, Beckwith, \&
  Armitage}]{SimonBeckwithEtAl2012}
Simon, J.~B., Beckwith, K., \& Armitage, P.~J. 2012, \mnras, 422, 2685,
  \dodoi{10.1111/j.1365-2966.2012.20835.x}

\bibitem[{Simon \& Hawley(2009)}]{SimonHawley2009}
Simon, J.~B., \& Hawley, J.~F. 2009, \apj, 707, 833,
  \dodoi{10.1088/0004-637X/707/1/833}

\bibitem[{{Sorathia} {et~al.}(2012){Sorathia}, {Reynolds}, {Stone}, \&
  {Beckwith}}]{SorathiaReynolds2012}
{Sorathia}, K.~A., {Reynolds}, C.~S., {Stone}, J.~M., \& {Beckwith}, K. 2012,
  \apj, 749, 189, \dodoi{10.1088/0004-637X/749/2/189}

\bibitem[{Spiteri \& Ruuth(2002)}]{Spiteri2002}
Spiteri, R.~J., \& Ruuth, S.~J. 2002, SIAM J. Numerical Analysis, 40, 469

\bibitem[{Stepney \& Guilbert(1983)}]{StepneyGuilbert1983}
Stepney, S., \& Guilbert, P.~W. 1983, \mnras, 204, 1269,
  \dodoi{10.1093/mnras/204.4.1269}

\bibitem[{{Stone} {et~al.}(2008){Stone}, {Gardiner}, {Teuben}, {Hawley}, \&
  {Simon}}]{2008ApJS..178..137S}
{Stone}, J.~M., {Gardiner}, T.~A., {Teuben}, P., {Hawley}, J.~F., \& {Simon},
  J.~B. 2008, \apjs, 178, 137, \dodoi{10.1086/588755}

\bibitem[{Takahashi {et~al.}(2016)Takahashi, Ohsuga, Kawashima, \&
  Sekiguchi}]{TakahashiOhsugaEtAl2016}
Takahashi, H.~R., Ohsuga, K., Kawashima, T., \& Sekiguchi, Y. 2016, \apj, 826,
  23, \dodoi{10.3847/0004-637X/826/1/23}

\bibitem[{{Takahashi} {et~al.}(1990){Takahashi}, {Nitta}, {Tatematsu}, \&
  {Tomimatsu}}]{TakahashiNittaEtAl1990}
{Takahashi}, M., {Nitta}, S., {Tatematsu}, Y., \& {Tomimatsu}, A. 1990, \apj,
  363, 206, \dodoi{10.1086/169331}

\bibitem[{{Takahashi}(2008)}]{Takahashi2008}
{Takahashi}, R. 2008, \mnras, 383, 1155,
  \dodoi{10.1111/j.1365-2966.2007.12612.x}

\bibitem[{{Tchekhovskoy} {et~al.}(2008){Tchekhovskoy}, {McKinney}, \&
  {Narayan}}]{Tchekhovskoy2008}
{Tchekhovskoy}, A., {McKinney}, J.~C., \& {Narayan}, R. 2008, \mnras, 388, 551,
  \dodoi{10.1111/j.1365-2966.2008.13425.x}

\bibitem[{{Tchekhovskoy} {et~al.}(2012){Tchekhovskoy}, {McKinney}, \&
  {Narayan}}]{TchekhovskoyMcKinney2012}
{Tchekhovskoy}, A., {McKinney}, J.~C., \& {Narayan}, R. 2012, in Journal of
  Physics Conference Series, Vol. 372, Journal of Physics Conference Series,
  012040

\bibitem[{{Tchekhovskoy} {et~al.}(2011){Tchekhovskoy}, {Narayan}, \&
  {McKinney}}]{TchekhovskoyNarayan2011}
{Tchekhovskoy}, A., {Narayan}, R., \& {McKinney}, J.~C. 2011, \mnras, 418, L79,
  \dodoi{10.1111/j.1745-3933.2011.01147.x}

\bibitem[{{T{\'o}th}(2000)}]{toth2000}
{T{\'o}th}, G. 2000, Journal of Computational Physics, 161, 605,
  \dodoi{10.1006/jcph.2000.6519}

\bibitem[{Uttley \& McHardy(2001)}]{UttleyMcHardy2001}
Uttley, P., \& McHardy, I.~M. 2001, \mnras, 323, L26,
  \dodoi{10.1046/j.1365-8711.2001.04496.x}

\bibitem[{van Leer(1977)}]{Leer1977}
van Leer, B. 1977, Journal of Computational Physics, 23, 276,
  \dodoi{10.1016/0021-9991(77)90095-X}

\bibitem[{van Leer(2006)}]{vanLeer2006}
van Leer, B.~J. 2006, Communications in Computational Physics, 1, 192

\bibitem[{{White} {et~al.}(2016){White}, {Stone}, \& {Gammie}}]{White2016}
{White}, C.~J., {Stone}, J.~M., \& {Gammie}, C.~F. 2016, Astrophys. J.s, 225,
  22, \dodoi{10.3847/0067-0049/225/2/22}

\bibitem[{White {et~al.}(2019)White, Stone, \& Quataert}]{WhiteStoneEtAl2019}
White, C.~J., Stone, J.~M., \& Quataert, E. 2019, arXiv e-prints.
\newblock \doarXiv{1903.01509}

\bibitem[{{Witzel} {et~al.}(2018){Witzel}, {Martinez}, {Hora}, {Willner},
  {Morris}, {Gammie}, {Becklin}, {Ashby}, {Baganoff}, {Carey}, {Do}, {Fazio},
  {Ghez}, {Glaccum}, {Haggard}, {Herrero-Illana}, {Ingalls}, {Narayan}, \&
  {Smith}}]{WitzelMartinez2018}
{Witzel}, G., {Martinez}, G., {Hora}, J., {et~al.} 2018, \apj, 863, 15,
  \dodoi{10.3847/1538-4357/aace62}

\bibitem[{Xia {et~al.}(2018)Xia, Teunissen, El~Mellah, Chan{\'e}, \&
  Keppens}]{XiaTeunissenEtAl2018}
Xia, C., Teunissen, J., El~Mellah, I., Chan{\'e}, E., \& Keppens, R. 2018,
  \apjs, 234, 30, \dodoi{10.3847/1538-4365/aaa6c8}

\bibitem[{{Younsi} \& et~al.(2019 in prep.)}]{YounsiEtAl2019}
{Younsi}, Z., \& et~al. 2019 in prep., \apjs, ?, ?

\bibitem[{{Yuan} {et~al.}(2002){Yuan}, {Markoff}, \&
  {Falcke}}]{YuanMarkoff2002a}
{Yuan}, F., {Markoff}, S., \& {Falcke}, H. 2002, \aap, 383, 854,
  \dodoi{10.1051/0004-6361:20011709}

\bibitem[{{Yuan} \& {Narayan}(2014)}]{YuanNarayan2014}
{Yuan}, F., \& {Narayan}, R. 2014, \araa, 52, 529,
  \dodoi{10.1146/annurev-astro-082812-141003}

\bibitem[{{Zanotti} \& {Dumbser}(2015)}]{Zanotti2015}
{Zanotti}, O., \& {Dumbser}, M. 2015, Computer Physics Communications, 188,
  110, \dodoi{10.1016/j.cpc.2014.11.015}

\bibitem[{{Zanotti} {et~al.}(2003){Zanotti}, {Rezzolla}, \& {Font}}]{Zanotti03}
{Zanotti}, O., {Rezzolla}, L., \& {Font}, J.~A. 2003, Mon. Not. R. Astron.
  Soc., 341, 832, \dodoi{10.1046/j.1365-8711.2003.06474.x}

\bibitem[{{Zilh{\~a}o} \& {Noble}(2014)}]{ZilhaoNoble2014}
{Zilh{\~a}o}, M., \& {Noble}, S.~C. 2014, Classical and Quantum Gravity, 31,
  065013, \dodoi{10.1088/0264-9381/31/6/065013}

\end{thebibliography}

\end{document}